\documentclass[11pt,a4paper]{article}
\usepackage{amsmath}
\usepackage{graphicx}
\usepackage{xcolor}
\usepackage[caption=false]{subfig}
\usepackage{mathrsfs,mathtools}
\usepackage{physics,amssymb}
\usepackage{siunitx}
\usepackage{bm}
\usepackage{float}
\usepackage{braket}
\usepackage{listings}

\usepackage{cases}
\usepackage{comment}
\usepackage{soul}
\usepackage{cancel}
\usepackage{cases}
\usepackage{booktabs}  % for better-looking tables
\usepackage[utf8]{inputenc}
\usepackage{url}
\usepackage{float}
\usepackage{longtable}
\usepackage[normalem]{ulem}
\usepackage{xspace}
\usepackage{aas_macros}
\setstcolor{red}
\usepackage{jcappub}
\hypersetup{colorlinks=true
,urlcolor=DARKBLUE
,anchorcolor=DARKBLUE
,citecolor=DARKBLUE
,filecolor=DARKBLUE
,linkcolor=DARKBLUE
,menucolor=DARKBLUE
,linktocpage=true
,pdfproducer=medialab
}

\usepackage{graphicx,epsfig,color,xcolor}
\usepackage[english]{babel}
\usepackage{amssymb,amsfonts,amsmath,physics}
\usepackage{mathtools}
\usepackage{siunitx}
\usepackage{verbatim}
\usepackage{ulem}
\usepackage{soul}
\usepackage{lipsum, babel}
\usepackage{acronym}
\usepackage{aas_macros}

\usepackage{hyperref}

\newcommand{\be}{\begin{equation}} \newcommand{\ee}{\end{equation}}
\newcommand{\bea}{\begin{eqnarray}} \newcommand{\eea}{\end{eqnarray}}

\newcommand{\Imag}{\textrm{Im}}
\newcommand{\Real}{\textrm{Re}}

\newcommand{\GW}{\mathrm{GW}}

\usepackage{fancyhdr}
\usepackage{amsmath}
\usepackage{mathrsfs}
\usepackage{float}
\usepackage{xcolor}
\usepackage{graphicx, epsfig, amssymb} 
\usepackage{amsmath, amsfonts}
\usepackage{bm} 
\usepackage{tensor}
\usepackage{soul}
\usepackage{hyperref}
\usepackage{physics}

\definecolor{MONZA}{HTML}{CF000F}
\definecolor{DARKBLUE}{HTML}{00008b}
\definecolor{DARKMAGENTA}{HTML}{8b008b}
\definecolor{DARKCYAN}{HTML}{008B8B}
\definecolor{DARKORANGE}{HTML}{FF8C00}
\definecolor{OBSERVATORY}{HTML}{049372}
\definecolor{GREENBAMBOO}{HTML}{006442}
\definecolor{TURQUOISE}{HTML}{36D7B7}
\definecolor{JUNGLEGREEN}{HTML}{26C281}

\newcommand{\albert}[1]{\textcolor{red}{\sffamily [Albert: #1]}}

\begin{document}

%\title{Relics from USR inflation}

\title{% \textcolor{blue}
{Inflationary relics from an Ultra-Slow-Roll plateau}}
\author[a,b]{Albert Escriv\`a}

\author[c]{, Jaume Garriga}

\author[d,e,f]{and Shi Pi}

\affiliation[a]{Institute for Advanced Research, Nagoya University, \\
Furo-cho Chikusa-ku, Nagoya 464-8601, Japan}
\affiliation[b]{Department of Physics, Nagoya University, \\
Furo-cho Chikusa-ku, Nagoya 464-8602, Japan}

\affiliation[c]{Departament de F\'isica Qu\`antica i Astrofi\'sica, i  Institut  de  Ci\`encies  del  Cosmos (ICCUB), Universitat de Barcelona. Mart\'i i Franqu\'es 1, 08028 Barcelona, Spain.}

\affiliation[d]{Institute of Theoretical Physics, Chinese Academy of Sciences, Beijing 100190, China}
\affiliation[e]{Center for High Energy Physics, Peking University, Beijing 100871, China}
\affiliation[f]{Kavli Institute for the Physics and Mathematics of the Universe (WPI), The University of Tokyo, Kashiwa, Chiba 277-8583, Japan}

\emailAdd{escriva.manas.alberto.k0@f.mail.nagoya-u.ac.jp}
\emailAdd{jaume.garriga@ub.edu}
\emailAdd{shi.pi@itp.ac.cn}

\date{\today}
\abstract{We investigate the formation of primordial black holes (PBHs) in inflationary scenarios featuring an ultra–slow-roll (USR) plateau with a sharp transition to slow roll. We focus on two coexisting production channels: PBHs originating from relic vacuum bubbles where the inflaton got trapped on the plateau, and PBHs arising from standard adiabatic density perturbations. From detailed numerical simulations we find that the bubbles are generically surrounded by type-II curvature fluctuations. Special attention is given to the distribution of initial conditions, including the relevant mean profiles and shape dispersion around them.
For the adiabatic channel, we extend the logarithmic template formula $\zeta[\zeta_G]$, which maps the Gaussian curvature perturbation to the fully non-Gaussian one while incorporating mode evolution, and we compare this with numerical results obtained using the $\delta N$ formalism. While the template departs from numerical results near its logarithmic divergence,
it still provides accurate threshold values for PBH formation in the parameter range relevant to our analysis.
Finally, we compute the PBH mass functions for both channels. We find that the adiabatic channel dominates over the bubble-induced channel by a factor $\sim \mathcal{O}(10-10^{2})$, and that both contributions are largely dominated by the mean profiles.} 

\maketitle
\flushbottom

\acresetall

\acrodef{GW}{gravitational wave}
\acrodef{CMB}{cosmic microwave background}
\acrodef{PBH}{primordial black hole}
\acrodef{DM}{Dark Matter}
\acrodef{FLRW}{Friedmann‐-Lema\^itre--Robertson--Walker}

\section{Introduction}
Primordial black holes (PBHs) may have formed the early Universe during the radiation-dominated epoch \cite{1967SvA....10..602Z,Hawking:1971ei,1974MNRAS.168..399C,1975ApJ...201....1C,1979A&A....80..104N} (see \cite{Escriva:2022duf} for an overview), and represent one of the most intriguing possibilities for the dark matter component of the Universe \cite{Chapline:1975ojl,2016PhRvD..94h3504C,2017JPhCS.840a2032G,2020ARNPS..70..355C,Carr:2020gox,2021JPhG...48d3001G,Carr:2021bzv}.
Their formation, governed by general relativity, depends sensitively on the statistical properties of primordial perturbations, originating during inflation or through other early-Universe mechanisms. Although no PBHs have been detected so far, upcoming gravitational-wave observations may significantly improve the prospects for testing their existence
%Their formation process is entirely governed by general relativity, yet their formation depends sensitively on the properties of the primordial fluctuations. These initial conditions could have been generated either during inflation or through processes associated with several mechanisms in the early Universe. Although it seems that no PBHs have been detected so far, upcoming gravitational-wave experiments offer a promising avenue to probe their possible existence 
\cite{2018CQGra..35f3001S,2016PhRvX...6d1015A,2021arXiv211103606T,Murgia:2019duy,Luo:2025ewp}.

A standard PBH formation mechanism is the collapse of large adiabatic curvature perturbations generated during inflation. This requires a substantial enhancement of the power spectrum at scales far smaller than those probed by the CMB. Once such overdensities re-enter the cosmological horizon, they may collapse into black holes. The PBH abundance depends exponentially on the threshold for critical collapse, which in turn depends on the profile of the perturbations. Accurately determining such thresholds is therefore crucial, and generally requires relativistic numerical simulations
%One of the most standard mechanisms for PBH formation is the collapse of adiabatic fluctuations generated during inflation, typically resulting from a sufficient enhancement of the power spectrum at scales much smaller than those probed by the CMB. Those fluctuations may collapse forming black holes once reenter the cosmological hoizon. The determination of the critical conditions for PBH formation is crucial because the abundance of the resulting PBHs is exponentially sensitive to the formation conditions. Moreover, determining these conditions generally requires relativistic numerical simulations 
\cite{2022Univ....8...66E}.

A particularly well-studied mechanism to enhance the amplitude of curvature perturbations during inflation involves a transient ultra-slow-roll (USR) phase \cite{PhysRevD.50.7173,Tsamis:2003px,Kinney:2005vj}. This regime occurs when the inflaton field encounters a nearly flat region of the potential, causing its velocity to decrease rapidly and the curvature perturbation to grow on super-horizon scales. As a result, the primordial power spectrum can be amplified by several orders of magnitude over a narrow range of scales, leading to the collapse of overdense regions into PBHs after horizon re-entry in the radiation era \cite{Garcia-Bellido:2017mdw,Motohashi:2017kbs,Germani:2017bcs,Byrnes:2018txb,Atal:2018neu,Figueroa:2021zah,Ragavendra:2023ret,Ballesteros:2024zdp,Fujita:2025imc,Namjoo:2025hrr}. Such an enhancement of the curvature perturbation 
generically leads to non-Gaussianities 
\cite{Namjoo:2012aa,Chen:2013eea,Cai:2018dkf,Biagetti:2018pjj,Passaglia:2018ixg,Atal:2019cdz,Atal:2019erb,Pi:2021dft,Pi:2022ysn,Tomberg:2023kli,Kawaguchi:2023mgk,Hooshangi:2023kss,Gow:2022jfb,Escriva:2023uko,Artigas:2024ajh,Jackson:2024aoo,Cruces:2024pni,Inui:2024sce,Iovino:2024sgs,Ballesteros:2024pbe,Wang:2024wxq,Caravano:2025diq,Cruces:2025typ} (see \cite{Pi:2024lsu} for a review), which can significantly alter the critical conditions for black hole formation, as shown by relativistic numerical studies \cite{2020JCAP...05..022A,2022JCAP...05..012E,Shimada:2024eec,Inui:2024fgk}.

%generically leads to non-Gaussianities \cite{} (see \cite{Pi:2024lsu} for a review), which can significantly alter the critical conditions for black hole formation, as shown by relativistic numerical studies \cite{2020JCAP...05..022A,2022JCAP...05..012E,Shimada:2024eec,Inui:2024fgk}.

Aside from the 
%mechanism of the collapse of 
adiabatic curvature fluctuations, another well-motivated mechanism for PBH formation involves relic vacuum bubbles produced during inflation through quantum tunneling \cite{Garriga:2015fdk, 2017JCAP...04..050D,2019JCAP...09..073A, 2020JCAP...09..023D,Kusenko:2020pcg,He:2023yvl,Deng:2018cxb,2020JCAP...05..022A,Escriva:2023uko}. In some of these scenarios \cite{Garriga:2015fdk, 2017JCAP...04..050D,2020JCAP...09..023D,Kusenko:2020pcg,He:2023yvl,Deng:2018cxb}, the tunneling rate is approximately constant throughout inflation, leading to a nearly scale-invariant distribution of bubble sizes. In other scenarios \cite{2019JCAP...09..073A,2020JCAP...05..022A,Escriva:2023uko,Kleban:2023ugf,Wang:2025hwc}, the tunneling rate may have a strong enhancement at a particular moment during inflation, producing a nearly monochromatic distribution of sizes. 
Once inflation ends, bubbles reenter the horizon and, if they are larger than a critical size, they undergo a peculiar form of gravitational collapse: their interiors continue inflating, and they form baby universes connected to the parent universe by transient wormholes. From the parent universe’s perspective, this process results in PBH's with masses of order $M \sim (GH)^{-1}$, where $H$ is the Hubble rate at the time of horizon re-entry. 

%On the other hand, v
Vacuum bubbles can naturally appear in single-field inflationary models where the potential contains a small barrier on its slope \cite{2019JCAP...09..073A,Mishra:2019pzq,ZhengRuiFeng:2021zoz,Wang:2021kbh,Rezazadeh:2021clf,Iacconi:2021ltm}. By using the $\delta N$ formalism, it was realized that local non-Gaussianities are well modeled by a logarithmic template, where the non-linear curvature perturbation $\zeta$ is a local function of the linearized perturbation $\zeta_G$ \cite{2019JCAP...09..073A},
\begin{equation}
\zeta\approx -\beta^{-1}\ln(1-\beta\zeta_G). \label{template}
\end{equation}
The value of $\beta$ is related to the second derivative of the potential at the top of the small barrier.
It was also realized that the relation is not invertible when the linear perturbation $\zeta_G$ exceeds $\beta^{-1}$. However, the probability for this to happen is finite, which hinted to a different channel for PBH formation. Indeed, while the barrier slows down the inflaton—enhancing curvature perturbations—it may also trap regions of the field that fluctuate backward, preventing them from crossing the barrier for $\zeta_G \gtrsim \beta^{-1}$. These trapped domains evolve into vacuum bubbles, providing an alternative channel of PBH formation distinct from the usual collapse of large adiabatic perturbations \cite{2019JCAP...09..073A,2020JCAP...05..022A,Escriva:2023uko}. Bubbles in this case form when the background field approaches the top of the barrier, leading to a sharply peaked distribution of bubble sizes. The precise form of this distribution was determined through dedicated numerical simulations \cite{Escriva:2023uko}, which found a scaling behavior for the bubble size. It was also shown that vacuum bubbles (surrounded by type-II fluctuations) can contribute higher PBH abundance than those forming from adiabatic perturbations, if the non-Gaussianity is large, i.e. $\beta \gtrsim 3.1$.

Eq. \eqref{template} is a special case of \textit{logaritimic duality} relation, which can be found in many different models when the nonlinear evolution of the curvature perturbation becomes important on superhorizon scales \cite{Pi:2022ysn}. Especially, in the USR inflation with a plateau-like potential and a sharp transition to slow roll, we have the same logarithmic relation \eqref{template} with $\beta=3$ \cite{Namjoo:2012aa,Chen:2013eea,Cai:2018dkf,Biagetti:2018pjj,Passaglia:2018ixg,Pi:2022ysn,Artigas:2024ajh}. As this is quite close to the critical value $\beta\approx3.1$ \cite{Escriva:2023uko}, a natural expectation will be that PBHs can also form via bubbles in such a USR inflation, which generates slightly less PBH abundance than the adiabatic channel. Physically, in some Hubble patches the velocities of the inflaton are smaller than the fiducial one due to the quantum fluctuations. Such patches are left behind by the fiducial trajectory and finally get stuck on the plateau by the Hubble friction. Then all the arguments we have for the bumpy potential can be applied here, which is the main task of this work.
%%%%%%%%%%%%%%%%%%%%%%%%%%%%
%%%%%%%%%%%%%%%%%%%%%%%%%%%%%
%With the above precedents, the purpose of this work is to study the formation of trapped vacuum bubbles in a single-field inflationary potential featuring a flat plateau. There,  the inflaton undergoes a brief USR phase before resuming slow roll. A naive argument suggests that, for the case with a sharp transition from USR to slow roll, this leads to a relation of the form Eq.(\ref{template}), with $\beta=3$  \cite{Namjoo:2012aa,Chen:2013eea,Cai:2018dkf,Biagetti:2018pjj,Passaglia:2018ixg,Pi:2022ysn,Artigas:2024ajh}. This is suggestive, because for potentials with a small barrier, the adiabatic channel and the bubble channel carry comparable  weights for $\beta\approx 3.1$ \cite{Escriva:2023uko}. It seems therefore interesting to analyze the present case, where there is no barrier in the potential. As we shall see, there are significant differences in the dynamics, but the field can still get trapped in the plateau due to occasional large backward fluctuations.

It should be noted that the case we consider in the present paper is {\em not} a $\beta\to 3$ limit of the potential considered in Ref.~\cite{2019JCAP...09..073A}. Instead, it is derived from USR dynamics without a barrier, and with a sharp transition to slow-roll. This is in itself an interesting example, where several features of the linearized power spectrum, and of its non-linear mapping to the curvature perturbation, can be handled semi-analytically and compared with numerical simulations. As we shall see, this leads to a rather tight picture of the dynamics. The sharp transition must be understood as an effective description of a situation in which the inflaton dynamics temporarily involves a scale much larger than the Hubble scale. Indeed, a steep “cliff” in the potential implies a
transient mass scale $M\equiv\sqrt{|V''|}\gg H$. Note that this is a strong inequality, and in this
sense the mass scale is not fine tuned. The emergence of $\beta = 3$ signals the presence of new physics that becomes relevant at the end of the USR phase. The associated phenomenology is robust and leads to distinctive observational
consequences.

The plan of the paper is the following. In Section \ref{model} we introduce the model and background dynamics. In Section \ref{statistics} we consider the mean profiles which are derived from the power spectrum. We will also discuss standard deviations from such profiles. In Section \ref{bubbleformation} we study the dynamics and size of the bubbles. Quantum diffusion will gradually pull some of the volume in the bubble into the slow-roll region. This process, and the global structure of the relics, is discussed in Section \ref{diffusion}. In Section \ref{adicha} we consider the adiabatic channel. We note that a logarithmic template of the form (\ref{template}) does not apply to the present case. The reason is that the separate universe picture does not really hold  during USR. Denoting the scale factor by $a$,  the time derivative of the background field decays as $a^{-3}$, while its perturbation decays slower, as $a^{-2}$, due to gradient terms. We derive a generalized template and compare it to the non-linear numerical evolution of the profiles. The critical thresholds for bubble formation, and for gravitational collapse of adiabatic perturbations, are determined for a relevant sample of profiles. In Section \ref{mafu} we compute the corresponding contributions to the mass function. Our conclusions are summarized in Section \ref{conclusions}.

\section{The model and background dynamics} \label{model}
\begin{figure}[H]
\centering
\includegraphics[width=3.0 in]{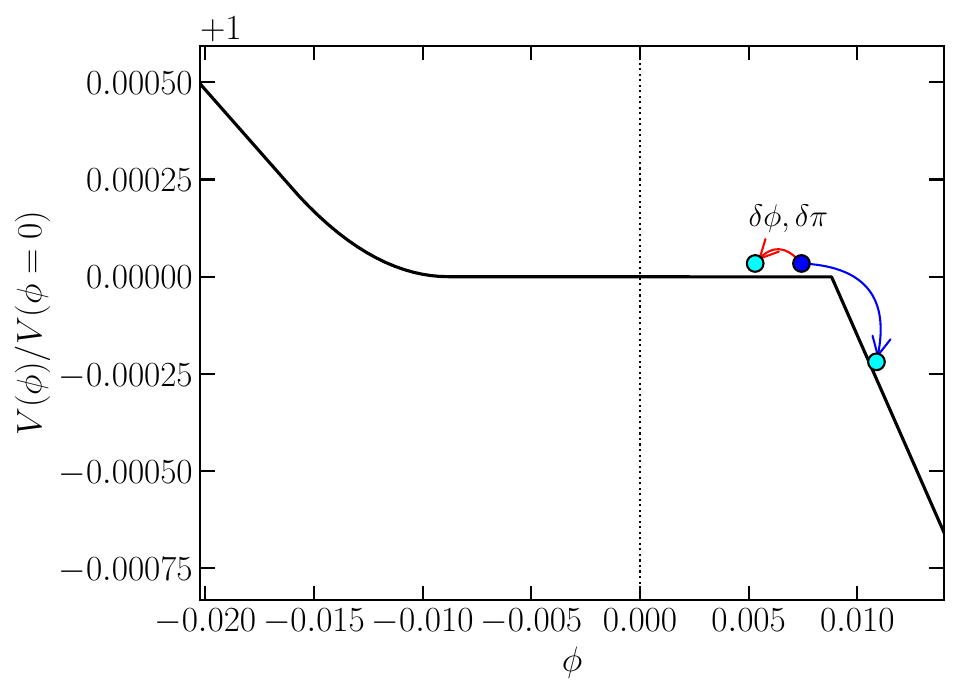}
\caption{Inflationary potential $V(\phi)$ as a function of $\phi$. The cyan circles represent inflationary trajectories, which can undergo a backward quantum fluctuation (red arrow) and become trapped, forming a vacuum relic, instead of rolling down from the flat plateau (blue arrow).}
\label{fig:V_phi}
\end{figure}

Let us consider an inflationary scenario driven by a single scalar field, whose action (in Planck units, where $M_{\rm pl}=1/\sqrt{8 \pi G}$ is the reduced Planck Mass) is given by
\begin{equation}
    S = \int d^4 x \sqrt{-g} \left[ \frac{R}{2}  -\frac{1}{2} \partial_{\mu} \phi \partial^{\mu} \phi -V(\phi) \right],
\end{equation}
where $R$ is the Ricci scalar, and $\phi$ is the inflaton. We consider the following piecewise template for the potential in a single field inflationary model, represented by a flat plateu of the inflaton potential in a specific range of $\phi$, and by two constant slopes below and after the plateu,
\[
   V(\phi)= \begin{dcases}
        V_0-\alpha_1(\phi+\tilde{L}_1) & -\infty < \phi <L'_1 \\
        V_0+\frac{m^2_1}{2}(\phi+L)^2 & -L'_1\leq \phi \leq -L \\
        V_0 & -L< \phi <L \\
        V_0-\frac{m^2_2}{2}(\phi-L)^2 & L<\phi<L'_2\\
        V_0-\alpha_2(\phi-\tilde{L}_2) & L'_2 < \phi < \infty
    \end{dcases}
\]
where $2L$ is approximately the total length of the flat region of the potential, $\alpha_1$, $\alpha_2$ is the constant linear slope of the potential (like in the Starobisnky model \cite{1992JETPL..55..489S}, see, for instance, \cite{Pi:2017gih,Gundhi:2020kzm,Pi:2022zxs} in the context of PBH formation) for $\phi<L'_1$ and $\phi>L'_2$ respectively. To connect the constant slope regions with the flat plateu we use a parabolic behaviour tunned by the masses $m_1,m_2$ which connect the two regions between $-L'_1\leq \phi \leq -L$ for $\phi<0$ and $L<\phi<L'_2$ for $\phi>0$. Therefore, our free parameters are $V_0,\alpha_1, \alpha_2 , m_1 , m_2 $ and we have applied to the values $\tilde{L}_1,\tilde{L}_2,L'_1,L'_2$ the corresponding junction conditions given by:

\begin{equation}
    L'_i = \frac{\alpha_i}{m^2_i}+L \, \, ; \qquad \tilde{L}_i = L + \frac{\alpha_i}{2 m^2_i}
\end{equation}
where the sub-index $i$ stands for the two branches of the potential. An schematic figure is shown in Fig.\ref{fig:V_phi}. For the purpose of our study, we have fixed the following parametters to be $V_0=2.85 \times 10^{-10} ,\alpha_1=1.6215\times 10^{-11},\alpha_2=3.423\times 10^{-11},L=0.008812,m_1=5 \times 10^{-4},m_2=1.0$.

In our framework, we take into account the possibility that quantum fluctuations may halt the inflaton field from exiting the flat region of the potential, thereby giving rise to regions of false vacuum trapping that remain stuck in the flat region of $V(\phi)$ without exit inflation. The homogeneous evolution of the inflaton is determined by the Klein–Gordon equation in a spatially flat Friedmann–Lemaître–Robertson–Walker (FLRW) background,
\begin{equation}
ds^{2} = -dt^{2} + a^{2}(t)(dr^{2} + r^{2} d\Omega^{2}),
\end{equation}
where $t$ denotes cosmic time and $a(t)$ the scale factor. The Hubble expansion rate is $H = d\ln a / dt$. For convenience, we employ the number of e-folds $N = \int H dt$ as the time variable. The background inflaton trajectory, $\phi_{\rm bkg}(N)$, then satisfies
\begin{equation}
\label{eq:phi_homogenea_new}
\ddot{\phi}_{\rm bkg} + 3\dot{\phi}_{\rm bkg} - \frac{1}{2}\dot{\phi}_{\rm bkg}^{3} +
\frac{V_{\phi}(\phi_{\rm bkg})}{V(\phi_{\rm bkg})}
\left(3 - \frac{\dot{\phi}_{\rm bkg}^{2}}{2}\right) = 0,
\end{equation}
where an overdot indicates differentiation with respect to $N$ (i.e. $\dot{\phi} \equiv d/dN$), and $V_{\phi} = dV/d\phi$. In deriving Eq.~\eqref{eq:phi_homogenea_new}, we have used $\dot{H} = -H\dot{\phi}_{\rm bkg}^{2}/2$ together with the Friedmann relation,
\begin{equation}
H^{2} = \frac{V(\phi_{\rm bkg})}{3 - \dot{\phi}_{\rm bkg}^{2}/2}.
\label{eq:H_new}
\end{equation}
To integrate Eq.~\eqref{eq:phi_homogenea_new}, we specify the initial conditions at $N = N_{\rm ini}$ as $\phi_{\rm bkg}(N_{\rm ini})$ and $\dot{\phi}_{\rm bkg}(N_{\rm ini})$, the latter estimated from the slow-roll limit:
\begin{equation}
\dot{\phi}_{\rm bkg}(N_{\rm ini}) = -\frac{V_{\phi}(\phi_{\rm bkg}(N_{\rm ini}))}{V(\phi_{\rm bkg}(N_{\rm ini}))}.
\end{equation}
The inflationary dynamics can be conveniently characterized in terms of the Hubble flow parameters, defined recursively as $\epsilon_{i+1} = d\ln \epsilon_{i}/dN$. In particular, the first two take the form $\epsilon_{1} = \dot{\phi}_{\rm bkg}^{2}/2$ and $\epsilon_{2} = \dot{\phi}_{\rm bkg}\ddot{\phi}_{\rm bkg}/\epsilon_{1}$. 
%The flat region of the potential allows for the existance of a (USR) period of inflation, where $\epsilon_2=-6$ in a specific range, that we fine-tunned to obtain the desired contribution of PBHs in the form of dark matter.

\begin{figure}[H]
\centering
\includegraphics[width=6.0 in]{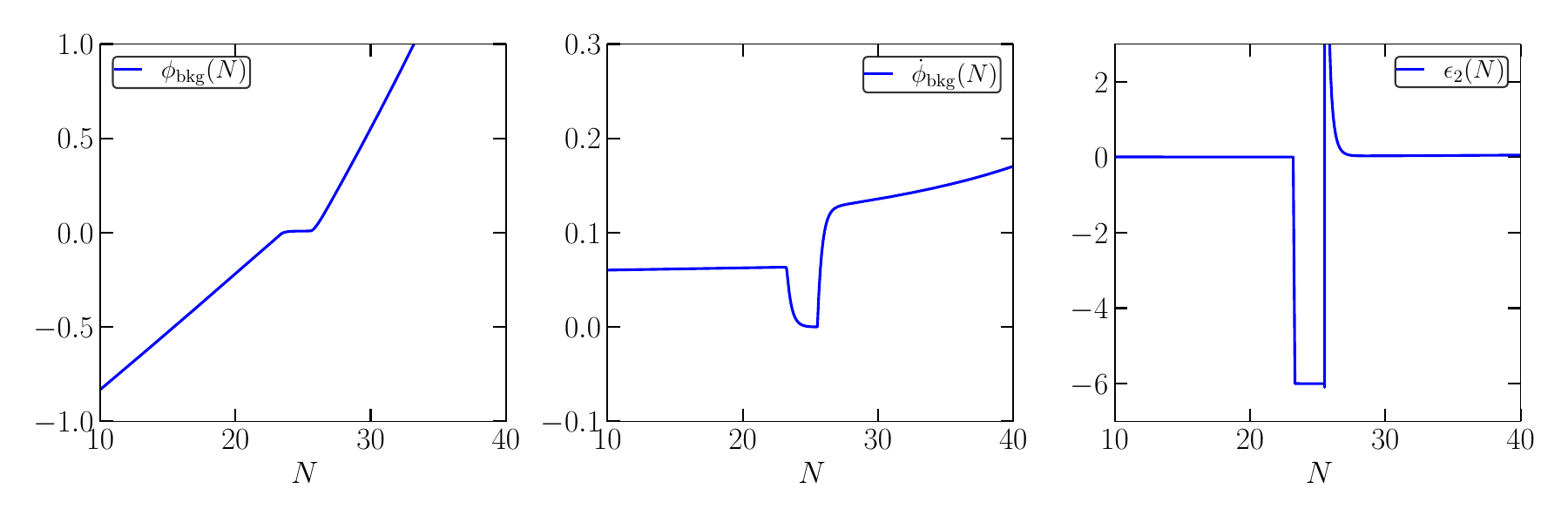}
\caption{Left-panel: Field $\phi_{\rm bkg}(N)$. Middle panel: Veolocity field $\dot{\phi}_{\rm bkg}(N)$. Right panel: $\epsilon_2(N)$ parametter.}
\label{fig:bkg_dynamics}
\end{figure}

In Fig.~\ref{fig:bkg_dynamics}, we show some of the background properties for the potential $V$. Between e-folds $N \approx [23.3,25.5]$, the inflaton field remains nearly constant for about $\sim 2$ e-folds. This stage corresponds to an (USR) phase, characterized by $\epsilon_2 = -6$. In this regime, the field velocity decreases to very small values, and after the USR phase, a strong acceleration occurs as the field resumes its roll down the potential.

On the other hand, choosing the constant mean curvature gauge (flat slicing), we define the spatial curvature of $\phi=\textrm{const}$ hypersurfaces, also called as Gaussian curvature flcutuation given by $\zeta_G \equiv -\delta \phi /\dot{\phi}_{\rm bkg}$. In Fourier space, the statistical properties of $\zeta_G$ are described by its power spectrum, defined through
\begin{equation}
\langle \zeta_G(N,\mathbf{k}),\zeta_G(N,\mathbf{k'}) \rangle
= \frac{2\pi^2}{k^3} \mathcal{P}_{\zeta_G}(N,k) (2\pi)^3 \delta^{(3)}(\mathbf{k}+\mathbf{k'}),
\end{equation}
where $\mathbf{k}$ is the comoving wavevector and $k = |\mathbf{k}|$. The mode evolution for each $k$ is governed by the Mukhanov–Sasaki (MS) equation,
\begin{equation}
\label{eq:MS_equation_new}
\ddot{\zeta}_{G} +
\left( 1 - \frac{1}{2}\dot{\phi}_{\rm bkg}^{2} + 2\frac{\dot{z}}{z} \right)\dot{\zeta}_{G}
+ \left( \frac{k}{aH} \right)^{2} \zeta_{G} = 0,
\end{equation}
where $z = a,\dot{\phi}_{\rm bkg}$ and $\dot{z} = a(\dot{\phi}_{\rm bkg} + \ddot{\phi}_{\rm bkg})$. The Hubble rate $H$ is expressed in terms of $\phi_{\rm bkg}$ by Eq.~\eqref{eq:H_new}.

The positive-frequency solutions of Eq.~\eqref{eq:MS_equation_new} are obtained numerically using the homogeneous background $\phi_{\rm bkg}(N)$ determined earlier from Eq.~\eqref{eq:phi_homogenea_new}, which is necessary to account for the enhancement of the curvature fluctuation due to the USR period. The integration begins deep inside the horizon ($k \gg aH$) at an initial e-fold $N_i$, where the Bunch–Davies vacuum is imposed:
\begin{align}
    \label{eq:BD_vacuum}
    \textrm{Re}[\zeta_{G}] &= \frac{1}{\sqrt{2k}} \frac{1}{z(N_{i})}, \,\,\,\, \,\,\,\,\,\,\,\,\,\,\,\,\,\,\,\,\,\,\,\,\,\, \,\,\,\, \,\,\,\, \,\,\,\,\,\,\,\,  \textrm{Im}[\zeta_{G}] = 0, \,\,\,\,  \\  \nonumber
    \textrm{Re}[\dot{\zeta}_{G}] &= -\frac{1}{2k}\frac{1}{z(N_{i})} \left(  \frac{\ddot{\phi}_{\rm bkg}(N_{i})}{\dot{\phi}_{\rm bkg}(N_{i})}+1 \right), \,\,\,\,  \textrm{Im}[\dot{\zeta}_{G}] = -\sqrt{\frac{k}{2}} \frac{1}{a(N_{i}) H(N_{i}) z(N_{i})}.
    \label{eq:BD_vacuum}
\end{align}

Finally, the dimensionless power spectrum of curvature fluctuations is then
\begin{equation}
\label{eq:PS_adiabatic_new}
\mathcal{P}_{\zeta_G}(N,k) = \frac{k^3}{2\pi^2} \left| \zeta_G(N,k) \right|^2,
\end{equation}
and becomes effectively constant once the mode $ k$ has sufficiently exited the horizon $ k \ll aH$.
%%%%%%%%%%%%%%%%%%%%%%%%%%%%%
%%%%%%%%%%%%%%%%%%%%%%%%%%%%%%%
%%%%%%%%%%%%%%%%%%%%%%%%%%%%%%%%
%%%%%%%%%%%%%%%%%%%%%%%%%%%%%%%%

It is also useful, for the purposes of the next section, to introduce the power spectra associated with the field perturbations $\delta \phi, \delta \pi$. Taking into account that $\delta \phi = -\zeta_G \dot{\phi}_{\rm bkg}$, we can define 
\begin{equation}
    \mathcal{P}_{\delta \phi} (N , k)= \frac{k^3}{2 \pi^2} \dot{\phi}_{\rm bkg}^2(N) \mid \zeta_G(N,k) \mid^2 . 
\end{equation}
Defining $\delta \pi \equiv \delta \dot{\phi}$, we have
\begin{equation}
    \delta \pi(N,k) = - \left[ \dot{\zeta}_{G}(N,k)\, \dot{\phi}_{\rm bkg}(N)   +\zeta_{G}(N,k) \, \ddot{\phi}_{\rm bkg}(N)\right].
\end{equation}
The corresponding power spectrum is then given by,
\begin{equation}
    \mathcal{P}_{\delta \pi} (N , k)= \frac{k^3}{2 \pi^2} \mid \delta \pi(N,k) \mid^2,
\end{equation}
where 
\begin{equation}
    \mid \delta \pi \mid^2 = \dot{\phi}^{2}_{\rm bkg}\mid \dot{\zeta}_G \mid^2 + \mid \zeta_G \mid^2  \ddot{\phi}^2_{\rm bkg}  + 2 \dot{\phi}_{\rm bkg} \ddot{\phi}_{\rm bkg} \left(\Imag(\zeta_G)\Imag(\dot{\zeta}_G)   + \Real(\zeta_G)\Real(\dot{\zeta}_G)\right) .
\end{equation}
We also define the power spectrum of the correlation between $\delta \phi$ and $\delta \pi$
\begin{equation}
    \mathcal{P}_{\delta \phi^{*} \delta \pi} (N , k)= \frac{k^3}{2 \pi^2} \textrm{Re}(\delta \phi^{*} \delta \pi),
\end{equation}
where
%\begin{eqnarray}
   % \delta \phi^{*} \delta \pi= &= -(\dot{\zeta}^{*}_G\dot{\phi}_{\rm bkg} +\zeta^{*}_G \ddot{\phi}_{\rm bkg}) (-\zeta_G \dot{\phi}_{\rm bkg}) = \zeta_G \dot{\zeta}^{*}_{G}\dot{\phi}^{2}_{\rm bkg}+\mid \zeta_G \mid^2 \ddot{\phi}_{\rm bkg}\dot{\phi}_{\rm bkg}=\\
   % &=(\Real(\zeta_G)+i \, \Imag(\zeta_G))(\Real(\dot{\zeta}_G)-i \, \Imag(\dot{\zeta}_G))\dot{\phi}^2_{\rm bkg}+\mid \zeta_G \mid^2 \ddot{\phi}_{\rm bkg}\dot{\phi}_{\rm bkg}
%\end{eqnarray}
%Taking the real part we obtain:
\begin{equation}
\textrm{Re}(\delta \phi^{*} \delta \pi) =\left[\Real(\zeta_G)\Real(\dot{\zeta}_G)+\Imag(\zeta_G)\Imag(\dot{\zeta}_G)\right]\dot{\phi}^2_{\rm bkg}+\mid \zeta_G \mid^2 \ddot{\phi}_{\rm bkg}\dot{\phi}_{\rm bkg}.
\end{equation}

\begin{figure}[!htbp]
\centering
\includegraphics[width=3.3 in]{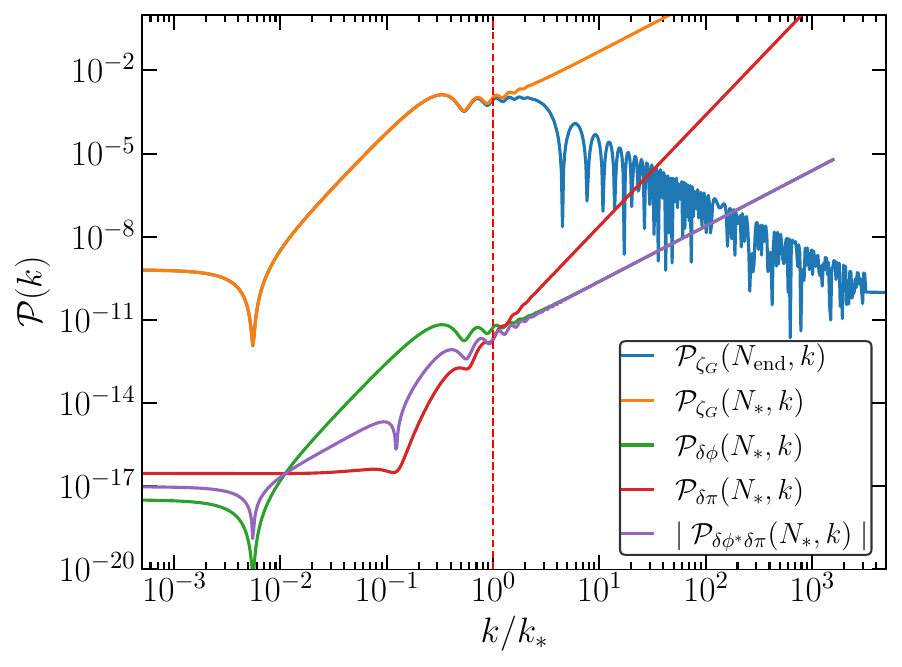}
\caption{Power spectra of the different previously defined quantities in terms of $k/k_*$, where $k_{*} = a(N_*) H(N_*)$. The spectra are evaluated at $N_*$, except for the blue curve, which is evaluated at the end of inflation, $N_{\rm end}$.}
\label{fig:power_spectrums}
\end{figure}

The different power spectra are shown in Fig.~\ref{fig:power_spectrums}. The power spectrum of the Gaussian curvature fluctuations exhibits a maximum at the scale $k_{\rm peak}$. The enhancement of the spectrum $\mathcal{P}_{\zeta_G}$ compared to the CMB scales ($\sim 2.2 \times 10^{-9}\, \textrm{Mpc}^{-1}$) arises from the amplification of the modes during the (USR) phase, characterized by $\epsilon_2 = -6$, which lasts for approximately two e-folds. This enhancement is essential to produce a sizable fraction of PBHs that can constitute dark matter, and we have fine-tuned the parameters of the potential to yield the desired PBH abundance of rouhgly equal to one.

The blue line corresponds to the spectrum of $\zeta_G$ once all modes $k$ have crossed the horizon at the end of inflation $N_{\rm end}$ with $\epsilon_{1}(N_{\rm end}=1)$. In contrast, when the modes are evaluated at a given time $N_*$ (before the end of inflation), not all of them have yet exited the cosmological horizon ($k \gg k_*$). In such cases, we expect the curvature modes to grow as in the Minkowski vacuum, i.e., $\sim k^2$. The same behavior applies to the power spectra associated with the perturbations $\delta\phi$, while the power spectrum for $\delta\pi$ behaves like $k^4$ due to the additional derivative.

\begin{figure}[!htbp]
\centering
\includegraphics[width=2.5 in]{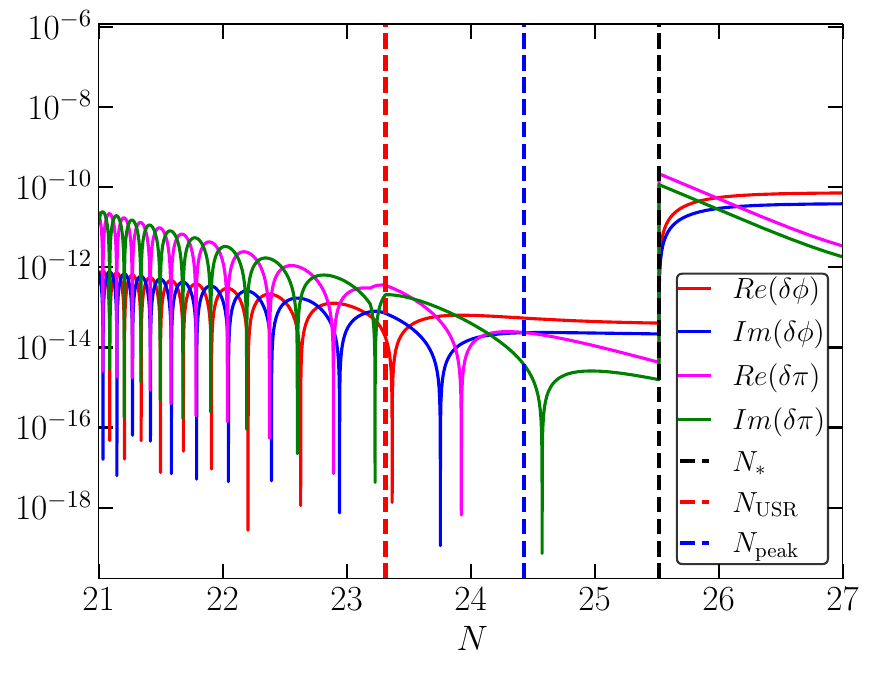}
\includegraphics[width=2.5 in]{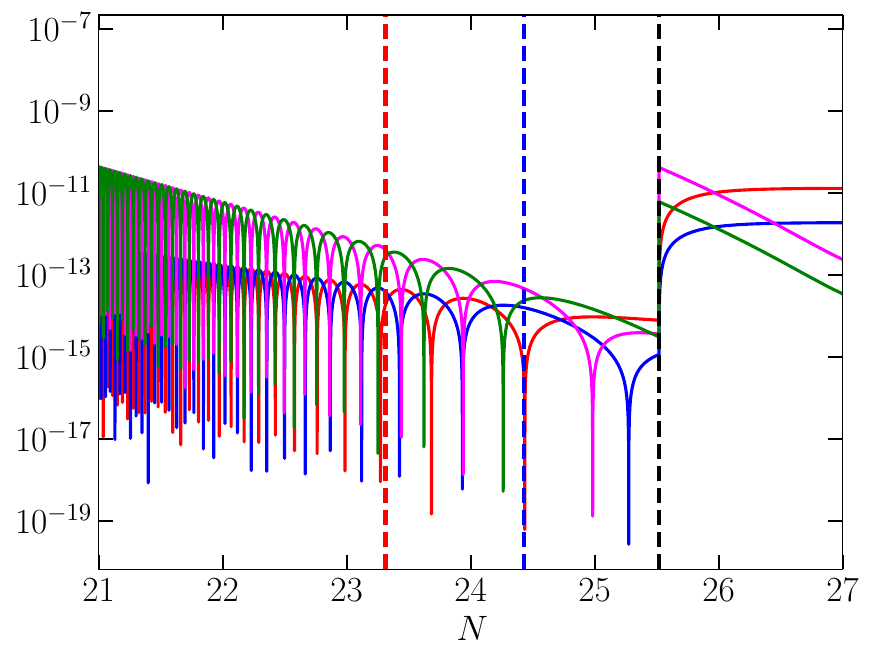}
\caption{Mode evolution of $\delta \phi$ and $\delta \pi$, showing their real and imaginary components as a function of the number of e-folds. The three vertical dashed lines correspond to the times $N_k$ when the wave mode $k$ re-enters the horizon, for the case of $N_{*}$ (black), the beginning of the USR phase $N_{\rm USR}$ (red line), and the location of the maximum of the power spectrum $\mathcal{P}_{\zeta_G}(k)$ at $k_{\rm peak}$ (blue). Left panel corresponds to $k=k_{\rm peak}$ and right panel $k=k_{*}$.}
\label{fig:deltaphi_pi}
\end{figure}

In Fig.~\ref{fig:deltaphi_pi}, we can see the evolution of the modes $\delta \phi$ and $\delta \pi$ as a function of the number of e-folds. For both cases, corresponding to mode evolutions with $k = k_{\rm peak}$ and $k = k_{*}$, it can be seen that the mode behavior changes drastically just after $N_*$, corresponding to the end of the USR stage ($\epsilon_2 \neq -6$). We therefore fix $N_* = 25.519$ as the limiting value marking the boundary of the end of the USR stage to account for the enhancement of the modes.

\section{Statistics of initial conditions and shape dispersion}\label{statistics}

Following \cite{Escriva:2023uko}, we will numerically study the formation of vacuum bubbles by solving the non-linear Klein-Gordon field equation in FLRW spacetime. Unlike the case of background evolution, this now includes the effect of gradient terms:

\begin{align}
%    \ddot{\phi}+\dot{\phi} \left( 3- \frac{1}{2} \dot{\phi}^2_{\rm bkg} \right) -\left(\frac{a_I H_I}{a(N) H(N)}\right)^{2} \Delta \phi +\frac{V_{\phi}(\phi)}{H(N)^{2}} = 0.    \label{eq:KG}  \\
        \ddot{\phi}+\dot{\phi} \left( 3- \frac{1}{2} \dot{\phi}^2_{\rm bkg} \right) -\frac{\nabla^2\phi}{a^2 H^2} +\frac{V_{\phi}(\phi)}{H^{2}} = 0.
    \label{eq:KG}
\end{align}
%{\color{red}(SP: Check the capital I subscript. $(k/aH)^2$?)}
We choose $N_*$ as the time at which we set up initial conditions for the subsequent classical evolution. Assuming spherical symmetry, such initial conditions take the form
\begin{equation}
%\phi(N_*)=\bar\phi(N_*)+\delta\phi(r),\quad \dot \phi(N_*)= \dot{\bar\phi}(N_*)+\delta\pi(r).
\phi(N_*,r)=\phi_{\rm bkg}(N_*)+\delta\phi(r),\quad \dot \phi(N_*,r)= \dot{\phi}_{\rm bkg}(N_*)+\delta\pi(r).
\label{eq:cosa}
\end{equation}
The shape of initial large perturbations has a significant impact on the thresholds for PBH formation and bubble formation. Therefore, we need to investigate in some detail the probability for different profiles.

\subsection{Mean profile $\bar\delta\phi(r)$, and deviations from it.}

The initial field profile can be written as
\begin{eqnarray}
    \delta\phi(r) &=& \bar\delta\phi(r) \pm \,\Delta_\phi(r),
    %\\
   %\delta\pi(r) &=& -\mu \Psi_\pi(r) \pm \, \Delta_\pi(r).
   \label{eq:IC}
\end{eqnarray}
The first term $\bar{\delta } \phi = \langle \delta \phi \rangle_{\mu}$ in the right hand side corresponds to the mean profile in the ensemble of realizations, subject to the constraint that $\delta\phi(0)=-\mu$. The negative sign is introduced for convenience, since we are interested in large backward fluctuations. The uncertainty $\Delta_\phi$ 
accounts for deviations around the mean, whose amplitude and shape we will now discuss.

Assuming spherical symmetry, the field perturbation can be expanded as 
$$\label{Bessel}\delta\phi(r)=\int B_k\ {\rm sinc}(kr) d\ln k.$$
For a Gaussian random field the PDF takes the form 
$P[\delta\phi]\propto \exp(-W[\delta\phi]/2),$
where
$$W=\int {B_k^2\over {\cal P}_{\delta\phi}(k)} d\ln k. $$
The power spectrum is evaluated at the time $N_*$, and all the integrals range from the infrared up to the value $k=k_*$.
The mean profile $\bar\delta\phi$ can be found by minimizing $W$ subject to the constraint $\delta\phi(0)=-\mu$. This immediately leads to the Bessel components 
\begin{equation}
\bar B_k= -\mu\ {\cal P}_{\delta\phi}(k)/\sigma_{\delta\phi}^2, \label{barbeka}
\end{equation}
where the variance can be expressed as
\begin{equation}
\sigma^2_{\delta\phi} = \int d\ln k \ {\cal P}_{\delta \phi}(k).
\end{equation}
Substituting in (\ref{Bessel}) we find
\begin{equation}
\bar\delta\phi = -\mu\Psi_\phi(r),
\end{equation}
where we have introduced the normalized correlator
\begin{equation}
\Psi_\phi(r)={\langle \delta\phi(0) \delta\phi(r) \rangle \over \sigma_{\delta\phi}^2}={1\over\sigma^2_{\delta\phi}}\int d\ln k \ \mathcal{P}_{\delta \phi}(k)\ 
{\rm sinc}(kr).
%\\
%\Psi_\pi(r)={\langle \delta\phi(0) \delta\pi(r) \rangle \over \sigma_{\delta\phi}^2},
\label{eq:Psi_phi}
\end{equation}

Let us now consider a generic realization around the mean value\footnote{At any specific distance $r$ from the center, and for a fixed value of $\delta\phi(0)$, the standard deviation $\tilde\Delta_\phi$ 
%and $\tilde\Delta_{\pi}$, 
characterizing the dispersion around the mean profile is given by
\begin{eqnarray}
\tilde \Delta_\phi(r)&=& \sigma_{\delta\phi} [1-\Psi^2_\phi(r)]^{1/2}, 
%\\
%\tilde\Delta_\pi(r)&=&
%\left[\sigma_{\delta\pi}^2-\sigma_{\delta\phi}^2\Psi^2_\pi(r)\right ]^{1/2}.
\end{eqnarray}
This holds at every $r$, and so we expect that $68\%$ of the realizations will be within the range $\delta\phi(r)=-\mu\Psi_{\delta\phi}(r) \pm \tilde\Delta_\phi(r)$. However, such realisations will generically oscillate around the mean value as we move in the radial direction. Note, in particular, that $\tilde\Delta_\phi(r)$ tends to a constant at large distances. If we consider a perturbation whose profile is given by the envelope of all realizations at a given sigma level $n$, $\delta\phi(r)=-\mu\Psi_{\delta\phi}(r) + n\ \tilde\Delta_\phi(r)$, with a fixed sign and value of $n$, then the probability of sustaining such deviation consistently to arbitrarily large $r$ would be zero. Hence, instead of considering the envelope, we will focus on deviations with finite probability, that die off at infinity. 
}, with
$B_k=\bar B_k + n\ \Delta B_k$. In position space we have
\begin{equation}\label{initialfield}
\delta\phi(r,n)= \bar\delta\phi(r) + n\ \Delta_\phi(r),
\end{equation}
with
\begin{equation}\label{deltapos}
\Delta_\phi(r) =\int \Delta B_k\ {\rm sinc}(kr) d\ln k.
%= C \int_0^{k_*} \sin(2 \pi k/k_*){\rm sinc}(kr) 
%\textcolor{red}
%{(k/k_*)}{d \ln k},
\end{equation}
The condition $\delta\phi(0)=\mu$ then requires
\begin{equation}\int\Delta B_k\ d\ln k =0,\label{zero}\end{equation}
%Since only the product $n \Delta B_k$ is relevant, we can always set the overall normalization of the $\Delta B_k$ as
%\begin{equation}
%\int {(\Delta B_k)^2\over {\cal P}_{\delta\phi}(k)} d\ln k=1.
%\end{equation}
and the change in the weight factor $W$ in the exponent of the Gaussian distribution is then given by 
\begin{equation}\label{Wn}
\Delta W(n) =n^2\int {(\Delta B_k)^2\over {\cal P}_{\delta\phi}(k)} d\ln k.
\end{equation}

%In other words, for the field configuration (\ref{initialfield}), the suppression relative to the mean will be $e^{-n^2/2}$, and
%can say that such configuration is $n$ standard deviations away from the mean. 
Our next task is to find a well motivated ansatz for the coefficients $\Delta B_k$. 

Consider a partition of the $k$ integration range $(0,k_*)$ into two subsets $U_+$ and $U_-$, with $B_k>0$ in $U_+$ and $B_k<0$ in $U_-$. Then, by Eq. (\ref{zero}), we have 
\begin{equation}\label{split}
\mu_\Delta \equiv\int_{U_+} n \Delta B_k d\ln k= -\int_{U_-} n \Delta B_k d\ln k.
\end{equation}
Minimizing $\Delta W =\Delta W_{U_+}+\Delta W_{U_-}$ subject to the constraint (\ref{split}), we readily find 
\begin{equation}
    \label{optimus}
  n\Delta B_k^{\pm}=\pm(\mu_\Delta/\sigma_\pm^2) {\cal P}_{\delta\phi},
\end{equation}    
    in $U_+$ and $U_-$ respectively. Here we have introduced $\sigma_+^2=\int_{U_+} {\cal P}_{\delta\phi}\ d\ln k \equiv \tilde{f} \sigma_{\delta\phi}^2$, and the complementary $\sigma_-^2 =(1-\tilde{f}) \sigma_{\delta\phi}^2$. 
This leads to 
$$\Delta W = {\mu_\Delta^2\over \sigma_{\delta\phi}^2}{1\over \tilde{f}(1-\tilde{f})}.$$ For fixed $\mu_\Delta$ the cost $\Delta W$ is minimized for $\tilde{f}=1/2$, when both subsets in the partition share equal weight in the variance of the field, i.e. $\sigma_+^2=\sigma_-^2$. In such case we have $\Delta W = 4\mu_\Delta^2\sigma_{\delta\phi}^{-2}$. 
For $\Delta W=n^2$, the suppression relative to the mean will be $e^{-n^2/2}$, and
we can say that such configuration is $n$ standard deviations away from the mean. Then, the $n=1$ deviation from the mean corresponds to $\mu_\Delta =(1/2)\sigma_{\delta\phi}$, and from (\ref{optimus}) we have 
\begin{equation}
\Delta B_k^{\pm}=\pm\sigma_{\delta\phi}^{-1} {\cal P}_{\delta\phi}.
\end{equation}

It then follows from (\ref{deltapos}) and (\ref{split})
that 
%$|\Delta_\phi(r)| \leq \mu_\Delta$, and therefore we have
\begin{equation} \label{bound}
|\Delta_\phi(r)| \leq{\sigma_{\delta\phi}/2}.
\end{equation}
The actual shape and amplitude of the standard deviation $\Delta_\phi(r)$ depends, however, on the choice of the partition $\{U_+,U_-\}$ in the momentum domain. For instance, if we choose the subsets $U_+$ and $U_-$ to be highly intertwinned, with relatively small domains densely alternating positive and negative values of $\Delta B_k$, these will interfere destructively in the integral (\ref{deltapos}), and we will have $|\Delta_\phi(r)| \ll{\sigma_{\delta\phi}/2}$.

Here, we will be interested in the opposite limit, where the standard deviation $\Delta_\phi(r)$ can be maximized. For that purpose, we consider a simple bisection where the range of $k$ is neatly divided into two sub-intervals: the long wavelength modes $U_+=(K_{IR},\bar k)$, and the short wavelength modes $U_-=(\bar k, k_*)$, so that the interference between positive and negative $\Delta B_k$ is mitigated. Here, the median $\bar k$ is defined by the equation $\int_{0}^{\bar k} {\cal P}(k) d\ln k = \int_{\bar k}^{k_*} {\cal P}(k) d\ln k$, and can be found numerically for any given power spectrum. This leads to the expression
\begin{equation}\label{ansatz}
%\Delta B_k = C (k/k_*) \sin(2 \pi k/k_*),
\Delta B_k =\sigma_{\delta\phi}^{-1}\ {\cal P}_{\delta\phi}(k)\ {\rm sign}(\bar k-k).
\end{equation} 
The sign function guarantees that (\ref{zero}) is satisfied. From Eqs. (\ref{ansatz}), (\ref{deltapos}), and using $2 \sigma_-^2=\sigma_{\delta\phi}^2$, we readily find
\begin{equation}\label{simple}
\Delta_\phi(r) = {\sigma_{\delta\phi}\over 2} \
[\Psi_{+}(r) - \Psi_{-}(r)].
\end{equation}
Here $\Psi_{\pm}$ are the correlation funcions, analogous to the one given in (\ref{eq:Psi_phi}), for the long wavelength ($k<\hat k$) and the short wavelenth ($k>\hat k$) halves of the spectrum, with normalization $\Psi_{\pm}(0)=1$.
Note that $U_+$ and $U_-$ give contributions of equal magnitude and opposite sign at $r=0$. However, as we move away from the origin, the effect of short wavelenth modes dies off quickly, while the effect of long wavelenths is relevant at larger distances. From (\ref{simple}), we expect the bound (\ref{bound}) to be nearly saturated, $|\Delta_\phi(r)| \sim {\sigma_{\delta\phi}/2}$, at intermediate distances $\bar k r\sim 1$.

  %An ansatz proportional to the power spectrum is natural, by analogy with (\ref{barbeka}), since from (\ref{Wn}) we may expect a larger amplitude $\Delta B_k$ at places where ${\cal P}_{\delta\phi}(k)$ is larger. 

Indeed, for the case of a power spectrum ${\cal P}_{\delta\phi}^{1/2}$ 
which is constant over an e-folding range $\Delta N=\ln(k_*/k_{IR})$, from some infrared scale $k_{IR}$ to the scale $k_*$, we find the analytic expression
\begin{equation}
\Delta_\phi(r) ={\sigma_{\delta\phi}\over \Delta N}\left[2I(\bar kr)-I(k_{IR}r)-I(k_*r)\right].
\end{equation}
Here $k_{IR}=e^{-\Delta N} k_*$, $\bar k= e^{-\Delta N/2} k_*$, and
\begin{equation}
I(x)={\rm Ci}(x)-{\rm sinc}(x),
\end{equation}
where ${\rm Ci}$ is the cosine integral function. For instance, for $\Delta N =2$, the value of the median is given by $\bar k=k_*/e \approx 0.367 k_*$. and $\Delta_\phi(r)$ has a maximum at $\bar k r \approx 2.1$, with $\Delta_\phi \approx 0.363\ \sigma_{\delta\phi}$, which is fairly close to the upper bound\footnote{We have also checked that, for a broad power spectrum $\Delta N\gg 1$, the upper bound $\Delta_\phi(r)\approx\sigma_{\delta\phi}/2$ is saturated at the intermediate scale $\bar k r\approx 1.9$.} (\ref{bound}).
We expect a similar result even if the power spectrum is not exactly constant. In the numerical example of our interest we find $\bar k \approx 0.360 k_*$, and a maximum of the deviation at $\bar k r \approx 1.98$, with $\Delta_\phi \approx 0.330\ \sigma_{\delta\phi}$.

 \subsection{Mean profile $\bar\delta\pi(r)$, and deviations from it.}
 
Let us now consider the initial perturbations of momentum, $\delta\pi$. We must take into consideration, however, that these are correlated with the perturbations $\delta\phi$. If we choose $\delta\phi(r,n)=\bar\delta\phi(r) + n \Delta_\phi(r)$, this will affect the mean value of the momentum perturbation
\begin{equation}
\bar\delta \pi(r,n) = \int{A_k(\bar B_k +n\ \Delta B_k)\ {\rm sinc}(kr)\ d\ln  k},
\end{equation}
where we have introduced
\begin{equation}
    A_k \equiv {Re[\delta\phi_k^* \delta\pi_k]\over |\delta\phi_k|^2}.
\end{equation}
On top of the mean value, we may consider deviations at $m$ sigma level. The initial perturbations will therefore be characterized by two integers $n$ and $m$:
\begin{equation}
\label{initialpi}
\delta\pi(r,n,m)= \bar\delta\pi(r,n) + m\ \Delta_\pi(r)= \int D_k\ {\rm sinc}(kr) d\ln k.
\end{equation}
with $D_k= A_k(\bar B_k + n\ \Delta B_k)+m\ \Delta D_k.$
For $n=m=0$, we find the mean value for the momentum perturbation, given by
\begin{eqnarray}
    \delta\pi(r,0,0) &=& -\mu \Psi_\pi(r),
    %\\
   %\delta\pi(r) &=& -\mu \Psi_\pi(r) \pm \, \Delta_\pi(r).
\end{eqnarray}
where $\mu=-\delta\phi(0)$ is the same as in Eq.(\ref{eq:Psi_phi}), and 
\begin{equation}
\Psi_\pi(r)={\langle \delta\phi(0) \delta\pi(r) \rangle \over \sigma_{\delta\phi}^2} ={1\over\sigma_{\delta\phi}^2} \int d\ln k \ \mathcal{P}_{\delta \phi^{*} \delta \pi}(k)\ {\rm sinc}(kr).
\end{equation}
%with
%\begin{eqnarray}
%    \langle \delta\phi(0) \delta\pi(r) \rangle =
%\int d\ln k \ \mathcal{P}_{\delta \phi^{*}_k %\delta \pi_k}(k)\ {\rm sinc}(kr).
%\end{eqnarray}
%More generally, the Bessel expansion of the initial perturbation (\ref{initialpi}) takes the form
%%   \delta\pi(r,n,m) = \int D_k\ {\rm sinc}(kr) d\ln k,
%\end{equation}
%where $D_k= A_k(\bar B_k + n\ \Delta B_k) + m\ \Delta D_k$.
%the Bessel coefficients for the mean profile are given by $\bar D_k = A_k \Bar B_k$, with 
The deviations $\Delta D_k$ are Gaussian distributed, with vanishing expectation value,  and the corresponding change in the weight factor is given by
\begin{equation}
\Delta W(m) =m^2 \int {(\Delta D_k)^2\over {\cal P}_{\delta\pi}(k)(1-\gamma_k^2)} d\ln k, 
\end{equation}
where the Pearson coefficient is given by
\begin{equation}
\gamma_k = A_k {|\delta \phi_k|\over |\delta \pi_k|}.
\end{equation}
Unlike the case of $\Delta_\phi(r)$, the $\Delta_\pi(r)$ are not constrained to vanish at the origin, and therefore the coefficients $\Delta D_k$ need not average to zero when integrated $d\ln k$. Minimizing $\Delta W$ for given $\Delta_\pi(0)$ we find,
\begin{equation}\label{ansatzd}
\Delta D_k = \tilde\sigma_\pi^{-1} {\cal P}_{\delta\pi}(k)(1-\gamma_k^2),
\end{equation}
 where
 \begin{equation}
 \tilde\sigma_\pi^2 =\int {\cal P}_{\delta\pi}(k)(1-\gamma_k^2) d\ln k.
 \end{equation}
The overall normalization of the $\Delta D_k$ in Eq. (\ref{ansatzd}) is chosen so that $\Delta W(m) = m^2$. Then, we can say that (\ref{initialpi}) is $m$ standard deviations away from the mean $\bar\delta\pi(r,n)$.
In position space, the standard deviation given by
\begin{equation}
    {\Delta_{\pi}(r) =\frac{1}{\tilde{\sigma}_{\pi}}\int_{0}^{k_*}  \mathcal{P}_{\delta \pi}(k)(1-\gamma^2_{k})\ {\rm sinc}(k r)d \ln k}.
\end{equation}

\vskip .2 cm

\begin{figure}[!htbp]
\centering
\includegraphics[width=2.5 in]{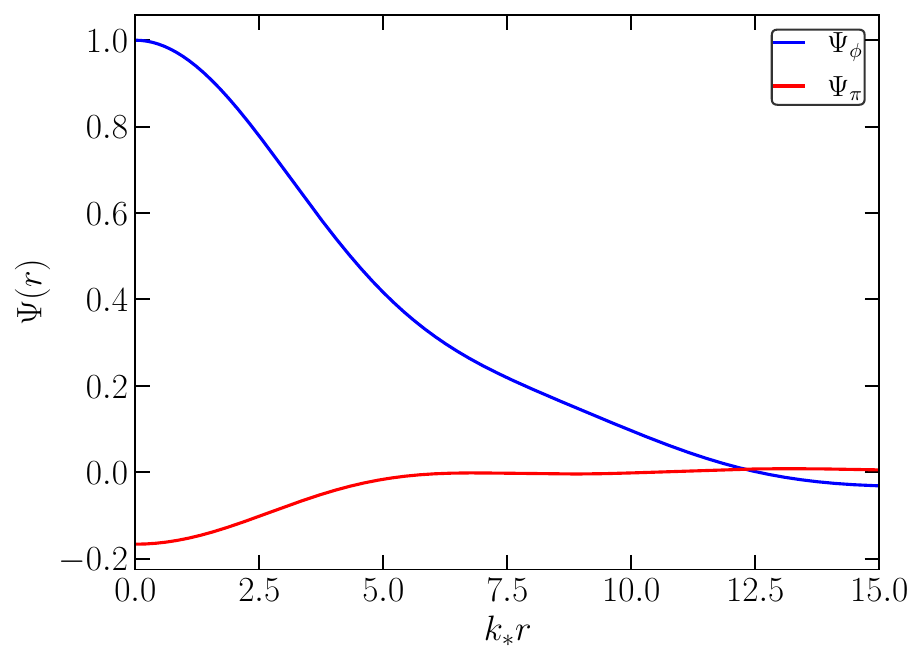}
\includegraphics[width=2.5 in]{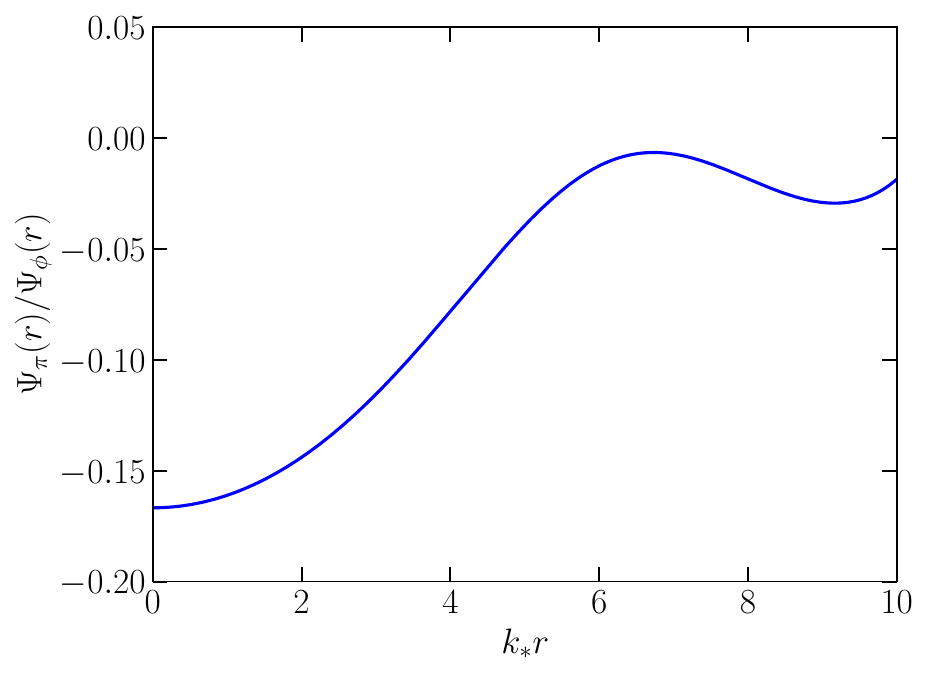}
\includegraphics[width=2.5 in]{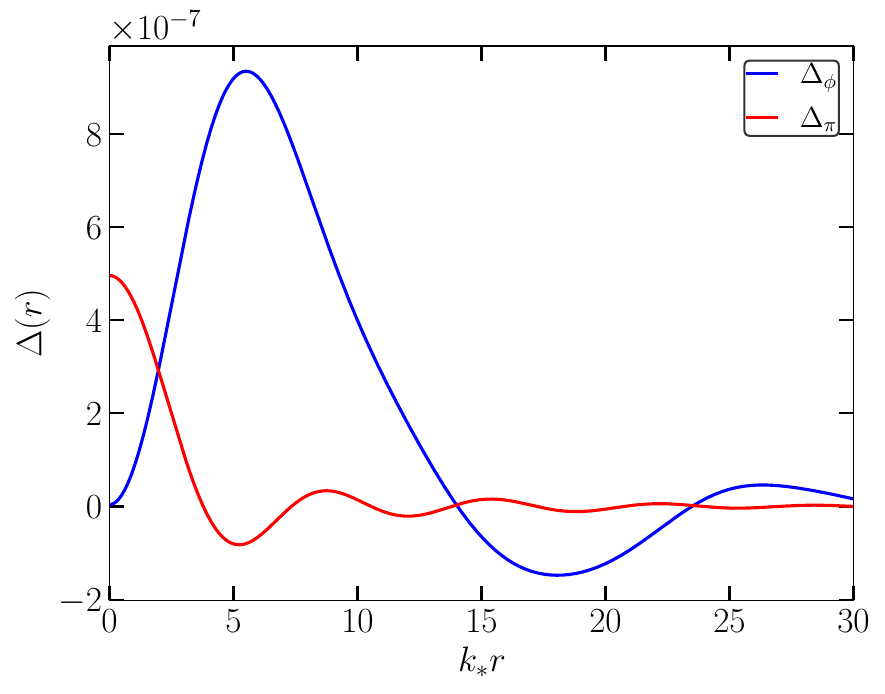}
\includegraphics[width=2.5 in]{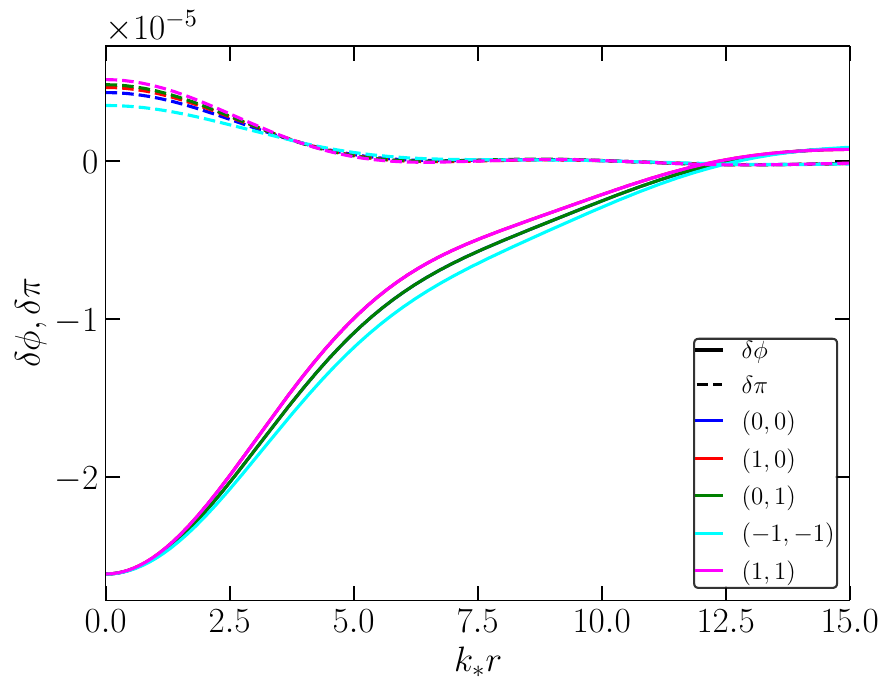}
\caption{Top-Left panel: Normalized correlators $\Psi_{\phi},\Psi_{\pi}$. Top-Right panel: Ratio between the correlators. Bottom-Left panel: Dispersion shapes $\Delta_{\phi}(r), \Delta_{\pi}(r)$. Bottom-Right panel: Shapes $\delta \phi(r,n,m),\delta \pi(r,n,m)$ taking $\mu \approx 2.613 \cdot 10^{-5}$. We have also used $\sigma_{\delta \phi}\approx 2.836 \cdot 10^{-6}$, $\tilde{\sigma}_{\pi} \approx 4.960 \cdot 10^{-7}$} from the power spectra numerically computed.
\label{fig:dispersions_correlations}
\end{figure}

Finally, to account for the likelihood of the shapes deviating from the mean profiles, we shall weigh those realizations by the factor
\begin{equation}
\Upsilon(n,m)=\textrm{Erfc}(\mid n \mid/\sqrt{2})\textrm{Erfc}(\mid m \mid/\sqrt{2}).
\label{eq:n_m_factors}
\end{equation}
An example of the shapes can be found in Fig.\ref{fig:dispersions_correlations}. 
There, we also display the anticorrelation between the two-point function $\Psi_{\phi}/\Psi_{\pi}$. 

\section{Bubble formation }\label{bubbleformation}
\begin{figure}[!htbp]
\centering
\includegraphics[width=2.5 in]{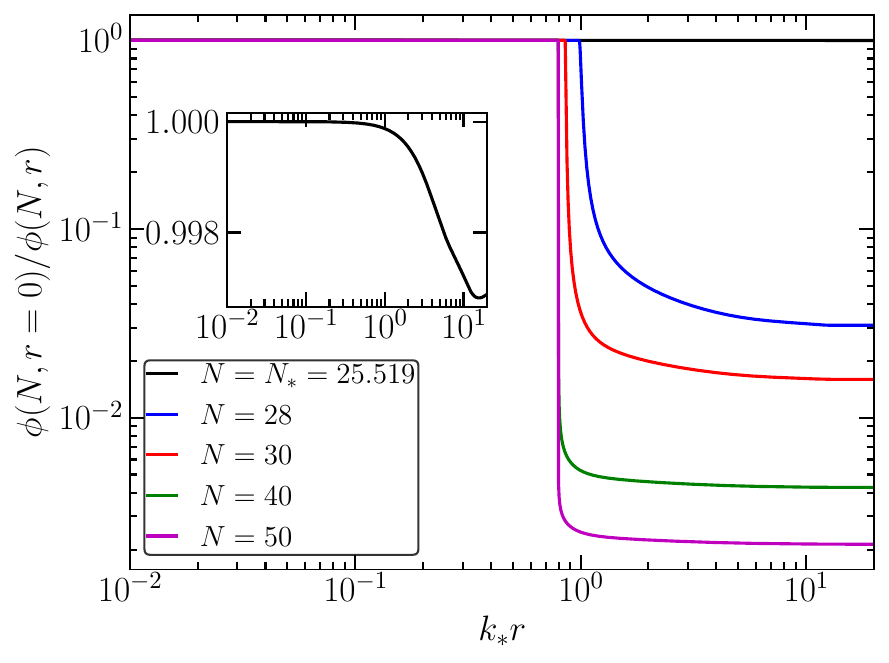}
\includegraphics[width=2.5 in]{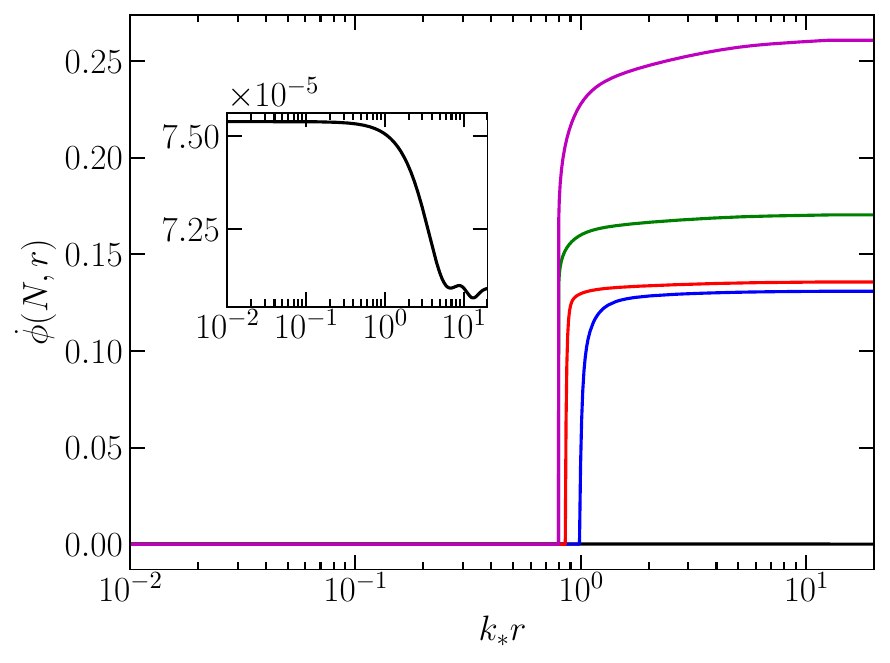}
\includegraphics[width=2.5 in]{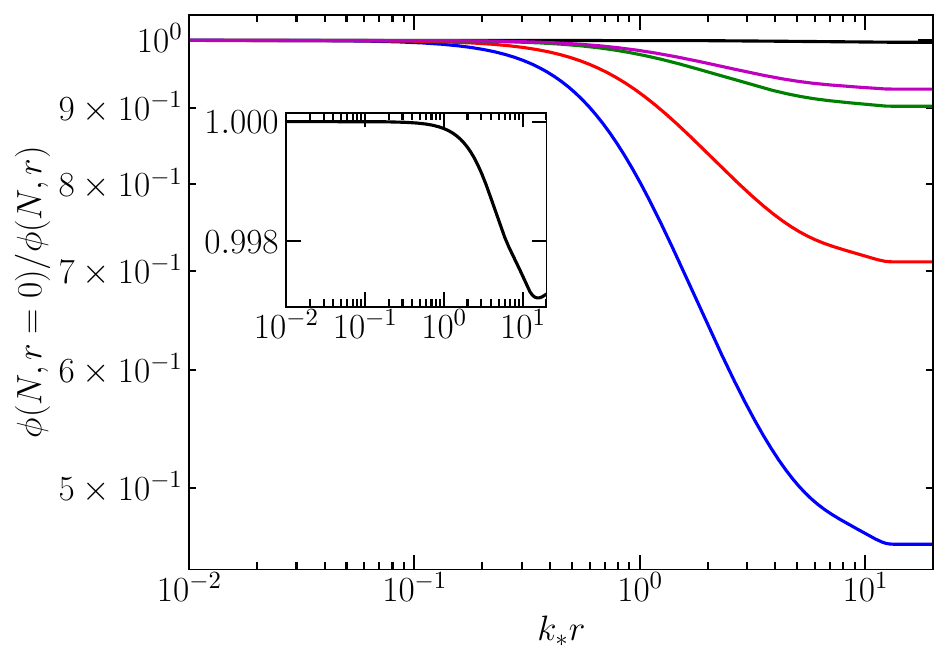}
\includegraphics[width=2.5 in]{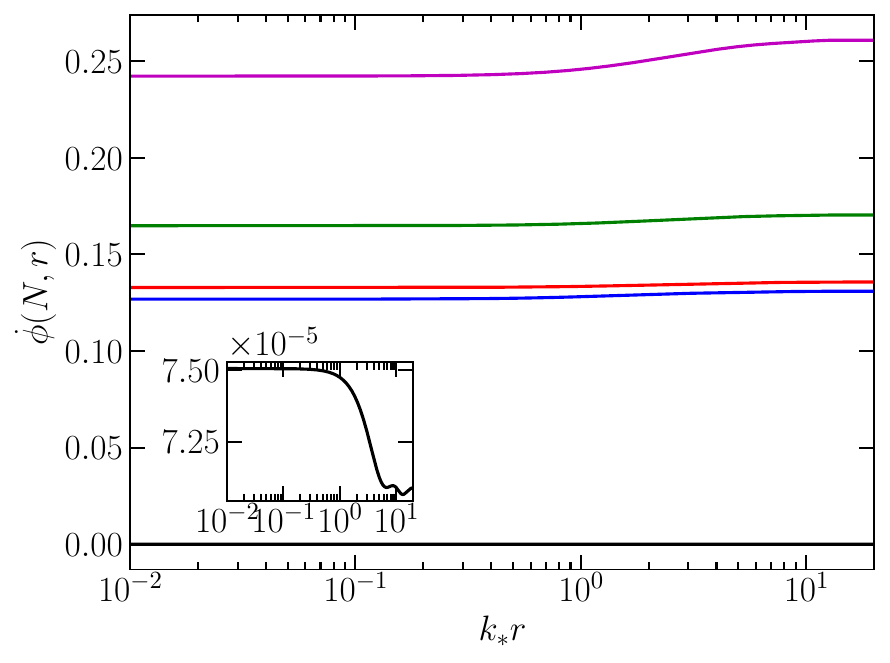}
\caption{Top-panels: Snapshots of the bubble dynamics for $\phi(N,r)$ (left-panel) and $\dot{\phi}(N,r)$ (right-panel) for the mean profile shape with $\mu=\mu^{\rm bub}_c+10^{-6}$. Bottom-panels: Same as top-panels but for subcritical amplitudes $\mu=\mu^{\rm bub}_c-10^{-6}$. The inner-sub plots show the initial shape at $N=N_*$}
\label{fig:dynamics_bubbles}
\end{figure}

%bubbleformation
We perform the numerical evolution of the scalar field $\phi(r, N)$ with radial dependence by numerically solving the Klein–Gordon field equation, Eq.~\eqref{eq:KG}. The numerical setup is similar to that used in \cite{Escriva:2023uko}, and we refer the reader for details. We rescale the radial coordinate in terms of the Hubble horizon at the scale $\tilde{r} = k_* r = r \cdot a(N_*)H(N_*) $. In particular, we discretize our grid from $\tilde{r} = 0$ up to a final grid point of $\tilde{r} \approx 20$. An example of the bubble formation dynamics can be found in Fig.~\ref{fig:dynamics_bubbles}.

Given an initial profile shape determined by Eq.~\eqref{initialfield}, \eqref{initialpi}, the fluctuation amplitude $\mu$ dictates the final outcome of the numerical evolution. There exists a critical threshold $\mu^{\rm bub}_c$ such that, for amplitudes above this value, vacuum bubbles are formed, whereas for amplitudes below it, the inflaton field does not become trapped and instead exits inflation. Two representative cases are shown in Fig.~\ref{fig:dynamics_bubbles} for $\mu = \mu^{\rm bub}_c \pm 10^{-6}$ with $n=m=0$. In the top panels, we display the case $\mu > \mu^{\rm bub}_c$, where a sufficiently large region of the inflaton field remains trapped, failing to roll down the potential slope and instead becoming stabilized with vanishing velocity $\dot{\phi} \simeq 0$. In contrast, the bottom panels illustrate the case $\mu < \mu^{\rm bub}_c$, where the fluctuations $\delta\phi$ and $\delta\pi$ are too small to prevent the inflaton from escaping the flat region, thereby ending inflation. In this sense, our results demonstrate that vacuum bubbles can indeed form during the USR stage with a flat plateu in the potential $V(\phi)$, similarly to what occurs in inflationary potentials containing a Gaussian bump~\cite{Escriva:2023uko}.

\begin{figure}[!htbp]
\centering
\includegraphics[width=3.5 in]{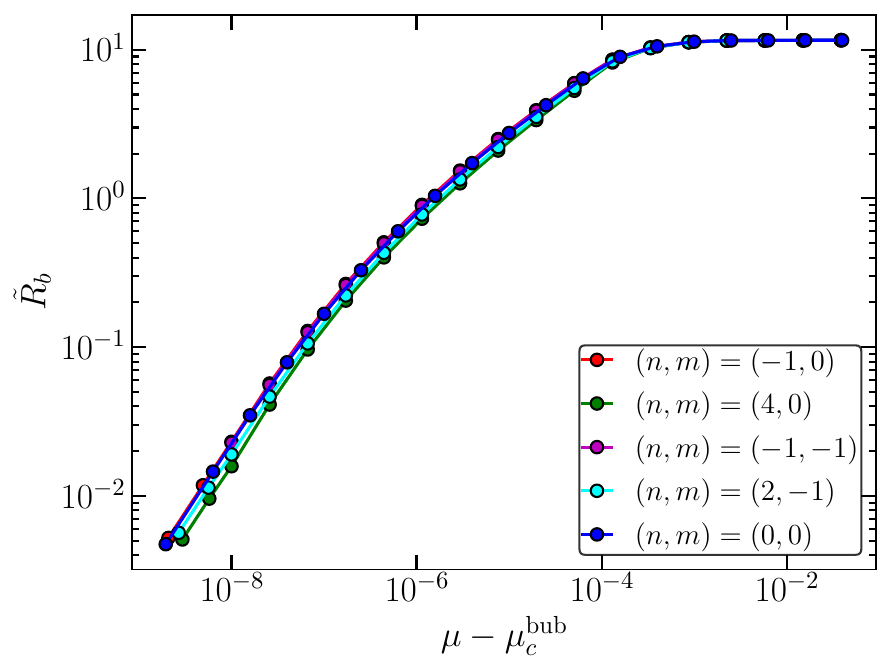}
\caption{Bubble size as a function of the fluctuation amplitude $\mu - \mu^{\rm bub}_c$ for the mean profile $(n,m)=(0,0)$ and other realizations $(n,m)$.}
\label{fig:bubble_size}
\end{figure}

Through a bisection method, we can determine the threshold values $\mu^{\rm bub}_c$ for which the inflaton field remains trapped and cannot exit inflation, forming trapped vacuum regions. The comoving size of the bubbles $\tilde{R}_b$ as a function of $\mu$ is shown in Fig.~\ref{fig:bubble_size}. For amplitudes $\mu \gg \mu^{\rm bub}_c$, the bubble size remains constant, and for even larger amplitudes, spherical shells are formed, as observed in \cite{Escriva:2023uko}. However, such cases correspond to fluctuations far from the critical thresholds and are statistically strongly suppressed. On the other hand, when $\mu-\mu^{\rm bub}_c \rightarrow 0$ becomes very small. A key qualitative difference between these results and those of \cite{Escriva:2023uko} is that we do not observe critical behavior\footnote{Specifically, the critical scaling behavior is given by $\tilde{R}_b \sim (\mu - \mu^{\rm bub}_c)^{\gamma_b}$, where $\gamma_b$ is a constant exponent.}. Instead the slope in the curves in Fig.~\ref{fig:bubble_size}  depends on the fluctuation amplitude $\mu$. A possible explanation for this difference is that, in the present case, there is no domain wall separating the bubble from the environment, as opposed to the case of a potential with a small barrier studied in \cite{Escriva:2023uko}. Once the field becomes trapped on the plateau, the portion of space remaining there can only expand (with perhaps some mild opposition due to gradient terms).\footnote{Critical behavior may still hold classically from initial conditions which are sufficiently close to $\mu^{\rm bub}_c$. From the figure, however, this would correspond to $\mu-\mu^{\rm bub}_c \ll 10^{-7}$. In this regime, quantum diffusion would completely dominate the fate of the bubble. We are interested in the range $\mu-\mu^{\rm bub}_c \gtrsim 10^{-5}$, which dominates the mass function. There, the effect of diffusion is marginal (we will briefly discuss the effects of difusion in Section \ref{diffusion}).} We also observe that the bubble size is weakly dependent on the deviation from the mean shapes, characterized by $(n,m)$.

In Table \ref{tab:num_results}, we present the numerical results for the threshold of vacuum bubble formation, $\mu^{\rm bub}_{c}$, and for the peak height, $\nu^{\rm bub}_c = \mu^{\rm bub}_{c}/\sigma_{\delta \phi}$, where we have considered both the mean profile $(n,m)=(0,0)$ and deviations from it $(n,m \neq 0)$. For positive $n$ with $m=0$, the threshold for vacuum bubble formation increases. This occurs because a positive perturbation through $\Delta_{\phi}$ is added to the mean profile, requiring a higher fluctuation amplitude $\mu$ in the backward direction to compensate and trap the field. Conversely, for $n<0$, the threshold decreases since the dispersion $\Delta_{\phi}$ has the same sign as $\bar{\delta}\phi$.

The same qualitative behaviour is found when $n=0$ with $m \neq 0$, because $\Delta_{\phi},\Delta_{\pi}$ have the same sign in the relevant range, as shown in Fig.\ref{fig:dispersions_correlations}, although the mean shapes $\delta\phi(r,0,0)$ and $\delta\pi(r,0,0)$ have opposite signs due to the anticorrelation. This implies that the largest reduction in the threshold of formation occurs when $n,m<0$, consistent with the numerical results obtained. We also note that the effect of the dispersion for $\delta\phi$ (sourced by $n$) is slightly more significant than for $\delta\pi$ (sourced by $m$); for instance, we find a larger increase (reduction) of the thresholds for $n$ ($-n$) with $m=0$ than for $m$ ($-m$) with $n=0$. When both $n$ and $m$ are nonzero and have opposite signs, the two effects combine, but the dispersion in $\delta\phi$ is slightly dominant.

\begin{figure}[!htbp]
\centering
\includegraphics[width=2.5 in]{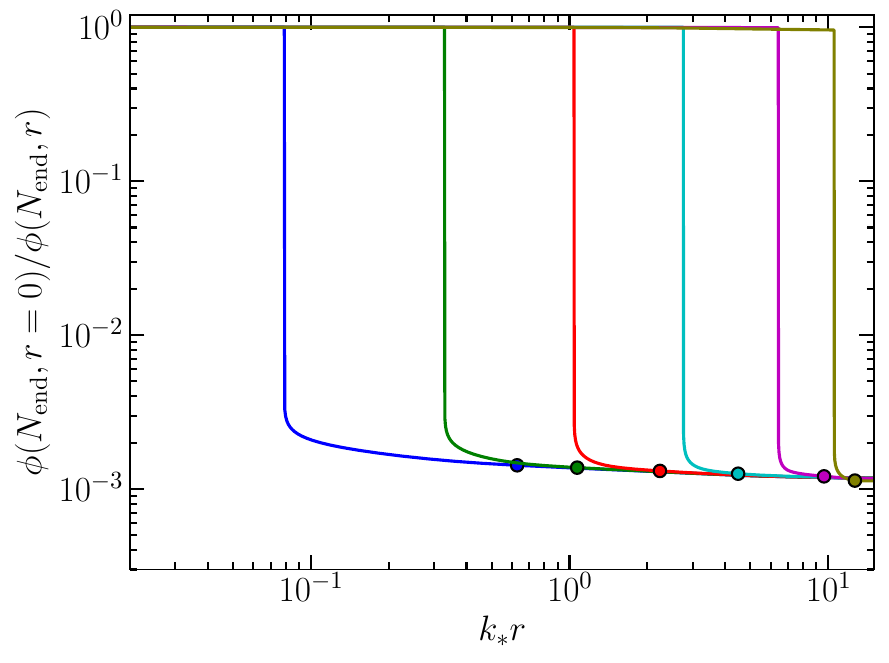}
\includegraphics[width=2.5 in]{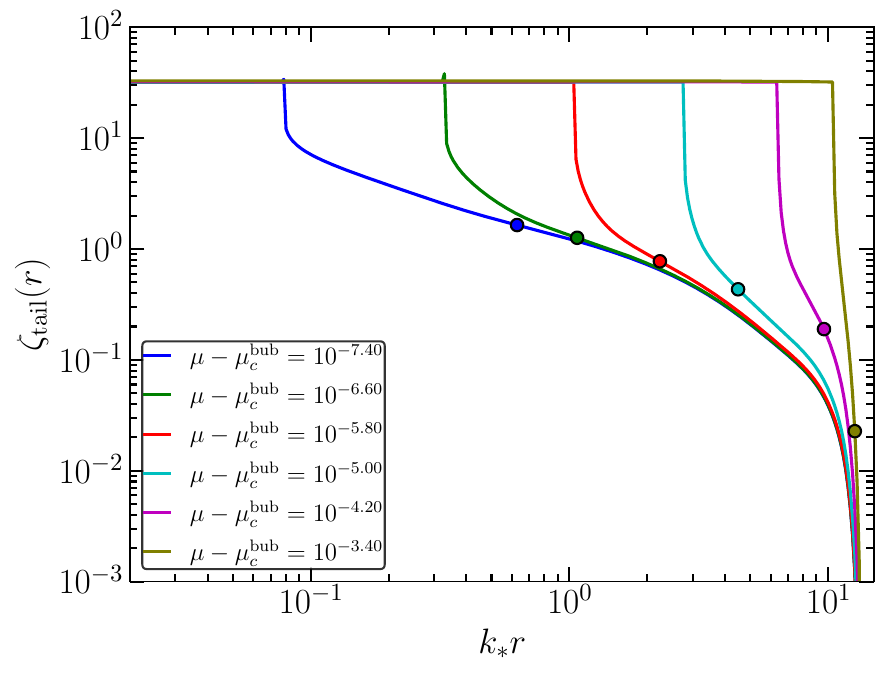}
\caption{Left-panel: Bubble shape profile at the end of inflation. Right-panel:Curvature flucation at the end of inflation from the $\delta N$ formalism. The dots represent the points where it is satisfied that $1+r_{\rm II}\zeta'(r_{\rm II})=0$.}
\label{fig:shapes_bubbles_end}
\end{figure}

By employing the $\delta N$ formalism, the curvature perturbation associated with the tail can be evaluated as $\zeta_{\rm tail} = \delta N$, where the local field value satisfies 
\begin{equation}
\phi(N_{\rm end}, r) = \phi_{\rm bkg}(N_{\rm end} - \delta N). \label{numdeltan}
\end{equation}
We find that type-II fluctuations surround the vacuum bubbles, characterized by a neck-like structure point which fulfills $1+r_{\rm II} \zeta'(r_{\rm II})=0$ \cite{Kopp:2010sh}, we define this points as the tail $r_{\rm II} \equiv r_{\rm tail}$, which are fluctuation amplitude dependent $\mu$ as can be observed in Fig.\ref{fig:shapes_bubbles_end}. This feature was also observed for vacuum bubbles formed in the scenario of \cite{Escriva:2023uko}, where the inflationary potential contains a {small barrier behind which the inflaton may be trapped. This suggests that the presence of type-II fluctuations surrounding regions where the field is trapped in an inflationary phase will be a generic phenomenon.}

\section{Quantum diffusion}\label{diffusion}

{Once a bubble is formed, quantum diffusion \cite{Vilenkin:1983xp,Starobinsky:1986fx} will cause small patches of it to escape from the inflationary plateau into the slow roll phase, and eventually thermalize. This process gradually perforates the eternally inflating domain, making it resemble a sponge. Progressively smaller holes form at increasingly later times, in a fractal-like pattern \cite{Aryal:1987vn}. Still, the overall co-moving size $\tilde R_b$ of the original region remains unchanged. Once this size falls within the horizon in the radiation era, the gravitational field of the relic sponge as a whole, carves a wormhole and migrates into a baby universe \cite{Garriga:2015fdk}. The sponge is eventually hidden behind the event horizon of a PBH, while the wormhole connecting the parent and the baby universe pinches off. Each small hole on the sponge enters slow roll and leads to an infinite thermalized region of its own, laying well within the baby universe. Such regions, stemming directly from diffusion, are very inhomogeneous in the infrared $k\lesssim k_*$, at all scales larger than the PBH scales in the power spectrum. In this sense, they correspond to rather harsh cosmological environments. On the other hand, such regions are completely inaccessible from the parent universe, to which they pose no threat.}

The fractal dimension of the sponge \cite{Aryal:1987vn,Vilenkin:1995yd} can easily be calculated by solving the Fokker-Planck equation (see e.g.\cite{Garriga:2005av} and references therein). In the present case, this reduces to the heat equation 
\begin{equation}
\partial_N F = k \ \partial^2_\phi F,
\end{equation}
for the field probability distribution $F(\phi, N) d\phi$ as a function time, with a constant diffusion coefficient $k=H^2/8\pi^2$ and absorbing boundary condition at $\phi=\phi_*$.  Consider an initial Hubble patch where the field value is given by $\phi=\phi_*-\mu$, with negligible time derivative, and $\mu \ll 2L$ . After $N$ e-foldings, the expectation value of the co-moving volume remaining in the plateau is given by the initial volume, times the survival probability,
\begin{equation}
\langle V_{com}(N) \rangle = {4\pi\over 3H^3}P_{sur}(N),
\end{equation}
where $P_{sur} =\int_{\phi_*-L}^{\phi_*} F d\phi$.
In our case, the length of the plateau satisfies $2L\gg H/2\pi$, and so the reflecting boundary condition at $\phi=\phi_*-2L$ is initially irrelevant. In this limit, we have
\begin{equation}
P_{sur}(N)= {\rm erf}\left[{\mu\over \sqrt{4kN}}\right]\approx {\mu\over\sqrt{\pi k N}}.\quad (\mu^2/4k\ll N \ll L^2/k)
\end{equation}
This transient behaviour lasts until diffusion brings the field near the reflecting boundary at $\phi=\phi_*-2L$. After that, the distribution is dominated by the lowest eigenvalue of the Laplacian $\partial_\phi^2$ in the diffusion equation, with reflecting and absorbing boundary conditions at the endpoints. The corresponding eigenvector is proportional to $\sin(\pi(\phi_*-\phi)/4L)$. The survival probability is then given by 
\begin{equation}
P_{sur}(N) \approx {4\over \pi} \sin(\pi\mu/4L) e^{-k(\pi/4L)^2 N}. \quad (N \gg L^2/k) 
\end{equation} 
Note that the co-moving volume goes to zero, decaying first as an inverse power of $N$ and then exponentially.
 On the other hand, the physical volume 
\begin{equation}
\langle V_{phy}(N)\rangle = \langle V_{com}(N) \rangle \ e^{3N},
\end{equation}
will grow as long as $H \ll L$. In this case the fractal dimension is given by 
\begin{equation}
\label{fdim}
D=3- H^2/(128 L^2),
\end{equation}
which in our case is very close to 3. In this sense, our sponges are very robust objects, growing at a healthy exponential rate.

Note, on the other hand, that here we considered the specific case of a potential with a flat plateau. For a plateau with some slope, the dynamics of vacuum-bubble formation can be significantly different, since the classical drift $v\equiv V'/3H^2$ will make it harder for the inflaton to stay trapped. However, our picture should still apply in the regime where the slope of the plateau is sufficiently small, $v\ll H/2\pi$, so that classical drift is slow compared with diffusion. In that case, when the slope is towards the absorbing boundary, and for a reflecting boundary which is far away, $L\gg H^2/(2\pi v) \gg H$, the fractal dimension of the inflating regions is given by $D=3-2\pi^2 v^2/H^2$, also very close to 3, while for $H\ll L\ll H^2/(2\pi v)$ we recover Eq.(\ref{fdim}).

%%%ordered table
%%%%%%%%%%%%%%%
\begin{table}[h!]
\centering
\scriptsize
\setlength{\tabcolsep}{4pt}
\renewcommand{\arraystretch}{1.1}
\begin{tabular}{rrrrrrrrrrr}
\hline
$n$ & $m$ & $\frac{\mu^{\rm bub}_c}{10^{-5}}$ & $\frac{\mu^{\rm adi}_{c}}{10^{-5}}$ & $\mu^{\zeta_G}_c$ & $\mu^{\rm div}_c$ & $\nu^{\rm bub}_c$ & $\nu^{\rm adi}_c$ & $\beta_1$ & $\beta_2$ & $\beta_3$\\
\hline
$0.0$ & $0.0$ & $2.61$ & $2.46$ & $0.346$ & $0.369$ & $9.21$ & $8.66$ & $1.00$ & $1.00$ & $125$ \\
$0.0$ & $-1.0$ & $2.60$ & $2.44$ & $0.344$ & $0.366$ & $9.16$ & $8.60$ & $5.40\times 10^{-1}$ & $5.25\times 10^{-1}$ & $124$ \\
$-1.0$ & $0.0$ & $2.59$ & $2.44$ & $0.345$ & $0.366$ & $9.14$ & $8.61$ & $6.18\times 10^{-1}$ & $4.79\times 10^{-1}$ & $99.0$ \\
$1.0$ & $0.0$ & $2.63$ & $2.46$ & $0.348$ & $0.372$ & $9.29$ & $8.69$ & $1.63\times 10^{-1}$ & $2.52\times 10^{-1}$ & $200$ \\
$0.0$ & $1.0$ & $2.63$ & $2.47$ & $0.348$ & $0.371$ & $9.28$ & $8.70$ & $1.86\times 10^{-1}$ & $2.17\times 10^{-1}$ & $150$ \\
$-1.0$ & $-1.0$ & $2.58$ & $2.43$ & $0.342$ & $0.363$ & $9.08$ & $8.56$ & $3.33\times 10^{-1}$ & $2.37\times 10^{-1}$ & $90.0$ \\
$0.0$ & $-2.0$ & $2.58$ & $2.42$ & $0.342$ & $0.364$ & $9.10$ & $8.55$ & $1.32\times 10^{-1}$ & $1.19\times 10^{-1}$ & $115$ \\
$1.0$ & $-1.0$ & $2.62$ & $2.45$ & $0.345$ & $0.369$ & $9.23$ & $8.63$ & $8.83\times 10^{-2}$ & $1.24\times 10^{-1}$ & $181$ \\
$-1.0$ & $1.0$ & $2.61$ & $2.46$ & $0.347$ & $0.368$ & $9.20$ & $8.66$ & $1.15\times 10^{-1}$ & $9.70\times 10^{-2}$ & $109$ \\
$-2.0$ & $0.0$ & $2.57$ & $2.43$ & $0.343$ & $0.363$ & $9.07$ & $8.57$ & $1.71\times 10^{-1}$ & $9.81\times 10^{-2}$ & $72.4$ \\
$-1.0$ & $-2.0$ & $2.56$ & $2.41$ & $0.340$ & $0.361$ & $9.02$ & $8.51$ & $8.07\times 10^{-2}$ & $5.27\times 10^{-2}$ & $80.9$ \\
$1.0$ & $1.0$ & $2.65$ & $2.48$ & $0.350$ & $0.374$ & $9.35$ & $8.74$ & $3.00\times 10^{-2}$ & $5.09\times 10^{-2}$ & $220$ \\
$-2.0$ & $-1.0$ & $2.55$ & $2.42$ & $0.341$ & $0.360$ & $9.01$ & $8.52$ & $9.15\times 10^{-2}$ & $4.84\times 10^{-2}$ & $65.2$ \\
$2.0$ & $0.0$ & $2.66$ & $2.47$ & $0.349$ & $0.375$ & $9.37$ & $8.72$ & $1.18\times 10^{-2}$ & $2.69\times 10^{-2}$ & $294$ \\
$1.0$ & $-2.0$ & $2.60$ & $2.43$ & $0.343$ & $0.367$ & $9.17$ & $8.58$ & $2.16\times 10^{-2}$ & $2.74\times 10^{-2}$ & $162$ \\
$0.0$ & $2.0$ & $2.65$ & $2.48$ & $0.350$ & $0.374$ & $9.34$ & $8.76$ & $1.55\times 10^{-2}$ & $2.01\times 10^{-2}$ & $168$ \\
$-2.0$ & $1.0$ & $2.59$ & $2.44$ & $0.345$ & $0.365$ & $9.13$ & $8.62$ & $3.21\times 10^{-2}$ & $2.03\times 10^{-2}$ & $80.5$ \\
$2.0$ & $-1.0$ & $2.64$ & $2.46$ & $0.347$ & $0.372$ & $9.31$ & $8.67$ & $6.49\times 10^{-3}$ & $1.29\times 10^{-2}$ & $257$ \\
$0.0$ & $-3.0$ & $2.56$ & $2.41$ & $0.340$ & $0.362$ & $9.04$ & $8.49$ & $1.32\times 10^{-2}$ & $1.12\times 10^{-2}$ & $106$ \\
$-2.0$ & $-2.0$ & $2.54$ & $2.40$ & $0.339$ & $0.358$ & $8.95$ & $8.46$ & $2.21\times 10^{-2}$ & $1.12\times 10^{-2}$ & $60.3$ \\
$-1.0$ & $2.0$ & $2.63$ & $2.47$ & $0.349$ & $0.371$ & $9.26$ & $8.71$ & $9.62\times 10^{-3}$ & $9.03\times 10^{-3}$ & $121$ \\
$-3.0$ & $0.0$ & $2.55$ & $2.42$ & $0.341$ & $0.360$ & $8.99$ & $8.52$ & $1.95\times 10^{-2}$ & $8.74\times 10^{-3}$ & $54.8$ \\
$-1.0$ & $-3.0$ & $2.54$ & $2.40$ & $0.338$ & $0.359$ & $8.97$ & $8.45$ & $8.03\times 10^{-3}$ & $5.00\times 10^{-3}$ & $74.8$ \\
$2.0$ & $1.0$ & $2.67$ & $2.49$ & $0.351$ & $0.377$ & $9.43$ & $8.78$ & $2.18\times 10^{-3}$ & $5.29\times 10^{-3}$ & $314$ \\
$-3.0$ & $-1.0$ & $2.53$ & $2.40$ & $0.339$ & $0.357$ & $8.94$ & $8.47$ & $1.04\times 10^{-2}$ & $4.42\times 10^{-3}$ & $49.9$ \\
$3.0$ & $0.0$ & $2.68$ & $2.48$ & $0.350$ & $0.378$ & $9.44$ & $8.76$ & $3.56\times 10^{-4}$ & $1.19\times 10^{-3}$ & $434$ \\
$-3.0$ & $1.0$ & $2.57$ & $2.43$ & $0.343$ & $0.362$ & $9.05$ & $8.57$ & $3.66\times 10^{-3}$ & $1.80\times 10^{-3}$ & $61.4$ \\
$-3.0$ & $-2.0$ & $2.52$ & $2.39$ & $0.337$ & $0.355$ & $8.88$ & $8.42$ & $2.51\times 10^{-3}$ & $1.03\times 10^{-3}$ & $44.9$ \\
$0.0$ & $3.0$ & $2.66$ & $2.50$ & $0.352$ & $0.376$ & $9.40$ & $8.81$ & $5.33\times 10^{-4}$ & $7.52\times 10^{-4}$ & $183$ \\
$-4.0$ & $0.0$ & $2.53$ & $2.40$ & $0.339$ & $0.357$ & $8.92$ & $8.47$ & $8.73\times 10^{-4}$ & $3.24\times 10^{-4}$ & $43.1$ \\
$-3.0$ & $2.0$ & $2.58$ & $2.44$ & $0.345$ & $0.365$ & $9.11$ & $8.62$ & $3.10\times 10^{-4}$ & $1.66\times 10^{-4}$ & $68.3$ \\
$1.0$ & $3.0$ & $2.69$ & $2.51$ & $0.354$ & $0.379$ & $9.47$ & $8.84$ & $8.52\times 10^{-5}$ & $1.76\times 10^{-4}$ & $268$ \\
$-4.0$ & $-1.0$ & $2.51$ & $2.39$ & $0.337$ & $0.355$ & $8.86$ & $8.42$ & $4.64\times 10^{-4}$ & $1.71\times 10^{-4}$ & $39.2$ \\
$-4.0$ & $1.0$ & $2.55$ & $2.42$ & $0.341$ & $0.359$ & $8.98$ & $8.52$ & $1.65\times 10^{-4}$ & $6.57\times 10^{-5}$ & $48.2$ \\
$0.0$ & $-4.0$ & $2.55$ & $2.39$ & $0.338$ & $0.359$ & $8.98$ & $8.44$ & $5.24\times 10^{-4}$ & $4.21\times 10^{-4}$ & $97.4$ \\
$4.0$ & $0.0$ & $2.70$ & $2.49$ & $0.352$ & $0.381$ & $9.51$ & $8.79$ & $4.23\times 10^{-6}$ & $2.09\times 10^{-5}$ & $641$ \\
$0.0$ & $4.0$ & $2.68$ & $2.51$ & $0.354$ & $0.378$ & $9.46$ & $8.86$ & $7.23\times 10^{-6}$ & $1.14\times 10^{-5}$ & $204$ \\
$0.0$ & $-5.0$ & $2.53$ & $2.38$ & $0.336$ & $0.357$ & $8.92$ & $8.39$ & $7.98\times 10^{-6}$ & $6.05\times 10^{-6}$ & $87.7$ \\
$-5.0$ & $0.0$ & $2.51$ & $2.38$ & $0.337$ & $0.354$ & $8.85$ & $8.41$ & $1.51\times 10^{-5}$ & $5.49\times 10^{-6}$ & $37.7$ \\
$0.0$ & $5.0$ & $2.70$ & $2.53$ & $0.357$ & $0.381$ & $9.52$ & $8.91$ & $3.76\times 10^{-8}$ & $6.63\times 10^{-8}$ & $228$ \\
$5.0$ & $0.0$ & $2.72$ & $2.50$ & $0.353$ & $0.384$ & $9.59$ & $8.82$ & $1.91\times 10^{-8}$ & $1.45\times 10^{-7}$ & $983$ \\
\hline
\end{tabular}
\caption{Results of the numerical simulations for the different $(n,m)$ realizations regarding the critical thresholds and the abundance of peaks for each channel of PBH production. {The definitions of $\beta_1, \beta_2,\beta_3$ are given in Eqs.~\eqref{eq:beta1}, \eqref{eq:beta2} and \eqref{eq:beta3}}.}
\label{tab:num_results}
\end{table}

\section{Adiabatic fluctuations  and generalized $\zeta[\zeta_G]$}\label{adicha}

Aside from the formation of vacuum bubbles discussed in the previous sections, another channel for PBH production exists, namely the standard adiabatic channel. In this case, density perturbations in the radiation fluid can undergo gravitational collapse if their amplitude exceeds a certain threshold.

Let us derive an expression for the curvature perturbation $\zeta=\delta N$ for given values of $\delta\phi_*$ and $\delta\pi_*$ at the conformal time $\eta=\eta_*$ on a flat slicing. Here, $\eta_*$ is the time when the background field (which determines the flat slicing) reaches the end of the USR plateau, and $\delta N$ is the additional duration of USR due to the field perturbations.

On the USR plateau, the potential is flat and the first slow roll parameter is very small. Therefore, we will assume that $H$ is constant. The general solution of the equation of motion for the field perturbation is then given by 
\begin{eqnarray}
\delta\phi&=& C_1 (c + x s) + C_2  (x c - s), \\
\delta\pi&=& C_1 (-x^2 c) + C_2 (x^2 s).
\end{eqnarray}
Here, $x=k\eta < 0$ and we have introduced the shorthand $s=\sin x, c=\cos x$.
We can invert this relation to express the coefficients $C_1$ and $C_2$ in terms of $\delta\phi_*$ and $\delta\pi_*$:
\begin{equation}
C_1 = {s_*\over x_*} \delta\phi_* + {s_*-x_* c_* \over x_*^3}\delta \pi_*%\label{C1}
, \quad
C_2 = {c_*\over x_*} \delta\phi_* +{c_*+x_*s_*\over x_*^3} \delta\pi_*.\label{C2}
\end{equation}
Here, we have introduced 
\begin{equation}
x_* \equiv k \eta_* = -(k/k_*),
\end{equation}
with $s_* = \sin x_*, c_*=\cos x_*$. Expanding in small $x^2\ll 1$ (corresponding to the long wavelength limit), we have
$c+x s = 1+{x^2/2} + O(x^4)$, and
$(s-xc)/x^3=1/3 -{x^2/30}+O(x^4).$
Hence, to leading order in $x^2$, we have
\begin{eqnarray}
\delta\phi&\approx& \left(1+{x^2\over 2}\right)C_1 - {x^3\over 3} C_2, \label{pertur1}\\
\delta\pi&\approx & -x^2 C_1 + x^3 C_2. \label{pertur2}
\end{eqnarray}
Note that $x$ is proportional to the inverse of the background scale factor $x\propto a^{-1}=e^{-N}$. The flat slicing corresponds to $N=const.$, and we can express 
\begin{equation}\label{xefold}
x= x_* e^{-(N-N_*)}.
\end{equation}
The most relevant feature of Eq. (\ref{pertur2}) is that  $\delta\pi \propto a^{-2}$ in the long wavelength limit, which is different from the background behaviour $\bar\pi\propto a^{-3}$. In this sense, long wavelength perturbations in USR do not behave like a homogeneous separate universe (at least, not a separate universe with flat spatial sections).

\begin{figure}[H]
\centering
\includegraphics[width=3 in]{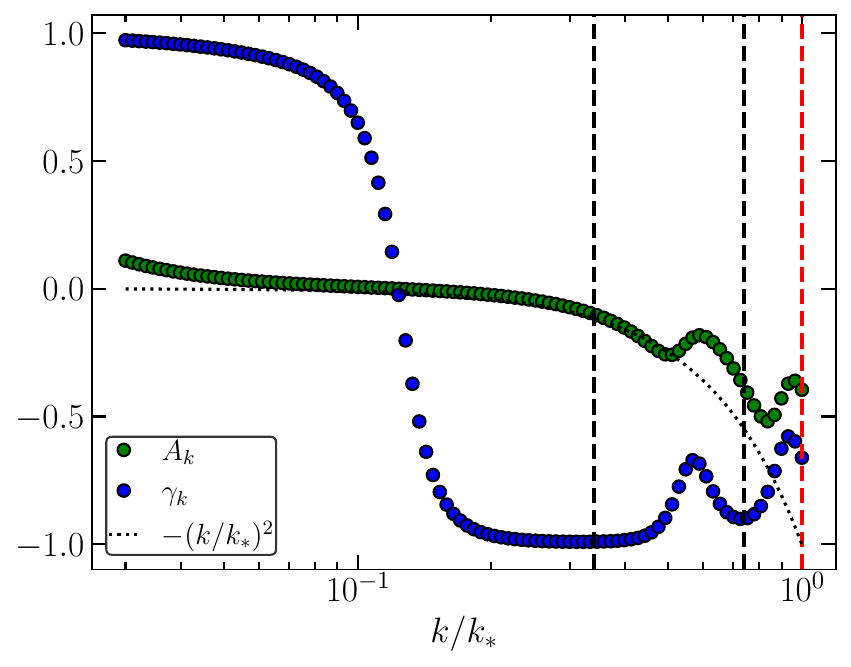}
\caption{Values of $A_k$ and $\gamma_k$ in terms of the wave-modes. The black vertical dashed lines denotes the location of the first peak of the power spectrum and the second one respectively. The vertical red line denotes the mode $k_*$.}
\label{fig:A_factor_gamma}
\end{figure}

It is also interesting to note that if the plateau is sufficiently long, we enter an attractor regime where $\delta\phi$ and $\delta\pi$ are strongly correlated, and proportional to each other:
\begin{equation}\label{pertattractor}
\delta\pi \approx -x^2 \delta\phi + O(x^3).
\end{equation}
As we shall see, the previous equation is also relevant when using Eq. (\ref{C2}) to find $C_1$ and $C_2$ from $\delta\phi_*$ and $\delta\pi_*$.
Indeed, from Fig.\ref{fig:A_factor_gamma} we see that for wavelengths $k$ which are near the peak of the power spectrum, $\delta\pi_{*k}$ and $\delta\phi_{*k}$ are strongly correlated. Therefore, in the range of $k$ where the Pearson coefficient is maximal $\gamma_k \approx -1$ (which includes the region near the maximum of the power spectrum), we can take them to be proportional:
\begin{equation}\label{proportionality}
\delta\pi_{*} \approx A_k \delta\phi_{*}.
\end{equation}
Furthermore, the figure shows that within this range, 
\begin{equation}\label{akisxx}
A_k \approx -x_*^2 = -(k/k_*)^2,
\end{equation}
in agreement with the attractor behavior (\ref{pertattractor}) for the field perturbations. The subindex $k$ indicates that the value of $A_k$ dependends on $k$. 

Let us now evaluate the curvature perturbation 
$\zeta=\delta N$ on hypersurfaces of constant $\phi=\phi_*$ 
(i.e. at the end of the USR plateau). Writing 
$$N=N_* +\delta N,$$
the excess duration $\delta N$ of USR inflation due to the field perturbation can be computed from the equation
\begin{equation}
\phi(N) = 
\phi_{\rm bkg}(N) + \delta\phi(N) = \phi_*.\label{phiisphistar}
\end{equation}
Here 
\begin{equation}
%\phi_{\rm bkg}(N) = \phi_* + (\bar\pi_*/3)(1-e^{-3(N-N_*)}), \label{bkgsol}
\phi_{\rm bkg}(N) = \phi_* + (\dot{\phi}_{\rm bkg,*}/3)(1-e^{-3(N-N_*)}), \label{bkgsol}
\end{equation}
is the background solution. Also, from (\ref{pertur1}) and (\ref{C2}) for small $x_*$, and using (\ref{proportionality}), we may write
\begin{equation}\label{deltaphin}
\delta\phi \approx [(1+x^2/2)(1-x_*^2/6+A_k/3)-(x^3/3)(1/x_* +A_k /x_*^3)] \delta\phi_*.
\end{equation}
Using (\ref{deltaphin}),(\ref{bkgsol}), and  (\ref{xefold}) in (\ref{phiisphistar}), we find
\begin{equation}
%\zeta_G(N_*) \equiv -{\delta\phi_*\over \bar\pi_*} ={1\over 3}{1-e^{-3\delta N}\over (1-x_*^2/6+A_k/3)(1+(x_*^2/2)e^{-2\delta N})-(1/3)(x_*^2+A_k)e^{-3\delta N}}.
\zeta_G(N_*) \equiv -{\delta\phi_*\over \dot{\phi}_{\rm bkg,*}} ={1\over 3}{1-e^{-3\delta N}\over (1-x_*^2/6+A_k/3)(1+(x_*^2/2)e^{-2\delta N})-(1/3)(x_*^2+A_k)e^{-3\delta N}}.
\end{equation}
The denominator simplifies considerably  by using (\ref{akisxx}). From $\zeta=\delta N$, we then have
\begin{equation}
\label{eq:delta_N}
    \zeta_G = \frac{1}{3} \frac{1- e^{-3 \zeta}}{1+\frac{A}{2}(1-e^{-2 \zeta})}.
\end{equation}
Here, and in what follows, we omit the argument $N_*$ in $\zeta_G$,  and we also drop the subindex $k$ in $A_k$. For practical purposes we may use the value of $A$ at the peak of the power spectrum, or its weighed average within a range of $k$ near the peak.

Our next task is to invert the analytic template (\ref{eq:delta_N}) so that we can express the actual curvature perturbation $\zeta$ as a function of the Gaussian variable $\zeta_G$. Note that $\zeta\to \infty$ corresponds to a finite value of $\zeta_G$, given by 
\begin{equation}
\zeta_{G}^{\rm div} \equiv \frac{1}{3} \frac{1}{(1+A/2)}.
\label{eq:mu_divergent}
\end{equation}
Values of $\zeta_G > \zeta_{\rm G}^{\rm div}$ do not map into finite adiabatic curvature perturbations $\zeta$. The corresponding probability in the Gaussian distribution goes into a different channel for PBH production, where in certain inflationary patches, the inflaton would classically be stuck in the USR plateau for an indefinite amount of time. Because of additional quantum fluctuations, a patch that is in the plateau enters a diffusion regime, which results in eternal inflation within the patch. From the point of view of outside observers, this will lead to a Type II perturbation. Once this perturbation enters the horizon in the radiation era, it most likely leads to a Type B PBH \cite{Uehara:2024yyp}, which features bifurcated trapping horizons in its formation process.

To solve Eq.~\eqref{eq:delta_N} for $\xi \equiv e^\zeta$, we note that this is a cubic equation, whose only real root can be found exactly through the \textit{Cardano} formula \cite{cardano1968}. The details are given in Appendix \ref{appendix:cardano_formula}, and here we provide the final result $\zeta[\zeta_G]$,

\begin{equation}
\label{zetanal}
%\zeta[\zeta_G] = \ln 2 - \ln\left( A \, \zeta_G + %g^{1/3}_{+} + g^{1/3}_- \right) ,
\zeta[\zeta_G] = -\ln \left[ \frac{\zeta_G A}{2}+\left(-\frac{q}{2}+\sqrt{\Delta}\right)^{1/3}+\left(-\frac{q}{2}-\sqrt{\Delta}\right)^{1/3} \right] ,
\end{equation}
where
\begin{eqnarray}
&q = -1+\zeta_G/\zeta^{\rm div}_G-A^3 \zeta^3_G/4,\\
&\Delta =  \left[6(A+2)(\zeta_G-\zeta_G^{\rm div}) \right]\left[ 6(A+2)(\zeta_G-\zeta_G^{\rm div}) -2A^3 \zeta^3_G\right]/64,
%&g_{\pm}(\zeta_G) = 4 - 6 (2 + A) \zeta_G + A^3 %\zeta_G^3 \pm 2 \sqrt{f},\\
%&f(\zeta_G) =(2 - 3 (2 + A) \zeta_G) (2 - 3 (2 + A) %\zeta_G + A^3 \zeta_G^3).
\end{eqnarray}
with $\zeta_G<\zeta^{\rm div}_G$. Expanding the exact expression Eq. (\ref{zetanal}) for small $A$ we find:
%\begin{equation}
%\label{eq:zeta_analitical}
%\zeta[\zeta_G] = \ln 2 - \ln\left( A \, \zeta_G + g^{1/3}_{+} + g^{1/3}_- \right) ,
%\end{equation}
%where
%\begin{eqnarray}
%&g_{\pm}(\zeta_G) = 4 - 6 (2 + A) 
%\zeta_G + A^3 \zeta_G^3 \pm 2 \sqrt{f},\%\
%&f(\zeta_G) =(2 - 3 (2 + A) \zeta_G) (2 %- 3 (2 + A) \zeta_G + A^3 \zeta_G^3)
%\end{eqnarray}
%Expanding in $A$, we have 
\footnote{To first order in $A$, the expansion can also be cast in the alternative form
\begin{equation}
\zeta[\zeta_G] =-{1\over 3}\ln[(1-3\zeta_G-{3\zeta_G A\over 2} (1-(1-3\zeta_G)^{2/3})+...],
\end{equation}
which may be useful for comparison with extended $\delta N$ \cite{Artigas:2024ajh}.}
\begin{equation}
\zeta[\zeta_G]=-\ln[(1 - 3 \zeta_G)^{1/3} +  
(1 -(1 - 3 \zeta_G)^{-2/3})\ (\zeta_G /2)\ A+ {
   (\zeta_G/2)^2 \over (1 - 3 \zeta_G)^{5/3}}\ A^2+...].\label{Aexp}
\end{equation}
In the limit $A\to 0$ this leads to the familiar result \cite{Biagetti:2018pjj,Passaglia:2018ixg,Pi:2022ysn,Artigas:2024ajh,Ballesteros:2024pbe,Wang:2024wxq,Caravano:2025diq,Cruces:2025typ} %{\color{red} Other references?}.
\begin{equation}
\zeta[\zeta_G]=-\frac{1}{3}\ln(1-3 \zeta_G).
\label{eq:template_USR_tipical}
\end{equation}
It has recently been suggested \cite{Iovino:2024sgs} that an absolute value should be used inside the logarithm in (\ref{eq:template_USR_tipical}). 
%That discussion is based on the analysis of Ref. \cite{Lyth:2004gb}, which applies to the case where the pressure is completely determined by the energy density, a condition which is indeed met in USR with a flat plateau. 
Consider, for the sake of argument, the differential number of e-foldings between a flat slicing and a final slicing with energy density given by $\rho({\bf x}, N)=\dot\phi^2({\bf x},N)/2 + V_0$, where $V_0$ is the constant height of the plateau. From the field equation of motion, with gradient terms neglected, we have $\dot\phi({\bf x},N) \approx \dot\phi({\bf x},N_*) e^{-3(N - N_*)}$. 
This leads to 
%a differential number of e-foldings 
\begin{equation}
    \delta N \equiv N-N_*\approx -(1/6) [\ln\{ \dot\phi^2({\bf x}, N)/\dot\phi^2({\bf x},N_*)\}].\label{squares}
\end{equation}
Writing $\phi({\bf x},N)=\phi_{\rm bkg}(N)+\delta\phi({\bf x},N),$
 and using the equations of motion, we have 
$
\delta\dot\phi({\bf x},N)+3\delta\phi({\bf x},N)=\delta\dot\phi({\bf x},N_*)+3\delta\phi({\bf x},N_*). 
$
We now choose the final hypersurface to be determined by $\phi({\bf x}, N)=\phi_*$, corresponding to the field value where the USR plateau ends. Then $\delta\phi({\bf x}, N)=0$, and we have 
$$
\zeta= \delta N \approx -{1\over 6} \ln\left(1+3{\delta\phi({\bf x},N_*)\over \dot\phi({\bf x},N_*)}\right)^2.$$
Neglecting the velocity perturbation at the time $N_*$ (which in the language above corresponds to setting $A=0$), we recover (\ref{eq:template_USR_tipical}). Written in this way, the argument of the logarithm is always positive, giving the false impression that the result should be valid for any value of $\delta\phi$. However, this is not the case. Note, first of all, that for $\delta \phi = -(1/3)\dot\phi$, the expression inside the logarithm vanishes, and consequently $\delta N$ diverges. From Eq. (\ref{squares}), it is clear that this  corresponds to the case where the field stops right at the edge of the USR plateau, without ever proceeding to the next phase of slow roll. Likewise, for $\delta \phi < -(1/3)\dot\phi$, the ratio $\dot\phi({\bf x}, N)/\dot\phi({\bf x},N_*)$ would have to be negative. This is simply not possible on the USR plateau, where, in the long wavelenth limit, the field velocity cannot change sign. We conclude that trajectories with $\delta \phi < -(1/3)\dot\phi$ never make it to the end of the USR plateau, becoming therefore trapped. This picture is in good agreement with our numerical simulations.

\begin{figure}[H]
\centering
\includegraphics[width=3 in]{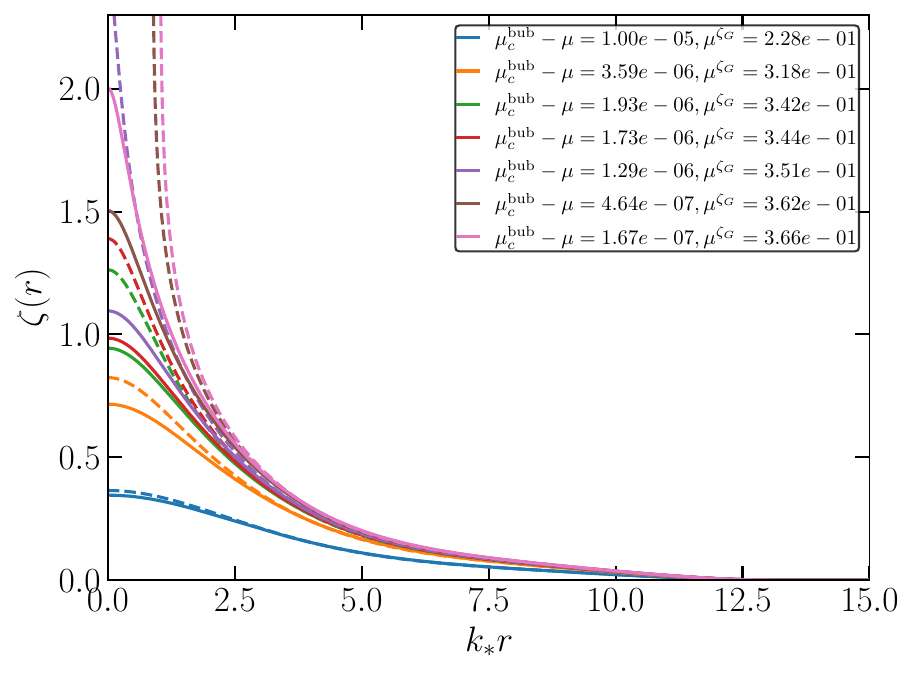}
\includegraphics[width=3 in]{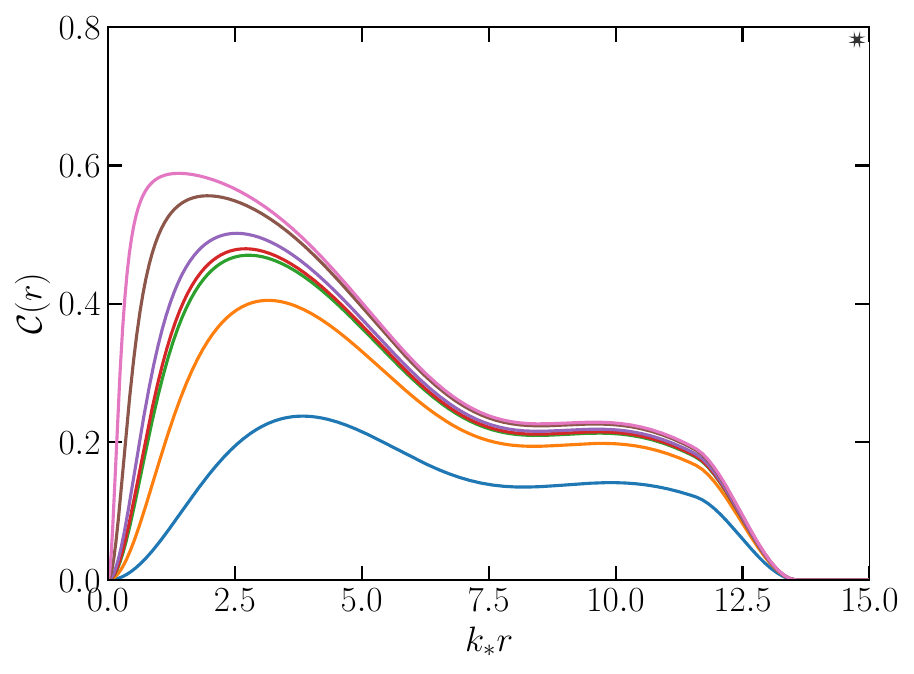}
\caption{Left-panel: Shapes of $\zeta$ for different amplitudes $\mu$ for the mean profile $(n,m)=(0,0)$. The solid line corresponds to the numerical $\zeta$ from $\delta N$, whereas the dashed line to the analytical template Eq.~\eqref{zetanal} with $A=-0.1$. Right-panel: Corresponding compaction functions $\mathcal{C}(r)$.}
\label{fig:shapes_C_zeta_deltaN}
\end{figure}

On the other hand, from Fig. \ref{fig:A_factor_gamma}, the value of $A\approx -0.1$ near the peak of the distribution is significant, and cannot be completely neglected.\footnote{Here, we are mostly interested in the non-perturbative regime, relevant for PBH production. However, a perturbative expansion may be useful for studying stochastic backgrounds of induced gravitational waves. Expanding  (\ref{Aexp}) in powers of $\zeta_G$, as $\zeta=\zeta_G + (3/5) f_\mathrm{NL}(k) \zeta_G^2 +...$ we find that the effective non-Gaussianity parameter
$f_\mathrm{NL}(k) = (5/2) -(5/3)(k/k_*)^2 -(5/12)(k/k_*)^4+...$ runs with $k$. This behaviour is expected to hold in the regime $(k/k_*)<0.5$, where the Pearson coefficient is maximal, $\gamma_k \approx -1$ (see Fig. \ref{fig:A_factor_gamma}). A qualitatively similar running of $f_\mathrm{NL}$, decreasing towards the UV, has been reported recently in \cite{Namjoo:2025hrr}. For earlier works on this topic, see \textit{e.g.} \cite{Passaglia:2018ixg,Ragavendra:2020sop,Ozsoy:2021pws,Motohashi:2023syh}.} 
In the left panel of Fig.~\ref{fig:shapes_C_zeta_deltaN}, we compare the analytical template Eq.~\eqref{eq:delta_N}  with the $\zeta$ computed via the numerical $\delta N$ method described around Eq. (\ref{numdeltan}), for $A=-0.1$ and $(n,m)=(0.0)$. Both are in good agreement when the amplitude $\mu^{\zeta_G}$ is low. Nonetheless, significant deviations appear as $\mu^{\zeta_G}$ approaches the singularity predicted by Eq.~\eqref{eq:mu_divergent} $\zeta_G^{\rm div}\approx 0.351$, where the nonlinear mapping would diverge. This value is approximately half-way between the naive value of $1/3$ which is obtained in the limit $A\to 0$, and the true divergent value 
$$\mu^{\rm div}_c \equiv \mu^{\rm bub}_c/\dot{\phi}_{\rm bkg}(N_*) \approx 0.369,$$ corresponding to the threshold for vacuum bubble formation, which is found numerically.
%from which we should expect
%\footnote{\Blue{Ref.~\cite{Iovino:2024sgs} suggests that an absolute value should be considered in the argument of the logarithmic relation $\zeta[\zeta_G]$ which would remove the divergence and therefore imply that vacuum bubble formation is not expected. In our work, however, we directly solve the field equations numerically and explicitly observe the formation of vacuum bubbles, thereby demonstrating the presence of this alternative PBH formation channel in our setup. This result is further supported by the close agreement between the critical threshold for vacuum-bubble formation, $\zeta_{\rm G}^{\rm div}$, and the critical value inferred from the divergence of the logarithmic relation, $\mu_{c}^{\rm div}$. For this reason, we do not impose or consider an absolute value in the logarithmic mapping in our analysis, as doing so would exclude a channel that is dynamically realized in our numerical solutions. We also remark that in our derivation for $\zeta[\zeta_G]$ in Eq.~\eqref{zetanal}, an absolute value for the logarithm argument is not realized.}} 
In the right panel of Fig.~\ref{fig:shapes_C_zeta_deltaN}, we also show the corresponding profiles of the compaction function \cite{Shibata:1999zs}, $\mathcal{C}(r) = (2/3)\left[1 - (1 + r\zeta'(r))^{2}\right]$, a useful quantity for characterizing the threshold for PBH formation. These profiles are characterized by a central peak followed by a gradual decay of the mass excess at sufficiently large radii $k_* r$, which resembles the shapes in \cite{Escriva:2023qnq}, with secondary peaks beyond the main one that triggers gravitational collapse.

For each curvature profile $\zeta(r)$ generated with an amplitude $\mu < \mu^{\rm bub}_c$ (which therefore does not lead to the formation of vacuum bubbles), we perform relativistic numerical simulations in spherical symmetry to track the collapse of adiabatic curvature fluctuations, from super-horizon scales up to the moment of PBH formation. In particular, we use the numerical code SPriBHoS-II \cite{Escriva:2025eqc} (see also \cite{escriva_solo}), which employs a new methodology based on the Misner–Sharp formalism \cite{PhysRev.136.B571} to follow the hydrodynamic collapse of generic curvature fluctuations (type-I and type-II) by using the trace of the extrinsic curvature. We refer the reader to that work for further details on the numerical procedure. We then iterate over different realizations of $\mu$ to determine the critical threshold conditions, defining this value as $\mu^{\rm adi}_{c}$. This can be converted into a threshold associated with the Gaussian curvature fluctuation $\zeta_G$, denoted by $\mu^{\zeta_G}_{c}= \mu^{\rm adi}_c/\dot{\phi}_{\rm bkg}(N_*)$, from which we define the peak height $\nu^{\rm adi}_{c}=\mu^{\zeta_G}_{c}/\sigma_{\zeta_G}(N_*)$, being $\sigma_{\zeta_G}(N_*)$ the width of the power spectrum of $\mathcal{P}_{\zeta_G}(N_*,k)$. The corresponding numerical results for these cases are shown in Table~\ref{tab:num_results}. For the mean profile $(n,m)=(0,0)$ we find that $\mu^{\rm adi}_c \approx 2.455 \cdot 10^{-5}$ and $\mathcal{C}_c(r_m) \approx 0.485$ (being $r_m$ the location of the maximum of the critical compaction function), and we also list the threshold for the other realizations $(n,m)$. 

%When comparing 
Let us now compare
the threshold obtained for the mean profile, $(n,m)=(0,0)$, with that of the analytical template, using the case $A = -0.1$.
%(motivated by Fig.\ref{fig:A_factor_gamma}, which is related to the maximum of the power spectrum). %we find good agreement with the numerical result from the $\delta N$, with a deviation in the threshold of about $\sim 3\%$, even the significant deviation when approaching the divergent point discussed previously.
%In spite of the limitations of the analytic template near the divergent point, which we mentioned above, 
%this still provides reasonably good agreement with the numerical $\delta N$ computation for lower values of $\zeta_G$, closer to the adiabatic threshold.
%For instance, for $(n,m)=(0,0)$ 
the full numerical simulation gives $\mu^{\zeta_G}_c = 0.346$, while using the template Eq.~\eqref{zetanal} we obtain $\mu^{\zeta_G}_c \approx 0.334$, which is about $\sim 3\%$ lower. The difference is even smaller ($\sim 0.6\%$) when comparing with the peak value of the compaction function, $\mathcal{C}_c(r_m) \approx 0.482$, using the same template. On the other hand, when taking $A = 0$, corresponding to the template often used for USR Eq.~\eqref{eq:template_USR_tipical}, we obtain $\mu^{\zeta_G}_c \approx 0.317$ and $\mathcal{C}_c(r_m) \approx 0.472$, with significantly larger deviations of $\sim 8\%$ and $\sim 3\%$, respectively, when compared with the numerical $\delta N$. We emphasize that an accurate determination of the PBH formation threshold, accounting for profile dependence, is crucial for reliably quantifying the resulting PBH abundance. In this work, such accuracy is achieved through dedicated numerical simulations.

For all cases considered $(n,m)$, the threshold lies in the type-I region of the collapse; i.e., the critical condition is characterized by type-I fluctuations (the areal radius $R \sim r e^{\zeta(r)}$ is a monotonically increasing function, and no neck points exist) and the numerical evolution corresponds to type-A PBHs (no formation of bifurcated horizons appear). When comparing our numerical results with the analytical estimations from \cite{universal1,Escriva:2025rja} to determine the threshold condition of the collapse, based on the curvatures of the peaks of the compaction function, we find good agreement, with deviations within $1\% \text{--} 3\%$, as expected. It is worth noting that these curvature profiles $\zeta(r)$ obtained from $\delta N$ formalism are a natural outcome of a realistic inflationary model, and not of a specific analytical template.

When evaluating the effect of the threshold in the adiabatic channel for different $(n,m)$, we find a qualitatively similar behavior to that of the bubble channel; i.e., positive values of $n,m$ tend to increase the threshold, whereas negative values tend to decrease it. We note that the threshold for bubble production is much larger that the quantum drift, $H /2\pi \approx 1.55 \times 10^{-6}$. On the other hand, the difference between the thresholds for bubble-induced and adiabatic PBH formation is of the order of a single quantum step of the inflaton, $\delta\phi \sim  H /2\pi$, which represents the fluctuation in one Hubble time. Forming an adiabatic PBH requires the field to linger for roughly one additional e-fold on the plateau relative to the background, during which it undergoes an extra fluctuation of order $H / 2\pi$. 
%In this sense, incorporating quantum drift within a fully quantum framework, rather than a purely classical one, may allow us to better quantify the impact of this effect on the results. This, however, goes beyond the scope of this work.
%This issue requires further study, and we plan to come back to it in future work.
A more comprehensive analysis of this aspect requires further investigation, and is left for future research.

To quantify the effect of the dispersion shapes on PBH abundance, in Table \ref{tab:num_results} we use the following definitions:
\begin{align}
    \label{eq:beta1}
    \beta_1(n,m) &= \frac{\mathcal{I}(\nu^{\rm bub}_{c}(n,m),\infty)}{\mathcal{I}(\nu^{\rm bub}_{c}(0,0),\infty)}\frac{\Upsilon(n,m)}{\Upsilon(0,0)},\\ 
    \label{eq:beta2}
    \beta_2(n,m) &= \frac{\mathcal{I}(\nu^{\rm adi}_{c}(n,m),\nu^{\rm bub}_{c}(n,m))}{\mathcal{I}(\nu^{\rm adi}_{c}(0,0),\nu^{\rm bub}_{c}(0,0))}\frac{\Upsilon(n,m)}{\Upsilon(0,0)}, \\ 
    \label{eq:beta3}
    \beta_3(n,m) &= \frac{\mathcal{I}(\nu^{\rm adi}_{c}(n,m),\nu^{\rm bub}_{c}(n,m))}{\mathcal{I}(\nu^{\rm bub}_{c}(n,m),\infty)},
\end{align}
where 
\begin{equation}
\mathcal{I}(\alpha,\gamma) = \int_{\alpha}^{\beta}e^{-\nu^2/2}(\nu^3-3\nu) d\nu,
\end{equation}
and we use the critical thresholds $\nu^{\rm adi}_c=\mu^{\zeta_G}_c/\sigma_{\zeta_G}$ and $\nu^{\rm bub}_c=\mu^{\rm bub}_c/\sigma_{\delta\phi}$ previously defined for both channels. 
Here, $\beta_1$ corresponds to the probability of forming bubbles from profiles which are more than $(n,m)$ standard deviations away from the mean, relative to the total probability of forming bubbles, and $\beta_2$ is the same concept applied to the adiabatic channel. Finally, $\beta_3$ is the probability of forming adiabatic PBH's divided by the probability of forming bubbles, from profiles which are $(n,m)$ standard deviations away from the mean.

%$\beta_1$ is the ratio of the integrated abundance of peaks in the adiabatic channel with fixed $(n,m)$ compared to the mean profile $(0,0)$. $\beta_2$ is defined similarly, but for the bubble channel, and $\beta_3$ is the ratio of the adiabatic to the bubble channel for a specific $(n,m)$ case. 

Our results indicate that the deviations away from the mean profile are small. Positive $n,m$ tend to increase the threshold for formation, and at the same time, these realizations are statistically suppressed relative to the mean shape; therefore, they hardly contribute compared with the mean profile $(0,0)$. The situation is different for negative $n,m$: in this case, there is a 
%chance that these realizations contribute to a higher abundance due to the 
reduction in the threshold, but we find that this is not enough to compensate for the statistical suppression.
%is stronger than the threshold reduction, 
Hence, the mean profile remains responsible for the largest contribution to the abundance of peaks, as reflected in the values of $\beta_1$ and $\beta_2$. For the different realizations, we also find that the contribution of the adiabatic channel is always higher than that of the bubble channel. Nonetheless, we observe a general trend that negative $n,m$ tend to make vacuum bubbles more abundant than positive $n,m$.

\section{PBH mass functions}\label{mafu}

Using the previous numerical results, we can determine the mass function contribution from each channel and for the different realizations of the dispersion shapes $(n, m)$. Following the methodology employed in \cite{Escriva:2023uko}, based on the statistics of curvature peaks of $\zeta$ \cite{1986ApJ...304...15B} to account for both channels of PBH production (see also \cite{Yoo:2018kvb,Yoo:2020dkz,Kitajima:2021fpq,Escriva:2023nzn,Pi:2024ert} in other scenarios), the mass function for the adiabatic channel is given by

\begin{equation}
\label{eq:mass_function_adiabatic}
    f_{\rm a}(M_a) = \frac{M_{\rm a}(\nu_a)\mathcal{N}_{\rm pk}(\nu_a(M_a))}{\rho_{\rm critical} \Omega_{\rm DM}} \Bigg | \frac{d \ln M_a(\nu_a)}{d \nu_a} \Bigg |^{-1},
\end{equation}

where $\Omega_{\rm DM}$ is the dark matter component of the Universe, $M_{a}$ is the mass function associated to the adiabatic PBHs that we numerically compute for different amplitudes $\nu_{a}>\nu_{a,c}$ with the relativistic numerical simulations (see \cite{Escriva:2025eqc} for details of the numerical procedure). We specifically find the expected critical behavior for the PBH mass \cite{Niemeyer:1999ak}, with 
$M_{a}(\mu) \sim \mathcal{K}_{a}(\mu-\mu^{\rm adi}_{c})^{\gamma_a}$, where $\gamma_{a} \approx 0.356$ and $\mathcal{K}_{a} \approx 16$ with the peak of the compaction function located at $k_* r_m \approx 2.8$ for the mean profile.

In the high-threshold regime, $\nu_c \gg 1$, the number density of peaks, $\mathcal{N}_{\rm pk}(\nu)$, can be approximated as \cite{1986ApJ...304...15B}
\begin{equation}
\label{eq:peak_number}
\mathcal{N}_{\rm pk}(\nu)  d\nu = \left( \frac{\sigma_1}{\sqrt{3} \sigma_0} \right)^3 e^{-\nu^2/2}(\nu^3 - 3 \nu)    d\nu,
\end{equation}
where $\sigma_l$ are the different spectral moments associated to the power spectrum $\mathcal{P}_{\zeta_G}(N_*,k)$ (adiabatic channel) and $\mathcal{P}_{\delta \phi}(N_*,k)$ (bubble channel) with 
\begin{equation}
\sigma^2_p = \int k^{2p} \mathcal{P}(N_*,k) d \ln k
\end{equation}
The same applies to the bubble channel for PBH formation, leading to 
\begin{equation}
\label{eq:mass_function_bubble}
    f_{\rm b}(M_b) = \frac{M_{ b}\mathcal{N}_{\rm pk}(\nu_b(M_b))}{\rho_{\rm critical} \Omega_{\rm DM}} \Bigg | \frac{d \ln M_b(\nu_b)}{d \nu_b} \Bigg |^{-1},
\end{equation}
We follow the analytical estimation from \cite{Escriva:2023uko} for the PBH mass associated with type-II fluctuations surrounding the vacuum bubble. In essence, we take into account the numerical results for the formation of PBHs from vacuum bubbles \cite{vacum_bubles,2017JCAP...04..050D}, where it was found that the mass of PBHs from the bubbles is a factor of $5.6$ times the mass of the cosmological horizon when the comoving size of the bubble reenters the cosmological horizon. We assume that the mass is given by the mass of the cosmological horizon when the comoving scale $k_{\rm tail}(\nu_b)=(r_{\rm tail}(\nu_b)e^{\zeta(r_{\rm tail}(\nu_b))})^{-1}$ reenters the horizon, multiplied by a factor $F = 3$,

\begin{equation}
 M_{b}(\nu_b) = F M_{k}(k_{\rm tail}(\nu_b)) .  
\end{equation}

A more precise determination of the mass associated with this type of type-II fluctuations surrounding the bubbles is left for future research. Then in total, the PBH abundance from both channels is computed as,

\begin{equation}
    f_{PBH}^{\rm tot} = \int_{-\infty}^{\infty} f_{a}(M_a)d \log M_a + \int_{-\infty}^{\infty} f_{b}(M_b) d \log M_b.
\end{equation}

\begin{figure}[!htbp]
\centering
\includegraphics[width=4 in]{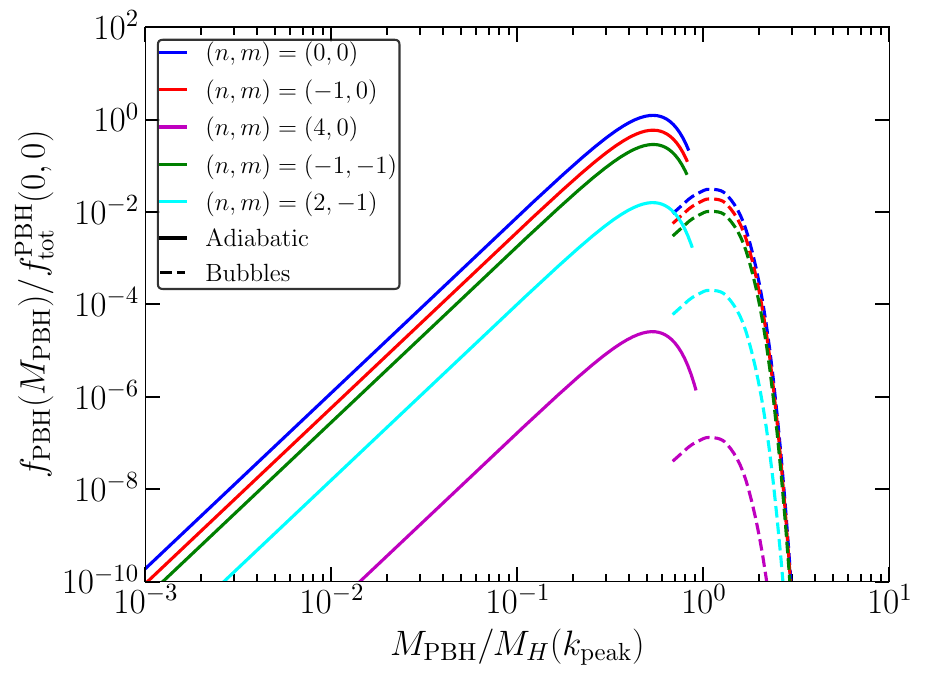}
\caption{PBH mass functions from the adiabatic and bubble channels for different realizations $(n,m)$. The different colors denote the $(n,m)$ combinations, with solid lines referring to the adiabatic channel and dashed lines to the bubble channel.}
\label{fig:pbh_mass_function}
\end{figure}

The numerical result is shown in Fig.~\ref{fig:pbh_mass_function}, where we can observe the different mass functions, and in Table~\ref{tab:num_results_fpbh}, where we report the values that represent their contribution to the dark matter abundance. The maximum PBH abundance is obtained for the mean profile $(n,m)=(0,0)$, with $f^{\rm tot}_{\rm PBH} \approx 0.492$ with $k_{\rm peak} \approx 1.09 \cdot 10^{13} \textrm{Mpc}^{-1}$, and we rescale the other cases relative to this value\footnote{The total PBH contribution to dark matter depends on the height of the peaks $\nu_c$ but also on the scale at which the power spectrum peaks, $k_{\rm peak}$ (see, for instance, Fig.15 in \cite{Escriva:2024lmm}). Achieving a desired PBH fraction can be done by shifting the peak wavenumber $k_{\rm peak}$ through the parameters of the inflationary potential with the same fixed $\nu_c$.}. The other realizations $(n,m)$ are suppressed with respect to the mean profile, giving a smaller contribution to the mass function; this also holds for the additional realizations listed in Table~\ref{tab:num_results}. This is consistent with the values of $\beta_1$ and $\beta_2$, which show a reduction in the integrated number of peaks, a trend that is directly reflected in the PBH mass function.

We also find that the adiabatic channel becomes dominant, giving—for the mean profile $(0,0)$—a contribution approximately $50$ times larger than that from vacuum bubbles. Consequently, only about $\sim 2\%$ of the total PBH dark matter abundance is contributed by vacuum bubbles.

\begin{table}[!htbp]
\centering
\label{tab:num_results_fpbh}
\renewcommand{\arraystretch}{1.2} % Adds vertical spacing between rows
\begin{tabular}{cccc}
\toprule
$n$ & $m$ & $f^{\rm tot}_{\rm PBH}(n,m)/f^{\rm tot}_{\rm PBH}(0,0)$ & $f^{\rm tot}_{\rm a}(n,m)/f^{\rm tot}_{\rm b}(n,m)$ \\
\midrule
$0.0$ & $0.0$  & $1.0$ & $51.68$ \\
$-1.0$ & $0.0$  & $4.78\times 10^{-1}$ & $39.71$ \\
$-1.0$ & $-1.0$ & $2.35\times 10^{-1}$ & $36.05$ \\
$2.0$ & $-1.0$   & $1.29\times 10^{-2}$ & $104.06$ \\
$4.0$ & $0.0$   & $2.08\times 10^{-5}$ & $260.07$ \\
\bottomrule
\end{tabular}
\caption{Numerical values of the PBH abundance and the relative contribution between adiabatic and bubble channels.}
\end{table}

\section{Conclusions}\label{conclusions}

We have studied PBH production in a single-field inflationary model containing a flat USR plateau, carrying out a joint analysis of two coexisting channels: (i) vacuum bubbles produced when sufficiently large backward fluctuations traps certain inflating regions on the on the plateau, and (ii) PBHs from the gravitational collapse of adiabatic curvature perturbations, which are enhanced due to USR. Our results combine linear-statistics arguments for initial shapes, a systematic exploration of shape dispersion, and fully non-linear numerical evolution.

For the bubble channel we find the following. There is a well-defined critical backward amplitude for the inflaton perturbation, $\mu^{\rm bub}_c$, above which a region is trapped on the plateau and a vacuum relic forms. The comoving bubble radius $\tilde R_b$ saturates for $\mu\gg\mu^{\rm bub}_c$; in the parameter range we studied we do not observe the classical critical scaling $\tilde R_b\propto(\mu-\mu^{\rm bub}_c)^{\gamma_b}$ at the scales that dominate the mass function. Bubbles produced in this setup are generically surrounded by type-II curvature profiles (neck-like points satisfying $1+r\zeta'(r)=0$), and the turning point radius $r_{II}(\mu)$ depends on the fluctuation amplitude. Allowing for the natural dispersion of initial shapes (the $(n,m)$ ensemble), thresholds shift modestly but the qualitative picture remains unchanged.

Quantum diffusion acting after bubble formation perforates the trapped region from the inside but does not change its comoving size: diffusion progressively converts subregions to slow-roll while the original co-moving extent of the trapped domain, and hence the relevant PBH scale, are essentially fixed by the initial bubble. The estimate in Sec.~5 shows that this perforation reduces the surviving comoving volume slowly (power-law and then exponential decay in time) while the physical volume grows, with a fractal dimension very close to 3. Therefore, the bubble relic still sources a PBH through the wormhole/baby-universe mechanism \cite{Garriga:2015fdk} after reentry.

For the adiabatic channel we extended the usual mapping between the linear Gaussian curvature and the non-linear $\zeta$ to account for USR dynamics. This extension is necessary, since even in the long wavelenght limit, 
the perturbation $\delta\pi$ does not evolve like the background solution, but has a slower decay in time due to gradient terms. The resulting generalized template reproduces the full non-linear evolution over the parameter range of interest, and provides reliable collapse thresholds $\mu^{\rm adi}_c$ except for amplitudes which are very close to the logarithmic divergence of the analytic template. 

%From numerical evolution, we find that $\mu^{\rm adi}_c$ is systematically smaller than $\mu^{\rm bub}_c$ for the family of profiles considered. Not correct, also not relevant.

Using the numerically determined thresholds, and the statistical weighting of profile deviations, we computed PBH mass functions for both channels. The main conclusions are: (i) the adiabatic channel dominates the PBH abundance in this model by roughly one to two orders of magnitude across the $(n,m)$ ensemble; (ii) both channels are largely governed by the mean profiles derived from the power spectrum, with shape dispersion giving only subleading corrections; and (iii) the mass distributions concentrate near the scale set by the USR enhancement.

Taken together, these results indicate that a flat USR plateau can produce PBHs by two distinct mechanisms even in the absence of an explicit potential barrier, but that the standard adiabatic collapse channel is generically the dominant contributor in the parameter region examined. The bubble channel remains a subdominant contribution which may nonetheless affect secondary features of the combined mass function, as is shown in Fig. \ref{fig:pbh_mass_function}. 

%{\color{red}(SP: I added this paragraph to discuss phenomenology.)}
The enhanced curvature perturbation that forms PBHs can also induce gravitational waves at the horizon reentry, which can be used to crosscheck this scenario. For instance, if asteroid–mass PBHs constitute all of the dark matter, the induced gravitational waves peak at mHz, with an amplitude of $\Omega_\GW\sim10^{-10}$, which is detectable by space-borne interferometers like LISA, Taiji, and TianQin. This prediction is robust against non-Gaussianity \cite{Cai:2018dig,Unal:2018yaa,Bartolo:2018evs}. The previous studies did not consider the PBHs from bubble channel, which, however, contributes sub-dominantly to the total abundance in the ultra-slow-roll model.

Our discussion here can be simply extended to a broader survey of plateau shapes (including small nonzero slopes), multifield generalizations, as well as an exploration of observational consequences of a mixed-channel mass function. Also, here we have focused in the case of a sharp transition to slow roll after the USR plateau. A smooth transition, on the other hand, introduces additional dynamics, since perturbations grow exponentially while the field decelerates towards the new attractor solution. The result is then sensitive to the second derivative of the potential in the region matching the plateau with the second slow roll slope. Consideration of such models requires a dedicated study, which is well beyond the scope of the present work, and is left for further research.
A more extensive consideration of stochastic effects may also be relevant \cite{Cruces:2021iwq,Tada:2021zzj,Figueroa:2021zah,Pattison:2021oen,Cruces:2022imf,Raatikainen:2023bzk,Tomberg:2025fku,Caravano:2024tlp,Caravano:2024moy,Mizuguchi:2024kbl,Raatikainen:2025gpd,Animali:2025pyf,Caravano:2025diq}. In addition, it would be interesting to evaluate the impact of deviations from sphericity on PBH production rates and on the resulting mass functions \cite{Escriva:2024aeo,Escriva:2024lmm}. We leave these for future works.

\begin{acknowledgments}
We thank Takahiro Tanaka for comments on the manuscript.
A.E. thanks the support from the YLC program at the Institute for Advanced Research, Nagoya University. 
This work is supported in part by the National Key Research and Development Program of China Grant No. 2021YFC2203004. 
JG is supported by MCIN/AEI/PID2022-136224NB-C22, by the “Center of Excellence Maria de Maeztu 2025-2029” grant CEX2024-001451-M, funded by MICIU/AEI/10.13039/501100011033, and by 2021-SGR00872 funded by AGAUR. 
S.P. is supported by the National Natural Science Foundation of China (NSFC) Grants Nos. 12475066 and 12447101, and JSPS KAKENHI grant No.~24K00624. 
S.P. acknowledges the APCTP symposium APCTP-2025-IW1117 and the 41st annual IAP symposium ``Inflation 2025'' for the helpful feedback from colleagues.
\end{acknowledgments}

\appendix
\section{\textit{Cardano} formula for cubic equation}
\label{appendix:cardano_formula}
In this appendix we show the necessary steps to obtain the final relation $\zeta[\zeta_G]$
First 
%let's us assume that we have a general equation with some constant coeficients $a,b,c,d$ with $a\xi^3+b \xi^2+c \xi+d=0$, 
let us write Eq.~\eqref{eq:delta_N} as  %(with $a=1$),
\begin{equation}
\xi^3+{b} \xi^2+{d}=0.
\end{equation}
The first step is remove the quadratic term with the change $\xi=y-b/3$,
\begin{align}
    y^3+py+q=0, \
    p = -\frac{b^2}{3} , \, \ q= \frac{2 b^3}{27}+d.
\end{align}
If the discriminant $\Delta = (q/2)^2+(p/3)^3$ is positive (diferent from zero), then we get a single real root, given by
\begin{equation}
    y = \left(-\frac{q}{2}+\sqrt{\Delta}\right)^{1/3} +\left(-\frac{q}{2}-\sqrt{\Delta}\right)^{1/3}
\end{equation}
from which we can obtain $\xi$. In our case, we have to solve
\begin{equation}
    \xi^3-\frac{3}{2}\zeta_G A \xi^2+ \left[3 \zeta_G(1+A/2)-1\right]=0.
\end{equation}
Then we identify the coefficients $ b=-3\zeta_G A/2, d=3 \zeta_G(1+A/2)-1$, and we have
\begin{equation}
    %\Delta = -\frac{1}{64} A^6 \zeta_G^6 + \frac{1}{4} \left( -1 + 3 \left(1 + \frac{A}{2}\right) \zeta_G - \frac{A^3 \zeta_G^3}{4} \right)^2.
    \Delta = \frac{1}{64} \left[6(A+2)(\zeta_G-\zeta_G^{\rm div}) \right]\left[ 6(A+2)(\zeta_G-\zeta_G^{\rm div}) -2A^3 \zeta^3_G\right]
\end{equation}
Taking into account that $\zeta_G< \zeta^{\rm div}_G$ (where $\zeta^{\rm div}_G$ is defined in Eq.~\eqref{eq:mu_divergent}), for $\Delta>0$ the solution is the real root indicated above. Making the change of variable and using $\zeta = -\ln\xi$ we finally obtain Eq.~\eqref{zetanal} of the main text. On the other hand, when $\zeta_G = \zeta^{\rm div}_G$ we have $\Delta = 0$, and the three possible solutions consist of a double real root $y_{2,3} = -3q/(2p)$, which gives $\xi_{2,3} = 0$, corresponding to the true divergent behaviour $\zeta(\zeta^{\rm div}_G)=\infty$ and a third solution $y_1 = 3q/p$, which gives $\xi_1 = A/(A+2)$ when evaluating directly from Eq.~\eqref{zetanal}. It should be noted that the physical solution corresponds to divergent behaviour $\zeta=\infty$.

\bibliographystyle{JHEP}
\bibliography{references}

@article{Pi:2021dft,
    author = "Pi, Shi and Sasaki, Misao",
    title = "{Primordial black hole formation in nonminimal curvaton scenarios}",
    eprint = "2112.12680",
    archivePrefix = "arXiv",
    primaryClass = "astro-ph.CO",
    reportNumber = "IPMU21-0075, YITP-21-131",
    doi = "10.1103/PhysRevD.108.L101301",
    journal = "Phys. Rev. D",
    volume = "108",
    number = "10",
    pages = "L101301",
    year = "2023"
}

@article{Wang:2024wxq,
    author = "Wang, Xinpeng and Ma, Xiao-Han and Sasaki, Misao",
    title = "{A complete analysis of inflation with piecewise quadratic potential}",
    eprint = "2412.16463",
    archivePrefix = "arXiv",
    primaryClass = "astro-ph.CO",
    reportNumber = "YITP-24-174",
    month = "12",
    year = "2024"
}

@article{Inui:2024sce,
    author = "Inui, Ryoto and Motohashi, Hayato and Pi, Shi and Tada, Yuichiro and Yokoyama, Shuichiro",
    title = "{Constant roll and non-Gaussian tail in light of logarithmic duality}",
    eprint = "2409.13500",
    archivePrefix = "arXiv",
    primaryClass = "astro-ph.CO",
    doi = "10.1088/1475-7516/2025/02/042",
    journal = "JCAP",
    volume = "02",
    pages = "042",
    year = "2025"
}

@article{Hooshangi:2023kss,
    author = "Hooshangi, Sina and Namjoo, Mohammad Hossein and Noorbala, Mahdiyar",
    title = "{Tail diversity from inflation}",
    eprint = "2305.19257",
    archivePrefix = "arXiv",
    primaryClass = "astro-ph.CO",
    doi = "10.1088/1475-7516/2023/09/023",
    journal = "JCAP",
    volume = "09",
    pages = "023",
    year = "2023"
}

@article{Kawaguchi:2023mgk,
    author = "Kawaguchi, Ryodai and Fujita, Tomohiro and Sasaki, Misao",
    title = "{Highly asymmetric probability distribution from a finite-width upward step during inflation}",
    eprint = "2305.18140",
    archivePrefix = "arXiv",
    primaryClass = "astro-ph.CO",
    reportNumber = "WUCG-23-07, YITP-23-69",
    doi = "10.1088/1475-7516/2023/11/021",
    journal = "JCAP",
    volume = "11",
    pages = "021",
    year = "2023"
}

@article{Tomberg:2023kli,
    author = "Tomberg, Eemeli",
    title = "{Stochastic constant-roll inflation and primordial black holes}",
    eprint = "2304.10903",
    archivePrefix = "arXiv",
    primaryClass = "astro-ph.CO",
    doi = "10.1103/PhysRevD.108.043502",
    journal = "Phys. Rev. D",
    volume = "108",
    number = "4",
    pages = "043502",
    year = "2023"
}

@article{Pi:2022ysn,
    author = "Pi, Shi and Sasaki, Misao",
    title = "{Logarithmic Duality of the Curvature Perturbation}",
    eprint = "2211.13932",
    archivePrefix = "arXiv",
    primaryClass = "astro-ph.CO",
    reportNumber = "IPMU22-0060, YITP-22-144",
    doi = "10.1103/PhysRevLett.131.011002",
    journal = "Phys. Rev. Lett.",
    volume = "131",
    number = "1",
    pages = "011002",
    year = "2023"
}

@article{Gow:2022jfb,
    author = "Gow, Andrew D. and Assadullahi, Hooshyar and Jackson, Joseph H. P. and Koyama, Kazuya and Vennin, Vincent and Wands, David",
    title = "{Non-perturbative non-Gaussianity and primordial black holes}",
    eprint = "2211.08348",
    archivePrefix = "arXiv",
    primaryClass = "astro-ph.CO",
    doi = "10.1209/0295-5075/acd417",
    journal = "EPL",
    volume = "142",
    number = "4",
    pages = "49001",
    year = "2023"
}

@article{Atal:2019erb,
    author = "Atal, Vicente and Cid, Judith and Escriv{\`a}, Albert and Garriga, Jaume",
    title = "{PBH in single field inflation: the effect of shape dispersion and non-Gaussianities}",
    eprint = "1908.11357",
    archivePrefix = "arXiv",
    primaryClass = "astro-ph.CO",
    doi = "10.1088/1475-7516/2020/05/022",
    journal = "JCAP",
    volume = "05",
    pages = "022",
    year = "2020"
}

@article{Atal:2019cdz,
    author = "Atal, Vicente and Garriga, Jaume and Marcos-Caballero, Airam",
    title = "{Primordial black hole formation with non-Gaussian curvature perturbations}",
    eprint = "1905.13202",
    archivePrefix = "arXiv",
    primaryClass = "astro-ph.CO",
    doi = "10.1088/1475-7516/2019/09/073",
    journal = "JCAP",
    volume = "09",
    pages = "073",
    year = "2019"
}

@article{Jackson:2024aoo,
    author = "Jackson, Joseph H. P. and Assadullahi, Hooshyar and Gow, Andrew D. and Koyama, Kazuya and Vennin, Vincent and Wands, David",
    title = "{Stochastic inflation beyond slow roll: noise modelling and importance sampling}",
    eprint = "2410.13683",
    archivePrefix = "arXiv",
    primaryClass = "astro-ph.CO",
    doi = "10.1088/1475-7516/2025/04/073",
    journal = "JCAP",
    volume = "04",
    pages = "073",
    year = "2025"
}

@article{Cruces:2024pni,
    author = "Cruces, Diego and Germani, Cristiano and Nassiri-Rad, Amin and Yamaguchi, Masahide",
    title = "{Small noise expansion of stochastic inflation}",
    eprint = "2410.17987",
    archivePrefix = "arXiv",
    primaryClass = "astro-ph.CO",
    doi = "10.1088/1475-7516/2025/04/090",
    journal = "JCAP",
    volume = "04",
    pages = "090",
    year = "2025"
}

@article{Ballesteros:2024pbe,
    author = "Ballesteros, Guillermo and Gamb{\'\i}n Egea, Jes{\'u}s and Konstandin, Thomas and P{\'e}rez Rodr{\'\i}guez, Alejandro and Pierre, Mathias and Rey, Juli{\'a}n",
    title = "{Intrinsic non-Gaussianity of ultra slow-roll inflation}",
    eprint = "2412.14106",
    archivePrefix = "arXiv",
    primaryClass = "astro-ph.CO",
    reportNumber = "IFT UAM-CSIC 24-183, DESY-24-205",
    doi = "10.1088/1475-7516/2026/01/012",
    journal = "JCAP",
    volume = "01",
    pages = "012",
    year = "2026"
}

@article{Caravano:2025diq,
    author = "Caravano, Angelo and Franciolini, Gabriele and Renaux-Petel, S{\'e}bastien",
    title = "{Ultraslow-roll inflation on the lattice. II. Nonperturbative curvature perturbation}",
    eprint = "2506.11795",
    archivePrefix = "arXiv",
    primaryClass = "astro-ph.CO",
    reportNumber = "CERN-TH-2025-116",
    doi = "10.1103/39qd-gdfm",
    journal = "Phys. Rev. D",
    volume = "112",
    number = "8",
    pages = "083508",
    year = "2025"
}

@article{Cruces:2025typ,
    author = "Cruces, Diego and Pi, Shi and Sasaki, Misao",
    title = "{$\delta n$ formalism: A new formulation for the probability density of the curvature perturbation}",
    eprint = "2505.24590",
    archivePrefix = "arXiv",
    primaryClass = "astro-ph.CO",
    month = "5",
    year = "2025"
}

@article{Namjoo:2012aa,
    author = "Namjoo, Mohammad Hossein and Firouzjahi, Hassan and Sasaki, Misao",
    title = "{Violation of non-Gaussianity consistency relation in a single field inflationary model}",
    eprint = "1210.3692",
    archivePrefix = "arXiv",
    primaryClass = "astro-ph.CO",
    reportNumber = "YITP-12-79, IPM-A-2012-015",
    doi = "10.1209/0295-5075/101/39001",
    journal = "EPL",
    volume = "101",
    number = "3",
    pages = "39001",
    year = "2013"
}

@article{Motohashi:2023syh,
    author = "Motohashi, Hayato and Tada, Yuichiro",
    title = "{Squeezed bispectrum and one-loop corrections in transient constant-roll inflation}",
    eprint = "2303.16035",
    archivePrefix = "arXiv",
    primaryClass = "astro-ph.CO",
    doi = "10.1088/1475-7516/2023/08/069",
    journal = "JCAP",
    volume = "08",
    pages = "069",
    year = "2023"
}

@article{Ozsoy:2021pws,
    author = {{\"O}zsoy, Ogan and Tasinato, Gianmassimo},
    title = "{Consistency conditions and primordial black holes in single field inflation}",
    eprint = "2111.02432",
    archivePrefix = "arXiv",
    primaryClass = "astro-ph.CO",
    doi = "10.1103/PhysRevD.105.023524",
    journal = "Phys. Rev. D",
    volume = "105",
    number = "2",
    pages = "023524",
    year = "2022"
}

@article{Ragavendra:2020sop,
    author = "Ragavendra, H. V. and Saha, Pankaj and Sriramkumar, L. and Silk, Joseph",
    title = "{Primordial black holes and secondary gravitational waves from ultraslow roll and punctuated inflation}",
    eprint = "2008.12202",
    archivePrefix = "arXiv",
    primaryClass = "astro-ph.CO",
    doi = "10.1103/PhysRevD.103.083510",
    journal = "Phys. Rev. D",
    volume = "103",
    number = "8",
    pages = "083510",
    year = "2021"
}

@article{Kleban:2023ugf,
    author = "Kleban, Matthew and Norton, Cameron E.",
    title = "{Monochromatic mass spectrum of primordial black holes}",
    eprint = "2310.09898",
    archivePrefix = "arXiv",
    primaryClass = "hep-th",
    doi = "10.1103/PhysRevD.111.023538",
    journal = "Phys. Rev. D",
    volume = "111",
    number = "2",
    pages = "023538",
    year = "2025"
}

@article{Chen:2013eea,
    author = "Chen, Xingang and Firouzjahi, Hassan and Komatsu, Eiichiro and Namjoo, Mohammad Hossein and Sasaki, Misao",
    title = "{In-in and $\delta N$ calculations of the bispectrum from non-attractor single-field inflation}",
    eprint = "1308.5341",
    archivePrefix = "arXiv",
    primaryClass = "astro-ph.CO",
    reportNumber = "YITP-13-81",
    doi = "10.1088/1475-7516/2013/12/039",
    journal = "JCAP",
    volume = "12",
    pages = "039",
    year = "2013"
}

@article{Cai:2018dkf,
    author = "Cai, Yi-Fu and Chen, Xingang and Namjoo, Mohammad Hossein and Sasaki, Misao and Wang, Dong-Gang and Wang, Ziwei",
    title = "{Revisiting non-Gaussianity from non-attractor inflation models}",
    eprint = "1712.09998",
    archivePrefix = "arXiv",
    primaryClass = "astro-ph.CO",
    reportNumber = "MIT-CTP-4974, YITP-17-133",
    doi = "10.1088/1475-7516/2018/05/012",
    journal = "JCAP",
    volume = "05",
    pages = "012",
    year = "2018"
}

@article{Biagetti:2018pjj,
    author = "Biagetti, Matteo and Franciolini, Gabriele and Kehagias, Alex and Riotto, Antonio",
    title = "{Primordial Black Holes from Inflation and Quantum Diffusion}",
    eprint = "1804.07124",
    archivePrefix = "arXiv",
    primaryClass = "astro-ph.CO",
    doi = "10.1088/1475-7516/2018/07/032",
    journal = "JCAP",
    volume = "07",
    pages = "032",
    year = "2018"
}

@article{Passaglia:2018ixg,
    author = "Passaglia, Samuel and Hu, Wayne and Motohashi, Hayato",
    title = "{Primordial black holes and local non-Gaussianity in canonical inflation}",
    eprint = "1812.08243",
    archivePrefix = "arXiv",
    primaryClass = "astro-ph.CO",
    reportNumber = "YITP-18-128",
    doi = "10.1103/PhysRevD.99.043536",
    journal = "Phys. Rev. D",
    volume = "99",
    number = "4",
    pages = "043536",
    year = "2019"
}

@article{Luo:2025ewp,
    author = "Luo, Jun and others",
    title = "{Fundamental Physics and Cosmology with TianQin}",
    eprint = "2502.20138",
    archivePrefix = "arXiv",
    primaryClass = "gr-qc",
    month = "2",
    year = "2025"
}

@article{Cai:2018dig,
    author = "Cai, Rong-gen and Pi, Shi and Sasaki, Misao",
    title = "{Gravitational Waves Induced by non-Gaussian Scalar Perturbations}",
    eprint = "1810.11000",
    archivePrefix = "arXiv",
    primaryClass = "astro-ph.CO",
    reportNumber = "IPMU18-0172, YITP-18-114",
    doi = "10.1103/PhysRevLett.122.201101",
    journal = "Phys. Rev. Lett.",
    volume = "122",
    number = "20",
    pages = "201101",
    year = "2019"
}

@article{Bartolo:2018evs,
    author = "Bartolo, N. and De Luca, V. and Franciolini, G. and Lewis, A. and Peloso, M. and Riotto, A.",
    title = "{Primordial Black Hole Dark Matter: LISA Serendipity}",
    eprint = "1810.12218",
    archivePrefix = "arXiv",
    primaryClass = "astro-ph.CO",
    doi = "10.1103/PhysRevLett.122.211301",
    journal = "Phys. Rev. Lett.",
    volume = "122",
    number = "21",
    pages = "211301",
    year = "2019"
}

@article{Escriva:2023uko,
    author = "Escriv\`a, Albert and Atal, Vicente and Garriga, Jaume",
    title = "{Formation of trapped vacuum bubbles during inflation, and consequences for PBH scenarios}",
    eprint = "2306.09990",
    archivePrefix = "arXiv",
    primaryClass = "astro-ph.CO",
    doi = "10.1088/1475-7516/2023/10/035",
    journal = "JCAP",
    volume = "10",
    pages = "035",
    year = "2023"
}

@article{Caravano:2024moy,
    author = "Caravano, Angelo and Franciolini, Gabriele and Renaux-Petel, S{\'e}bastien",
    title = "{Ultraslow-roll inflation on the lattice: Backreaction and nonlinear effects}",
    eprint = "2410.23942",
    archivePrefix = "arXiv",
    primaryClass = "astro-ph.CO",
    reportNumber = "CERN-TH-2024-181",
    doi = "10.1103/PhysRevD.111.063518",
    journal = "Phys. Rev. D",
    volume = "111",
    number = "6",
    pages = "063518",
    year = "2025"
}

@article{Pattison:2021oen,
    author = "Pattison, Chris and Vennin, Vincent and Wands, David and Assadullahi, Hooshyar",
    title = "{Ultra-slow-roll inflation with quantum diffusion}",
    eprint = "2101.05741",
    archivePrefix = "arXiv",
    primaryClass = "astro-ph.CO",
    doi = "10.1088/1475-7516/2021/04/080",
    journal = "JCAP",
    volume = "04",
    pages = "080",
    year = "2021"
}

@article{Caravano:2024tlp,
    author = "Caravano, Angelo and Inomata, Keisuke and Renaux-Petel, S{\'e}bastien",
    title = "{Inflationary Butterfly Effect: Nonperturbative Dynamics from Small-Scale Features}",
    eprint = "2403.12811",
    archivePrefix = "arXiv",
    primaryClass = "astro-ph.CO",
    doi = "10.1103/PhysRevLett.133.151001",
    journal = "Phys. Rev. Lett.",
    volume = "133",
    number = "15",
    pages = "151001",
    year = "2024"
}

@article{Tsamis:2003px,
    author = "Tsamis, N. C. and Woodard, Richard P.",
    title = "{Improved estimates of cosmological perturbations}",
    eprint = "astro-ph/0307463",
    archivePrefix = "arXiv",
    reportNumber = "UFIFT-HEP-03-16, CRETE-03-13",
    doi = "10.1103/PhysRevD.69.084005",
    journal = "Phys. Rev. D",
    volume = "69",
    pages = "084005",
    year = "2004"
}

@article{Kinney:2005vj,
    author = "Kinney, William H.",
    title = "{Horizon crossing and inflation with large eta}",
    eprint = "gr-qc/0503017",
    archivePrefix = "arXiv",
    doi = "10.1103/PhysRevD.72.023515",
    journal = "Phys. Rev. D",
    volume = "72",
    pages = "023515",
    year = "2005"
}

@article{1974MNRAS.168..399C,
	title        = {{Black holes in the early Universe}},
	author       = {{Carr}, B. and {Hawking}, S.},
	year         = 1974,
	month        = aug,
	journal      = {\mnras},
	volume       = 168,
	pages        = {399--416},
	doi          = {10.1093/mnras/168.2.399},
	adsurl       = {https://ui.adsabs.harvard.edu/abs/1974MNRAS.168..399C}
}

@article{PhysRevD.50.7173,
	title        = {Inflation and primordial black holes as dark matter},
	author       = {Ivanov, P. and Naselsky, P. and Novikov, I.},
	year         = 1994,
	month        = {Dec},
	journal      = {Phys. Rev. D},
	publisher    = {American Physical Society},
	volume       = 50,
	pages        = {7173--7178},
	doi          = {10.1103/PhysRevD.50.7173},
	url          = {https://link.aps.org/doi/10.1103/PhysRevD.50.7173},
	issue        = 12,
	numpages     = {0}
}

@article{Cruces:2021iwq,
    author = "Cruces, Diego and Germani, Cristiano",
    title = "{Stochastic inflation at all order in slow-roll parameters: Foundations}",
    eprint = "2107.12735",
    archivePrefix = "arXiv",
    primaryClass = "gr-qc",
    doi = "10.1103/PhysRevD.105.023533",
    journal = "Phys. Rev. D",
    volume = "105",
    number = "2",
    pages = "023533",
    year = "2022"
}

@article{Mizuguchi:2024kbl,
    author = "Mizuguchi, Yurino and Murata, Tomoaki and Tada, Yuichiro",
    title = "{STOLAS: STOchastic LAttice Simulation of cosmic inflation}",
    eprint = "2405.10692",
    archivePrefix = "arXiv",
    primaryClass = "astro-ph.CO",
    reportNumber = "RUP-24-10",
    doi = "10.1088/1475-7516/2024/12/050",
    journal = "JCAP",
    volume = "12",
    pages = "050",
    year = "2024"
}

@article{Tada:2021zzj,
    author = "Tada, Yuichiro and Vennin, Vincent",
    title = "{Statistics of coarse-grained cosmological fields in stochastic inflation}",
    eprint = "2111.15280",
    archivePrefix = "arXiv",
    primaryClass = "astro-ph.CO",
    doi = "10.1088/1475-7516/2022/02/021",
    journal = "JCAP",
    volume = "02",
    number = "02",
    pages = "021",
    year = "2022"
}

@article{Animali:2025pyf,
    author = "Animali, Chiara and Auclair, Pierre and Blachier, Baptiste and Vennin, Vincent",
    title = "{Harvesting primordial black holes from stochastic trees with FOREST}",
    eprint = "2501.05371",
    archivePrefix = "arXiv",
    primaryClass = "astro-ph.CO",
    doi = "10.1088/1475-7516/2025/05/019",
    journal = "JCAP",
    volume = "05",
    pages = "019",
    year = "2025"
}

@article{Yoo:2018kvb,
    author = "Yoo, Chul-Moon and Harada, Tomohiro and Garriga, Jaume and Kohri, Kazunori",
    title = "{Primordial black hole abundance from random Gaussian curvature perturbations and a local density threshold}",
    eprint = "1805.03946",
    archivePrefix = "arXiv",
    primaryClass = "astro-ph.CO",
    reportNumber = "RUP-18-15, KEK-Cosmo-225, KEK-TH-2052",
    doi = "10.1093/ptep/pty120",
    journal = "PTEP",
    volume = "2018",
    number = "12",
    pages = "123E01",
    year = "2018",
    note = "[Erratum: PTEP 2024, 049202 (2024)]"
}

@article{Artigas:2024ajh,
    author = "Artigas, Danilo and Pi, Shi and Tanaka, Takahiro",
    title = "{Extended {\ensuremath{\delta}}N Formalism: Nonspatially Flat Separate-Universe Approach}",
    eprint = "2408.09964",
    archivePrefix = "arXiv",
    primaryClass = "astro-ph.CO",
    doi = "10.1103/PhysRevLett.134.221001",
    journal = "Phys. Rev. Lett.",
    volume = "134",
    number = "22",
    pages = "221001",
    year = "2025"
}

@article{Yoo:2020dkz,
    author = "Yoo, Chul-Moon and Harada, Tomohiro and Hirano, Shin'ichi and Kohri, Kazunori",
    title = "{Abundance of Primordial Black Holes in Peak Theory for an Arbitrary Power Spectrum}",
    eprint = "2008.02425",
    archivePrefix = "arXiv",
    primaryClass = "astro-ph.CO",
    reportNumber = "RUP-20-25, KEK-Cosmo-261, KEK-TH-2245",
    doi = "10.1093/ptep/ptaa155",
    journal = "PTEP",
    volume = "2021",
    number = "1",
    pages = "013E02",
    year = "2021",
    note = "[Erratum: PTEP 2024, 049203 (2024)]"
}

@article{Kitajima:2021fpq,
    author = "Kitajima, Naoya and Tada, Yuichiro and Yokoyama, Shuichiro and Yoo, Chul-Moon",
    title = "{Primordial black holes in peak theory with a non-Gaussian tail}",
    eprint = "2109.00791",
    archivePrefix = "arXiv",
    primaryClass = "astro-ph.CO",
    reportNumber = "TU-1130",
    doi = "10.1088/1475-7516/2021/10/053",
    journal = "JCAP",
    volume = "10",
    pages = "053",
    year = "2021"
}

@article{Pi:2024ert,
    author = "Pi, Shi and Sasaki, Misao and Takhistov, Volodymyr and Wang, Jianing",
    title = "{Primordial Black Hole formation from power spectrum with finite-width}",
    eprint = "2501.00295",
    archivePrefix = "arXiv",
    primaryClass = "astro-ph.CO",
    reportNumber = "YITP-24-184, KEK-QUP-2024-0028, KEK-TH-2676, KEK-Cosmo-0369",
    doi = "10.1088/1475-7516/2025/09/045",
    journal = "JCAP",
    volume = "09",
    pages = "045",
    year = "2025"
}

@article{Escriva:2023nzn,
    author = "Escriva, Albert and Tada, Yuichiro and Yoo, Chul-Moon",
    title = "{Primordial black holes and induced gravitational waves from a smooth crossover beyond standard model theories}",
    eprint = "2311.17760",
    archivePrefix = "arXiv",
    primaryClass = "astro-ph.CO",
    doi = "10.1103/PhysRevD.110.063521",
    journal = "Phys. Rev. D",
    volume = "110",
    number = "6",
    pages = "063521",
    year = "2024"
}

@article{1967SvA....10..602Z,
	title        = {{The Hypothesis of Cores Retarded during Expansion and the Hot Cosmological Model}},
	author       = {{Zel'dovich}, Ya. B. and {Novikov}, I.~D.},
	year         = 1967,
	month        = feb,
	journal      = {\sovast},
	volume       = 10,
	pages        = 602,
	adsurl       = {https://ui.adsabs.harvard.edu/abs/1967SvA....10..602Z}
}

@article{2019JCAP...09..073A,
	title        = {{Primordial black hole formation with non-Gaussian curvature perturbations}},
	author       = {Atal, Vicente and Garriga, Jaume and Marcos-Caballero, Airam},
	year         = 2019,
	journal      = {JCAP},
	volume       = {09},
	pages        = {073},
	doi          = {10.1088/1475-7516/2019/09/073},
	eprint       = {1905.13202},
	archiveprefix = {arXiv},
	primaryclass = {astro-ph.CO}
}

@article{1975ApJ...201....1C,
	title        = {{The primordial black hole mass spectrum.}},
	author       = {{Carr}, B.},
	year         = 1975,
	month        = oct,
	journal      = {\apj},
	volume       = 201,
	pages        = {1--19},
	doi          = {10.1086/153853},
	adsurl       = {https://ui.adsabs.harvard.edu/abs/1975ApJ...201....1C}
}

@article{Cruces:2022imf,
    author = "Cruces, Diego",
    title = "{Review on Stochastic Approach to Inflation}",
    eprint = "2203.13852",
    archivePrefix = "arXiv",
    primaryClass = "gr-qc",
    doi = "10.3390/universe8060334",
    journal = "Universe",
    volume = "8",
    number = "6",
    pages = "334",
    year = "2022"
}

@article{Pi:2017gih,
    author = "Pi, Shi and Zhang, Ying-li and Huang, Qing-Guo and Sasaki, Misao",
    title = "{Scalaron from $R^2$-gravity as a heavy field}",
    eprint = "1712.09896",
    archivePrefix = "arXiv",
    primaryClass = "astro-ph.CO",
    reportNumber = "YITP-17-135",
    doi = "10.1088/1475-7516/2018/05/042",
    journal = "JCAP",
    volume = "05",
    pages = "042",
    year = "2018"
}

@article{Garcia-Bellido:2017mdw,
    author = "Garcia-Bellido, Juan and Ruiz Morales, Ester",
    title = "{Primordial black holes from single field models of inflation}",
    eprint = "1702.03901",
    archivePrefix = "arXiv",
    primaryClass = "astro-ph.CO",
    reportNumber = "IFT-UAM-CSIC-17-007, CERN-TH-2017-196",
    doi = "10.1016/j.dark.2017.09.007",
    journal = "Phys. Dark Univ.",
    volume = "18",
    pages = "47--54",
    year = "2017"
}

@article{Raatikainen:2025gpd,
    author = "Raatikainen, Sami and Rasanen, Syksy and Tomberg, Eemeli",
    title = "{Effect of stochastic kicks on primordial black hole abundance and mass via the compaction function}",
    eprint = "2510.09303",
    archivePrefix = "arXiv",
    primaryClass = "astro-ph.CO",
    reportNumber = "HIP-2025-28/TH",
    month = "10",
    year = "2025"
}

@article{Tomberg:2025fku,
    author = "Tomberg, Eemeli and Dimopoulos, Konstantinos",
    title = "{Eternal inflation near inflection points: a challenge to primordial black hole models}",
    eprint = "2507.15522",
    archivePrefix = "arXiv",
    primaryClass = "astro-ph.CO",
    month = "7",
    year = "2025"
}

@article{Raatikainen:2023bzk,
    author = {Raatikainen, Sami and R{\"a}s{\"a}nen, Syksy and Tomberg, Eemeli},
    title = "{Primordial Black Hole Compaction Function from Stochastic Fluctuations in Ultraslow-Roll Inflation}",
    eprint = "2312.12911",
    archivePrefix = "arXiv",
    primaryClass = "astro-ph.CO",
    reportNumber = "HIP-2023-18/TH",
    doi = "10.1103/PhysRevLett.133.121403",
    journal = "Phys. Rev. Lett.",
    volume = "133",
    number = "12",
    pages = "121403",
    year = "2024"
}

@article{Figueroa:2021zah,
    author = "Figueroa, Daniel G. and Raatikainen, Sami and Rasanen, Syksy and Tomberg, Eemeli",
    title = "{Implications of stochastic effects for primordial black hole production in ultra-slow-roll inflation}",
    eprint = "2111.07437",
    archivePrefix = "arXiv",
    primaryClass = "astro-ph.CO",
    reportNumber = "HIP-2021-31/TH",
    doi = "10.1088/1475-7516/2022/05/027",
    journal = "JCAP",
    volume = "05",
    number = "05",
    pages = "027",
    year = "2022"
}

@article{Gundhi:2020kzm,
    author = "Gundhi, Anirudh and Ketov, Sergei V. and Steinwachs, Christian F.",
    title = "{Primordial black hole dark matter in dilaton-extended two-field Starobinsky inflation}",
    eprint = "2011.05999",
    archivePrefix = "arXiv",
    primaryClass = "hep-th",
    reportNumber = "FR-PHENO-2020-018, IPMU20-0118",
    doi = "10.1103/PhysRevD.103.083518",
    journal = "Phys. Rev. D",
    volume = "103",
    number = "8",
    pages = "083518",
    year = "2021"
}

@article{2020JCAP...05..022A,
	title        = {{PBH in single field inflation: the effect of shape dispersion and non-Gaussianities}},
	author       = {{Atal}, Vicente and {Cid}, Judith and {Escriv{\`a}}, Albert and {Garriga}, Jaume},
	year         = 2020,
	month        = may,
	journal      = {\jcap},
	volume       = 2020,
	number       = 5,
	pages        = {022},
	doi          = {10.1088/1475-7516/2020/05/022},
	eid          = {022},
	archiveprefix = {arXiv},
	eprint       = {1908.11357},
	primaryclass = {astro-ph.CO},
	adsurl       = {https://ui.adsabs.harvard.edu/abs/2020JCAP...05..022A}
}

@article{Byrnes:2018txb,
    author = "Byrnes, Christian T. and Cole, Philippa S. and Patil, Subodh P.",
    title = "{Steepest growth of the power spectrum and primordial black holes}",
    eprint = "1811.11158",
    archivePrefix = "arXiv",
    primaryClass = "astro-ph.CO",
    doi = "10.1088/1475-7516/2019/06/028",
    journal = "JCAP",
    volume = "06",
    pages = "028",
    year = "2019"
}

@article{Atal:2018neu,
    author = "Atal, Vicente and Germani, Cristiano",
    title = "{The role of non-gaussianities in Primordial Black Hole formation}",
    eprint = "1811.07857",
    archivePrefix = "arXiv",
    primaryClass = "astro-ph.CO",
    reportNumber = "ICCUB-18-022",
    doi = "10.1016/j.dark.2019.100275",
    journal = "Phys. Dark Univ.",
    volume = "24",
    pages = "100275",
    year = "2019"
}

@article{Germani:2017bcs,
    author = "Germani, Cristiano and Prokopec, Tomislav",
    title = "{On primordial black holes from an inflection point}",
    eprint = "1706.04226",
    archivePrefix = "arXiv",
    primaryClass = "astro-ph.CO",
    reportNumber = "ICCUB-17-012",
    doi = "10.1016/j.dark.2017.09.001",
    journal = "Phys. Dark Univ.",
    volume = "18",
    pages = "6--10",
    year = "2017"
}

@article{universal1,
	title        = {Universal threshold for primordial black hole formation},
	author       = {Escriv\`a, Albert and Germani, Cristiano and Sheth, Ravi K.},
	year         = 2020,
	month        = feb,
	journal      = {Phys. Rev. D},
	publisher    = {American Physical Society},
	volume       = 101,
	pages        = {044022},
	doi          = {10.1103/PhysRevD.101.044022},
	url          = {https://link.aps.org/doi/10.1103/PhysRevD.101.044022},
	issue        = 4,
	numpages     = 5
}

@article{Escriva:2025rja,
    author = "Escriv{\`a}, Albert",
    title = "{Threshold for PBH formation in the type-II region and its analytical estimation}",
    eprint = "2504.05814",
    archivePrefix = "arXiv",
    primaryClass = "astro-ph.CO",
    doi = "10.1103/mq67-bbvj",
    journal = "Phys. Rev. D",
    volume = "112",
    number = "10",
    pages = "103527",
    year = "2025"
}

@article{PhysRev.136.B571,
  title = {Relativistic Equations for Adiabatic, Spherically Symmetric Gravitational Collapse},
  author = {Misner, Charles W. and Sharp, David H.},
  journal = {Phys. Rev.},
  volume = {136},
  issue = {2B},
  pages = {B571--B576},
  numpages = {0},
  year = {1964},
  month = {Oct},
  publisher = {American Physical Society},
  doi = {10.1103/PhysRev.136.B571},
  url = {https://link.aps.org/doi/10.1103/PhysRev.136.B571}
}

@article{Escriva:2025eqc,
    author = "Escriv{\`a}, Albert",
    title = "{A new approach for simulating PBH formation from generic curvature fluctuations with the Misner-Sharp formalism}",
    eprint = "2504.05813",
    archivePrefix = "arXiv",
    primaryClass = "astro-ph.CO",
    doi = "10.1016/j.dark.2025.102177",
    journal = "Phys. Dark Univ.",
    volume = "50",
    pages = "102177",
    year = "2025"
}

@article{escriva_solo,
	title        = {Simulation of primordial black hole formation using pseudo-spectral methods},
	author       = {Albert Escrivà},
	year         = 2020,
	journal      = {Physics of the Dark Universe},
	volume       = 27,
	pages        = 100466,
	doi          = {https://doi.org/10.1016/j.dark.2020.100466},
	issn         = {2212-6864},
	url          = {http://www.sciencedirect.com/science/article/pii/S2212686419302845}
}

@article{1986ApJ...304...15B,
	title        = {{The Statistics of Peaks of Gaussian Random Fields}},
	author       = {{Bardeen}, J.~M. and {Bond}, J.~R. and {Kaiser}, N. and {Szalay}, A.~S.},
	year         = 1986,
	month        = may,
	journal      = {\apj},
	volume       = 304,
	pages        = 15,
	doi          = {10.1086/164143},
	adsurl       = {https://ui.adsabs.harvard.edu/abs/1986ApJ...304...15B}
}

@article{vacum_bubles,
	title        = {{Primordial black hole formation by vacuum bubbles}},
	author       = {Deng, Heling and Vilenkin, Alexander},
	year         = 2017,
	journal      = {JCAP},
	volume       = 12,
	pages        = {044},
	doi          = {10.1088/1475-7516/2017/12/044},
	eprint       = {1710.02865},
	archiveprefix = {arXiv},
	primaryclass = {gr-qc}
}

@article{2020ARNPS..70..355C,
	title        = {{Primordial Black Holes as Dark Matter: Recent Developments}},
	author       = {{Carr}, Bernard and {K{\"u}hnel}, Florian},
	year         = 2020,
	month        = oct,
	journal      = {Annual Review of Nuclear and Particle Science},
	volume       = 70,
	pages        = {355--394},
	doi          = {10.1146/annurev-nucl-050520-125911},
	archiveprefix = {arXiv},
	eprint       = {2006.02838},
	primaryclass = {astro-ph.CO},
	adsurl       = {https://ui.adsabs.harvard.edu/abs/2020ARNPS..70..355C}
}

@article{Wang:2025hwc,
    author = "Wang, Haonan and Zhang, Ying-li and Suyama, Teruaki",
    title = "{Nearly Monochromatic Primordial Black Holes as total Dark Matter from Bubble Collapse}",
    eprint = "2510.19233",
    archivePrefix = "arXiv",
    primaryClass = "astro-ph.CO",
    month = "10",
    year = "2025",
    author = "Kleban, Matthew and Norton, Cameron E.",
    title = "{Monochromatic mass spectrum of primordial black holes}",
    eprint = "2310.09898",
    archivePrefix = "arXiv",
    primaryClass = "hep-th",
    doi = "10.1103/PhysRevD.111.023538",
    journal = "Phys. Rev. D",
    volume = "111",
    number = "2",
    pages = "023538",
    year = "2025"
}

@article{Garriga:2015fdk,
	title        = {{Black holes and the multiverse}},
	author       = {Garriga, Jaume and Vilenkin, Alexander and Zhang, Jun},
	year         = 2016,
	journal      = {JCAP},
	volume       = {02},
	pages        = {064},
	doi          = {10.1088/1475-7516/2016/02/064},
	eprint       = {1512.01819},
	archiveprefix = {arXiv},
	primaryclass = {hep-th}
}

@article{Aryal:1987vn,
    author = "Aryal, Mukunda and Vilenkin, Alexander",
    title = "{The Fractal Dimension of Inflationary Universe}",
    reportNumber = "TUTP-87-11",
    doi = "10.1016/0370-2693(87)90932-4",
    journal = "Phys. Lett. B",
    volume = "199",
    pages = "351--357",
    year = "1987"
}

@article{Vilenkin:1995yd,
    author = "Vilenkin, Alexander",
    title = "{Making predictions in eternally inflating universe}",
    eprint = "gr-qc/9505031",
    archivePrefix = "arXiv",
    doi = "10.1103/PhysRevD.52.3365",
    journal = "Phys. Rev. D",
    volume = "52",
    pages = "3365--3374",
    year = "1995"
}

@article{Starobinsky:1986fx,
    author = "Starobinsky, Alexei A.",
    title = "{STOCHASTIC DE SITTER (INFLATIONARY) STAGE IN THE EARLY UNIVERSE}",
    doi = "10.1007/3-540-16452-9_6",
    journal = "Lect. Notes Phys.",
    volume = "246",
    pages = "107--126",
    year = "1986"
}

@article{Vilenkin:1983xp,
    author = "Vilenkin, Alexander",
    title = "{Quantum Fluctuations in the New Inflationary Universe}",
    reportNumber = "TUTP-83-2",
    doi = "10.1016/0550-3213(83)90208-0",
    journal = "Nucl. Phys. B",
    volume = "226",
    pages = "527--546",
    year = "1983"
}

@article{Garriga:2005av,
    author = "Garriga, Jaume and Schwartz-Perlov, Delia and Vilenkin, Alexander and Winitzki, Sergei",
    title = "{Probabilities in the inflationary multiverse}",
    eprint = "hep-th/0509184",
    archivePrefix = "arXiv",
    doi = "10.1088/1475-7516/2006/01/017",
    journal = "JCAP",
    volume = "01",
    pages = "017",
    year = "2006"
}

@article{Shibata:1999zs,
	title        = {{Black hole formation in the Friedmann universe: Formulation and computation in numerical relativity}},
	author       = {Shibata, Masaru and Sasaki, Misao},
	year         = 1999,
	journal      = {Phys. Rev. D},
	volume       = 60,
	pages        = {084002},
	doi          = {10.1103/PhysRevD.60.084002},
	eprint       = {gr-qc/9905064},
	archiveprefix = {arXiv},
	reportnumber = {OU-TAP-93}
}

@article{Unal:2018yaa,
	title        = {{Imprints of Primordial Non-Gaussianity on Gravitational Wave Spectrum}},
	author       = {Unal, Caner},
	year         = 2019,
	journal      = {Phys. Rev. D},
	volume       = 99,
	number       = 4,
	pages        = {041301},
	doi          = {10.1103/PhysRevD.99.041301},
	eprint       = {1811.09151},
	archiveprefix = {arXiv},
	primaryclass = {astro-ph.CO}
}

@article{2017JCAP...04..050D,
	title        = {{Primordial black hole and wormhole formation by domain walls}},
	author       = {Deng, Heling and Garriga, Jaume and Vilenkin, Alexander},
	year         = 2017,
	journal      = {JCAP},
	volume       = {04},
	pages        = {050},
	doi          = {10.1088/1475-7516/2017/04/050},
	eprint       = {1612.03753},
	archiveprefix = {arXiv},
	primaryclass = {gr-qc}
}

@article{2018CQGra..35f3001S,
	title        = {{Primordial black holes{\textemdash}perspectives in gravitational wave astronomy}},
	author       = {{Sasaki}, Misao and {Suyama}, Teruaki and {Tanaka}, Takahiro and {Yokoyama}, Shuichiro},
	year         = 2018,
	month        = mar,
	journal      = {Classical and Quantum Gravity},
	volume       = 35,
	number       = 6,
	pages        = {063001},
	doi          = {10.1088/1361-6382/aaa7b4},
	eid          = {063001},
	archiveprefix = {arXiv},
	eprint       = {1801.05235},
	primaryclass = {astro-ph.CO},
	adsurl       = {https://ui.adsabs.harvard.edu/abs/2018CQGra..35f3001S}
}

@article{2021JPhG...48d3001G,
	title        = {{Primordial black holes as a dark matter candidate}},
	author       = {{Green}, Anne M. and {Kavanagh}, Bradley J.},
	year         = 2021,
	month        = apr,
	journal      = {Journal of Physics G Nuclear Physics},
	volume       = 48,
	number       = 4,
	pages        = {043001},
	doi          = {10.1088/1361-6471/abc534},
	eid          = {043001},
	archiveprefix = {arXiv},
	eprint       = {2007.10722},
	primaryclass = {astro-ph.CO},
	adsurl       = {https://ui.adsabs.harvard.edu/abs/2021JPhG...48d3001G}
}

@article{2020JCAP...09..023D,
	title        = {{Primordial black hole formation by vacuum bubbles. Part II}},
	author       = {{Deng}, Heling},
	year         = 2020,
	month        = sep,
	journal      = {\jcap},
	volume       = 2020,
	number       = 9,
	pages        = {023},
	doi          = {10.1088/1475-7516/2020/09/023},
	eid          = {023},
	archiveprefix = {arXiv},
	eprint       = {2006.11907},
	primaryclass = {astro-ph.CO},
	adsurl       = {https://ui.adsabs.harvard.edu/abs/2020JCAP...09..023D}
}

@article{Kusenko:2020pcg,
	title        = {{Exploring Primordial Black Holes from the Multiverse with Optical Telescopes}},
	author       = {Kusenko, Alexander and Sasaki, Misao and Sugiyama, Sunao and Takada, Masahiro and Takhistov, Volodymyr and Vitagliano, Edoardo},
	year         = 2020,
	journal      = {Phys. Rev. Lett.},
	volume       = 125,
	pages        = 181304,
	doi          = {10.1103/PhysRevLett.125.181304},
	eprint       = {2001.09160},
	archiveprefix = {arXiv},
	primaryclass = {astro-ph.CO},
	reportnumber = {IPMU20-0006, YITP-20-11}
}

@article{Niemeyer:1999ak,
    author = "Niemeyer, Jens C. and Jedamzik, K.",
    title = "{Dynamics of primordial black hole formation}",
    eprint = "astro-ph/9901292",
    archivePrefix = "arXiv",
    doi = "10.1103/PhysRevD.59.124013",
    journal = "Phys. Rev. D",
    volume = "59",
    pages = "124013",
    year = "1999"
}

@article{Kopp:2010sh,
	title        = {{Separate Universes Do Not Constrain Primordial Black Hole Formation}},
	author       = {Kopp, Michael and Hofmann, Stefan and Weller, Jochen},
	year         = 2011,
	journal      = {Phys. Rev. D},
	volume       = 83,
	pages        = 124025,
	doi          = {10.1103/PhysRevD.83.124025},
	eprint       = {1012.4369},
	archiveprefix = {arXiv},
	primaryclass = {astro-ph.CO}
}

@article{Mishra:2019pzq,
	title        = {{Primordial Black Holes from a tiny bump/dip in the Inflaton potential}},
	author       = {Mishra, Swagat S. and Sahni, Varun},
	year         = 2020,
	journal      = {JCAP},
	volume       = {04},
	pages        = {007},
	doi          = {10.1088/1475-7516/2020/04/007},
	eprint       = {1911.00057},
	archiveprefix = {arXiv},
	primaryclass = {gr-qc}
}

@article{Escriva:2022duf,
	title        = {{Primordial Black Holes}},
	author       = {Escriv\`a, Albert and Kuhnel, Florian and Tada, Yuichiro},
	year         = 2022,
	month        = 11,
	eprint       = {2211.05767},
	archiveprefix = {arXiv},
	primaryclass = {astro-ph.CO}
}

@inproceedings{2017JPhCS.840a2032G,
	title        = {{Massive Primordial Black Holes as Dark Matter and their detection with Gravitational Waves}},
	author       = {{Garc{\'\i}a-Bellido}, Juan},
	year         = 2017,
	month        = may,
	booktitle    = {Journal of Physics Conference Series},
	series       = {Journal of Physics Conference Series},
	volume       = 840,
	pages        = {012032},
	doi          = {10.1088/1742-6596/840/1/012032},
	eid          = {012032},
	archiveprefix = {arXiv},
	eprint       = {1702.08275},
	primaryclass = {astro-ph.CO},
	adsurl       = {https://ui.adsabs.harvard.edu/abs/2017JPhCS.840a2032G}
}

@article{2016PhRvD..94h3504C,
	title        = {{Primordial black holes as dark matter}},
	author       = {{Carr}, Bernard and {K{\"u}hnel}, Florian and {Sandstad}, Marit},
	year         = 2016,
	month        = oct,
	journal      = {\prd},
	volume       = 94,
	number       = 8,
	pages        = {083504},
	doi          = {10.1103/PhysRevD.94.083504},
	adsurl       = {https://ui.adsabs.harvard.edu/abs/2016PhRvD..94h3504C},
	archiveprefix = {arXiv},
	eid          = {083504},
	eprint       = {1607.06077},
	primaryclass = {astro-ph.CO}
}

@article{Carr:2021bzv,
	title        = {{Primordial black holes as dark matter candidates}},
	author       = {Carr, Bernard and K{\"u}hnel, Florian},
	year         = 2022,
	journal      = {SciPost Phys. Lect. Notes},
	volume       = 48,
	pages        = 1,
	doi          = {10.21468/SciPostPhysLectNotes.48},
	eprint       = {2110.02821},
	archiveprefix = {arXiv},
	primaryclass = {astro-ph.CO}
}

@article{Murgia:2019duy,
	title        = {{Lyman-\ensuremath{\alpha} Forest Constraints on Primordial Black Holes as Dark Matter}},
	author       = {Murgia, Riccardo and Scelfo, Giulio and Viel, Matteo and Raccanelli, Alvise},
	year         = 2019,
	journal      = {Phys. Rev. Lett.},
	volume       = 123,
	number       = 7,
	pages        = {071102},
	doi          = {10.1103/PhysRevLett.123.071102},
	eprint       = {1903.10509},
	archiveprefix = {arXiv},
	primaryclass = {astro-ph.CO},
	reportnumber = {CERN-TH-2019-029}
}

@article{Carr:2020gox,
	title        = {{Constraints on primordial black holes}},
	author       = {Carr, Bernard and Kohri, Kazunori and Sendouda, Yuuiti and Yokoyama, Jun'ichi},
	year         = 2021,
	journal      = {Rept. Prog. Phys.},
	volume       = 84,
	number       = 11,
	pages        = 116902,
	doi          = {10.1088/1361-6633/ac1e31},
	eprint       = {2002.12778},
	archiveprefix = {arXiv},
	primaryclass = {astro-ph.CO},
	reportnumber = {RESCEU-03/20; KEK-Cosmo-249; KEK-TH-2199; IPMU20-0024}
}

@article{Hawking:1971ei,
	title        = {{Gravitationally collapsed objects of very low mass}},
	author       = {Hawking, Stephen},
	year         = 1971,
	journal      = {Mon. Not. Roy. Astron. Soc.},
	volume       = 152,
	pages        = 75
}

@article{Chapline:1975ojl,
	title        = {{Cosmological effects of primordial black holes}},
	author       = {Chapline, George F.},
	year         = 1975,
	journal      = {Nature},
	volume       = 253,
	number       = 5489,
	pages        = {251--252},
	doi          = {10.1038/253251a0}
}

@article{2021arXiv211103606T,
    author = "Abbott, R. and others",
    collaboration = "KAGRA, VIRGO, LIGO Scientific",
    title = "{GWTC-3: Compact Binary Coalescences Observed by LIGO and Virgo during the Second Part of the Third Observing Run}",
    eprint = "2111.03606",
    archivePrefix = "arXiv",
    primaryClass = "gr-qc",
    reportNumber = "LIGO-P2000318",
    doi = "10.1103/PhysRevX.13.041039",
    journal = "Phys. Rev. X",
    volume = "13",
    number = "4",
    pages = "041039",
    year = "2023"
}

@article{2016PhRvX...6d1015A,
	title        = {{Binary Black Hole Mergers in the First Advanced LIGO Observing Run}},
	author       = {Abbott, B. and others and {LIGO Scientific Collaboration} and {Virgo Collaboration}},
	year         = 2016,
	month        = oct,
	journal      = {Physical Review X},
	volume       = 6,
	number       = 4,
	pages        = {041015},
	doi          = {10.1103/PhysRevX.6.041015},
	eid          = {041015},
	archiveprefix = {arXiv},
	eprint       = {1606.04856},
	primaryclass = {gr-qc},
	adsurl       = {https://ui.adsabs.harvard.edu/abs/2016PhRvX...6d1015A}
}

@article{2022JCAP...05..012E,
	title        = {{Simulation of primordial black holes with large negative non-Gaussianity}},
	author       = {{Escriv{\`a}}, Albert and {Tada}, Yuichiro and {Yokoyama}, Shuichiro and {Yoo}, Chul-Moon},
	year         = 2022,
	month        = may,
	journal      = {\jcap},
	volume       = 2022,
	number       = 5,
	pages        = {012},
	doi          = {10.1088/1475-7516/2022/05/012},
	eid          = {012},
	archiveprefix = {arXiv},
	eprint       = {2202.01028},
	primaryclass = {astro-ph.CO},
	adsurl       = {https://ui.adsabs.harvard.edu/abs/2022JCAP...05..012E}
}

@article{2022Univ....8...66E,
	title        = {{PBH Formation from Spherically Symmetric Hydrodynamical Perturbations: A Review}},
	author       = {{Escriv{\`a}}, Albert},
	year         = 2022,
	month        = jan,
	journal      = {Universe},
	volume       = 8,
	number       = 2,
	pages        = 66,
	doi          = {10.3390/universe8020066},
	archiveprefix = {arXiv},
	eprint       = {2111.12693},
	primaryclass = {gr-qc},
	adsurl       = {https://ui.adsabs.harvard.edu/abs/2022Univ....8...66E}
}

@article{ZhengRuiFeng:2021zoz,
	title        = {{On primordial black holes and secondary gravitational waves generated from inflation with solo/multi-bumpy potential *}},
	author       = {Ruifeng Zheng and Shi Jiaming and Taotao Qiu},
	year         = 2022,
	journal      = {Chin. Phys. C},
	volume       = 46,
	number       = 4,
	pages        = {045103},
	doi          = {10.1088/1674-1137/ac42bd},
	eprint       = {2106.04303},
	archiveprefix = {arXiv},
	primaryclass = {astro-ph.CO}
}

@article{Wang:2021kbh,
	title        = {{Primordial black holes from the perturbations in the inflaton potential in peak theory}},
	author       = {Wang, Qing and Liu, Yi-Chen and Su, Bing-Yu and Li, Nan},
	year         = 2021,
	journal      = {Phys. Rev. D},
	volume       = 104,
	number       = 8,
	pages        = {083546},
	doi          = {10.1103/PhysRevD.104.083546},
	eprint       = {2111.10028},
	archiveprefix = {arXiv},
	primaryclass = {astro-ph.CO}
}

@article{Rezazadeh:2021clf,
	title        = {{Non-Gaussianity and secondary gravitational waves from primordial black holes production in $\alpha $-attractor inflation}},
	author       = {Rezazadeh, Kazem and Teimoori, Zeinab and Karimi, Saeid and Karami, Kayoomars},
	year         = 2022,
	journal      = {Eur. Phys. J. C},
	volume       = 82,
	number       = 8,
	pages        = 758,
	doi          = {10.1140/epjc/s10052-022-10735-w},
	eprint       = {2110.01482},
	archiveprefix = {arXiv},
	primaryclass = {gr-qc}
}

@article{Iacconi:2021ltm,
	title        = {{Revisiting small-scale fluctuations in \ensuremath{\alpha}-attractor models of inflation}},
	author       = {Iacconi, Laura and Assadullahi, Hooshyar and Fasiello, Matteo and Wands, David},
	year         = 2022,
	journal      = {JCAP},
	volume       = {06},
	number       = {06},
	pages        = {007},
	doi          = {10.1088/1475-7516/2022/06/007},
	eprint       = {2112.05092},
	archiveprefix = {arXiv},
	primaryclass = {astro-ph.CO}
}

@book{cardano1968,
  author = {Cardano, Girolamo},
  title = {Ars Magna, or, The Rules of Algebra},
  year = {1968},
  edition = {MIT Press edition},
  publisher = {MIT Press},
  address = {Cambridge, MA},
  editor = {Witmer, T. Richard},
  note = {Translated and edited from the 1545 edition of Artis magnae, sive, de regulis algebraicis. Lib. unus, with additions from the 1570 and 1663 eds.}
}

@article{Pi:2022zxs,
    author = "Pi, Shi and Wang, Jianing",
    title = "{Primordial black hole formation in Starobinsky's linear potential model}",
    eprint = "2209.14183",
    archivePrefix = "arXiv",
    primaryClass = "astro-ph.CO",
    reportNumber = "IPMU22-0047",
    doi = "10.1088/1475-7516/2023/06/018",
    journal = "JCAP",
    volume = "06",
    pages = "018",
    year = "2023"
}

@article{Namjoo:2025hrr,
    author = "Namjoo, Mohammad Hossein and Nikbakht, Bahar",
    title = "{Geometry of non-Gaussianity in transient non-attractor inflation}",
    eprint = "2512.11020",
    archivePrefix = "arXiv",
    primaryClass = "astro-ph.CO",
    month = "12",
    year = "2025"
}

@article{He:2023yvl,
    author = "He, Jibin and Deng, Heling and Piao, Yun-Song and Zhang, Jun",
    title = "{Implications of GWTC-3 on primordial black holes from vacuum bubbles}",
    eprint = "2303.16810",
    archivePrefix = "arXiv",
    primaryClass = "astro-ph.CO",
    doi = "10.1103/PhysRevD.109.044035",
    journal = "Phys. Rev. D",
    volume = "109",
    number = "4",
    pages = "044035",
    year = "2024"
}

@article{Deng:2018cxb,
    author = "Deng, Heling and Vilenkin, Alexander and Yamada, Masaki",
    title = "{CMB spectral distortions from black holes formed by vacuum bubbles}",
    eprint = "1804.10059",
    archivePrefix = "arXiv",
    primaryClass = "gr-qc",
    doi = "10.1088/1475-7516/2018/07/059",
    journal = "JCAP",
    volume = "07",
    pages = "059",
    year = "2018"
}

@ARTICLE{1979A&A....80..104N,
       author = {{Novikov}, I.~D. and {Polnarev}, A.~G. and {Starobinskii}, A.~A. and {Zeldovich}, Ia. B.},
        title = "{Primordial black holes}",
      journal = {\aap},
         year = 1979,
        month = nov,
       volume = {80},
       number = {1},
        pages = {104-109},
       adsurl = {https://ui.adsabs.harvard.edu/abs/1979A&A....80..104N}
}

@article{Escriva:2024lmm,
    author = "Escriv{\`a}, Albert and Yoo, Chul-Moon",
    title = "{Simulations of ellipsoidal primordial black hole formation}",
    eprint = "2410.03452",
    archivePrefix = "arXiv",
    primaryClass = "gr-qc",
    doi = "10.1103/PhysRevD.112.083518",
    journal = "Phys. Rev. D",
    volume = "112",
    number = "8",
    pages = "083518",
    year = "2025"
}

@article{Iovino:2024sgs,
    author = "Iovino, A. J. and Matarrese, S. and Perna, G. and Ricciardone, A. and Riotto, A.",
    title = "{How well do we know the scalar-induced gravitational waves?}",
    eprint = "2412.06764",
    archivePrefix = "arXiv",
    primaryClass = "astro-ph.CO",
    doi = "10.1016/j.physletb.2025.140039",
    journal = "Phys. Lett. B",
    volume = "872",
    pages = "140039",
    year = "2026"
}

@article{Escriva:2024aeo,
    author = "Escriv{\`a}, Albert and Yoo, Chul-Moon",
    title = "{Nonspherical effects on the mass function of primordial black holes}",
    eprint = "2410.03451",
    archivePrefix = "arXiv",
    primaryClass = "gr-qc",
    doi = "10.1103/4jbp-87wc",
    journal = "Phys. Rev. D",
    volume = "112",
    number = "8",
    pages = "L081304",
    year = "2025"
}

@article{Uehara:2024yyp,
    author = "Uehara, Koichiro and Escriv{\`a}, Albert and Harada, Tomohiro and Saito, Daiki and Yoo, Chul-Moon",
    title = "{Numerical simulation of type II primordial black hole formation}",
    eprint = "2401.06329",
    archivePrefix = "arXiv",
    primaryClass = "gr-qc",
    reportNumber = "RUP-24-1",
    doi = "10.1088/1475-7516/2025/01/003",
    journal = "JCAP",
    volume = "01",
    pages = "003",
    year = "2025"
}

@article{Ballesteros:2024zdp,
    author = "Ballesteros, Guillermo and Gamb{\'\i}n Egea, Jes{\'u}s",
    title = "{One-loop power spectrum in ultra slow-roll inflation and implications for primordial black hole dark matter}",
    eprint = "2404.07196",
    archivePrefix = "arXiv",
    primaryClass = "astro-ph.CO",
    doi = "10.1088/1475-7516/2024/07/052",
    journal = "JCAP",
    volume = "07",
    pages = "052",
    year = "2024"
}

@article{Motohashi:2017kbs,
    author = "Motohashi, Hayato and Hu, Wayne",
    title = "{Primordial Black Holes and Slow-Roll Violation}",
    eprint = "1706.06784",
    archivePrefix = "arXiv",
    primaryClass = "astro-ph.CO",
    doi = "10.1103/PhysRevD.96.063503",
    journal = "Phys. Rev. D",
    volume = "96",
    number = "6",
    pages = "063503",
    year = "2017"
}

@article{Ragavendra:2023ret,
    author = "Ragavendra, H. V. and Sriramkumar, L.",
    title = "{Observational Imprints of Enhanced Scalar Power on Small Scales in Ultra Slow Roll Inflation and Associated Non-Gaussianities}",
    eprint = "2301.08887",
    archivePrefix = "arXiv",
    primaryClass = "astro-ph.CO",
    doi = "10.3390/galaxies11010034",
    journal = "Galaxies",
    volume = "11",
    number = "1",
    pages = "34",
    year = "2023"
}

@inbook{Pi:2024lsu,
    author = "Pi, Shi",
    title = "{Non-Gaussianities and Primordial Black Holes}",
    eprint = "2404.06151",
    archivePrefix = "arXiv",
    primaryClass = "astro-ph.CO",
    doi = "10.1007/978-981-97-8887-3_7",
    year = "2025"
}

@article{Escriva:2023qnq,
    author = "Escriv{\`a}, Albert and Yoo, Chul-Moon",
    title = "{Primordial Black hole formation from overlapping cosmological fluctuations}",
    eprint = "2310.16482",
    archivePrefix = "arXiv",
    primaryClass = "gr-qc",
    doi = "10.1088/1475-7516/2024/04/048",
    journal = "JCAP",
    volume = "04",
    pages = "048",
    year = "2024"
}

@article{Shimada:2024eec,
    author = "Shimada, Masaaki and Escriv{\'a}, Albert and Saito, Daiki and Uehara, Koichiro and Yoo, Chul-Moon",
    title = "{Primordial black hole formation from type II fluctuations with primordial non-Gaussianity}",
    eprint = "2411.07648",
    archivePrefix = "arXiv",
    primaryClass = "gr-qc",
    doi = "10.1088/1475-7516/2025/02/018",
    journal = "JCAP",
    volume = "02",
    pages = "018",
    year = "2025"
}

@article{Inui:2024fgk,
    author = "Inui, Ryoto and Joana, Cristian and Motohashi, Hayato and Pi, Shi and Tada, Yuichiro and Yokoyama, Shuichiro",
    title = "{Primordial black holes and induced gravitational waves from logarithmic non-Gaussianity}",
    eprint = "2411.07647",
    archivePrefix = "arXiv",
    primaryClass = "astro-ph.CO",
    doi = "10.1088/1475-7516/2025/03/021",
    journal = "JCAP",
    volume = "03",
    pages = "021",
    year = "2025"
}

@ARTICLE{1992JETPL..55..489S,
       author = {{Starobinskij}, A.~A.},
        title = "{Spectrum of adiabatic perturbations in the universe when there are singularities in the inflationary potential.}",
      journal = {Soviet Journal of Experimental and Theoretical Physics Letters},
         year = 1992,
        month = may,
       volume = {55},
       number = {9},
        pages = {489-494},
       adsurl = {https://ui.adsabs.harvard.edu/abs/1992JETPL..55..489S}
}

@article{Fujita:2025imc,
    author = "Fujita, Tomohiro and Kawaguchi, Ryodai and Sasaki, Misao and Tada, Yuichiro",
    title = "{Dip and non-linearity in the curvature perturbation from inflation with a transient non-slow-roll stage}",
    eprint = "2503.19744",
    archivePrefix = "arXiv",
    primaryClass = "astro-ph.CO",
    reportNumber = "WUCG-25-03, YITP-25-21",
    doi = "10.1088/1475-7516/2025/09/046",
    journal = "JCAP",
    volume = "09",
    pages = "046",
    year = "2025"
}

\end{document}